\documentclass[aps,floatfix,pra,reprint,superscriptaddress,nofootinbib]{revtex4-2}

\usepackage{amsmath,amssymb,bm}
\usepackage{graphicx}
\usepackage{dcolumn}   
\usepackage{hyperref}  
\usepackage{braket}
\usepackage{comment}
\usepackage{cancel}
\hypersetup{colorlinks=true,allcolors=blue}

\begin{document}

\title{Elastic and Spin-Changing Cross Sections of Spin-Polarized Atomic Tritium}
\author{M. G. Elliott}
  \email{morgan.elliott@uta.edu} 
\author{B.J.P. Jones}
  \email{ben.jones@uta.edu}
  \affiliation{Department of Physics, University of Texas at Arlington, Arlington, TX 76019, USA}
  
\date{\today}

\begin{abstract}
    The rates of elastic and inelastic scattering processes of spin-polarized atomic tritium  are vital inputs for the design and operation of experiments using cold, magnetically trapped tritium atoms.  Elastic scattering dominates the total cross section and dictates the thermophysical properties of the vapor, determining the efficiency of magnetic evaporative cooling and the fluid dynamical properties of the trapped atom cloud.  Spin-changing cross sections in both exchange and dipolar channels determine the trap lifetime of the various hyperfine states, imposing constraints on the required atomic tritium supply rate to maintain a given trap density.  Motivated by the needs of next-generation spectrometers that will study the tritium beta endpoint to infer the mass of the neutrino, we  present new calculations of the elastic, spin-exchange and dipolar cross sections for spin polarized atomic tritium.  Cross sections and rate constants are reported for magnetic field strengths of 0–10 T and temperatures from 0-100~K in all relevant Zeeman-hyperfine channels for both tritium and hydrogen.  Results are bench-marked against past calculations for atomic hydrogen and the limited available results for tritium, and extend far into the regimes where past calculations have not been available.  
\end{abstract}

\maketitle
\section{Introduction}

Hydrogen-like atoms are among the simplest quantum systems that can be studied in the laboratory, each comprising a single electron bound to a nearly stationary nucleus in the center-of-mass frame.  Experimental and theoretical studies of hydrogen have accordingly laid many of the foundations for much of modern physics, with results such as the Balmer series~\cite{bohr1913constitution}, the hyperfine splitting~\cite{Ramsey1963}, and the Lamb shift~\cite{lamb1947fine,bethe1950numerical} producing crucial input to the early development and subsequent verification of quantum theory. More than 140 years after Balmer introduced his empirical formula for the hydrogen spectrum, precision studies of hydrogen and its isotopes are still contributing to our understanding of fundamental physics.  Comparisons of the quantum properties of hydrogen and anti-hydrogen are  being conducted~\cite{baker2025precision,alpha2020investigation} to test for charge–parity–time asymmetry~\cite{kostelecky2015lorentz} and determine the antiproton charge radius~\cite{crivelli2016antiproton}; the isotope shift between the energy levels of deuterium and hydrogen atoms provides a precision test of physics beyond the standard model~\cite{potvliege2023deuterium}; both muonium~\cite{jungmann2016precision,hughes2001test} and positronium~\cite{adkins2022precision} spectroscopy offer a wealth of opportunities to search for new physics; and magnetically trapped atomic tritium sources are being pursued in order to measure the absolute value of the mass of the neutrino~\cite{esfahani2022project,aker2020first}.

Systems comprising two hydrogen-like atoms (two electrons and two nuclei) are still amenable to first-principles analysis, but are considerably more complicated than their single atom counterparts.  As the simplest possible molecule, H$_2$ has been exploited as a laboratory to search for new physics through precision molecular spectroscopy~\cite{liu2023precision,schiller2022precision,lai2025precision}.  Much theoretical work has therefore been undertaken to precisely calculate the spectrum of excited states of the H$_2$ molecule, typically using  non-relativistic quantum mechanics with relativistic and quantum electrodynamic corrections~\cite{wolniewicz1993relativistic, lai2022precision,silkowski2023leading,puchalski2019nonadiabatic,salumbides2011qed,saly2023pre,komasa2011quantum,pachucki2014accurate}. Complementary studies focused on the molecular spectroscopy of both deuterium~\cite{Etters1975,veirs1987raman,wcislo2018accurate} and tritium  molecules~\cite{hollik2020study,lai2020precision,cozijn2024precision} provide opportunities to validate these predictions with a suite of isotopologues, offering a recipe to reduce atomic and molecular physics uncertainties by studying a collection of closely related four-body systems.

The radioactive beta instability of tritium, with an especially low Q-value of 18.592071(22)~keV~\cite{medina2023mass}, makes it a compelling source for precision nuclear physics.  Experiments that aim to measure the absolute mass of the neutrino~\cite{kraus2005final,belesev2013search}, for example, reconstruct the shape of the end-point of the emitted beta electron spectrum to infer the mass of the invisible recoil~\cite{formaggio2021direct}. The strongest constraint today derives from the KATRIN experiment, which uses a Magnetic Adiabatic Collimation combined with Electrostatic (MAC-E)
filter \cite{Aker:2021ty}, delivering the upper limit $m_{\beta}\leq0.45\,\mathrm{eV}$ at 90\% confidence level \cite{katrin2025direct}.  

The sensitivity of experiments using molecular tritium sources, such as that developed for KATRIN, will soon become limited by smearing of the spectral end-point due to molecular vibrational and rotational modes of the final-state T-$^3$He molecule~\cite{Bodine:2015aa,Schneidewind:2023xmj}.   To advance neutrino mass sensitivity to the ambitious 40~meV goal of the Project 8 experiment~\cite{esfahani2022project}, a tritium source based on atomic (T) rather than molecular (T$_2$) must be realized.  This presents a set of entangled technological challenges: atomic tritium must be  produced at high fluxes, cooled~\cite{esfahani2025dynamics},  trapped for suitably long times, and monitored with a highly precise electron spectroscopy method~\cite{Monreal:2009za} in order to reconstruct the shape of the  $\beta$ end-point with sub-eV precision.  This requires a deep understanding of the properties of trapped, hyper-polarized, mK-temperature atomic T, a novel material that has not yet been realized at scale in the laboratory.  Additional potential applications of magnetically-trapped tritium vapor include the production of tritium Bose-Einstein condensates~\cite{blume2002formation}, and a platform for quantum enhancement of neutrino emission through superradiance~\cite{jones2024superradiant}.  

Cold and spin-polarized T is a complex fluid with numerous internal degrees of freedom that support transitions between trappable and un-trappable states, all at rates that depend non-trivially on the applied magnetic field and temperature.   Much effort has been dedicated to cooling and manipulation of polarized atomic hydrogen~\cite{walraven2020atomic,silvera1982stabilization,hess1986evaporative,masuhara1988evaporative,fried1998bose}, driven significantly by the quest for Bose Einstein condensates of hydrogen atoms, a feat achieved by Greytak {\em et. al.} in Ref.~\cite{greytak2000bose}. The atomic tritium system bears some similarities to atomic hydrogen, though as we will discuss, the scattering properties of the two isotopes differ dramatically, in  ways that carry significant experimental implications.

To introduce the important scattering interactions in magnetically trapped tritium vapor, we briefly summarize some basic concepts.  The interaction between internal angular momenta of atoms and an applied magnetic field leads to a Zeeman shift in the atomic energy levels, as shown in Fig.~\ref{Zeeman_Levels_of_T}.  The interaction with the external magnetic field is overwhelmingly dominated by the electron spin rather than the nuclear spin due to the relative sizes of the Bohr and nuclear magnetons.  In accordance with Earnshaw's theorem~\cite{earnshaw1848nature},  trapped atoms in static magnetic fields must necessarily be spin polarized into low-field seeking states, as Maxwell's equations do not admit static local magnetic field maxima.  This means that in any trapped scattering interaction, the electrons will predominantly be polarized parallel to the magnetic field, in the so-called $c$ and $d$ states.  

Since the trapped states all have predominantly parallel electron spins, the spin wave function of any electron pair on any two atoms will typically occupy a triplet $|\uparrow \uparrow\rangle$ configuration, which is symmetrical under electron exchange. This in turn requires an antisymmetric spatial electron wave function, due to the required exchange symmetry for fermions. Unlike the singlet wave functions that support the molecular states of hydrogen and tritium, this triplet wave function does not produce any bound states for either species.   The behavior of spin-polarized tritium and hydrogen is thus restricted to scattering interactions in a mostly repulsive triplet-state potential~\cite{Silvera1986}.

Polarized atomic hydrogen or tritium collisions can be separated into elastic and inelastic channels.   Elastic scattering interactions such as $dd\rightarrow dd$ and $cc\rightarrow cc$  are those that preserve both the electron and nuclear spins between initial and final states.  These  dominate the total scattering cross section and therefore the thermodynamic behavior of the polarized gas, making them important inputs for the design of evaporative cooling methods~\cite{esfahani2025dynamics} and for understanding the fluid properties of magnetically trapped tritium and hydrogen vapor.   Inelastic scattering, on the other hand, leads to changes of the electron and nuclear spin.  Inelastic collisions can be further delineated into spin exchange and dipolar categories.   Spin exchange collisions such as $cc\rightarrow aa$ and $cc\rightarrow bd$ change the hyperfine states of each atom while preserving the total electron spin projection $m_s$, and are relatively fast because they are mediated by the triplet potential itself.  Dipolar interactions, which are required for processes that change $m_s$ of the atom pair including $dd\rightarrow aa$ and $dd\rightarrow ad$, emerge from the much weaker magnetic dipole interaction that couples orbital and spin angular momenta. Both types of inelastic interactions drive trapped states into un-trapped ones, with dipolar collisions dictating the trap lifetime of $d$  states and spin-exchange interaction dictating the trap lifetime of $c$ states. Because spin exchange interactions are so much faster than dipolar interactions, the $d$ states are expected to have the longest trappable lifetime and be the primary species of study in magnetic trapping experiments.

In this paper we present calculations of the cross sections for the elastic, spin exchange and dipolar cross sections of atomic tritium, over a range of magnetic fields from 0 to 10~T and temperatures from 10$^{-3}$~K to 100~K.  Our results advance beyond those previously presented in the literature in several important ways. The elastic scattering cross section of spin-polarized tritium has been reported by several authors~\cite{blume2002formation, Stwalley2004, Al-Maaitah2012,sandouqa2018thermophysical}.  We follow the method outlined in Ref.~\cite{o2000optical} and find results that are broadly consistent with past works.  The wave functions generated in the elastic scattering analysis provide basis states for the Distorted Wave Born Approximation (DWBA) that we then use to calculate inelastic cross sections.  Our inelastic calculations are indebted to the seminal analysis of Stoof {\em et. al.} of atomic hydrogen~\cite{Stoof1987}, which evaluated scattering rates using both coupled-channel and DWBA perturbative formalisms and found consistency between the two approaches.  We benchmark our calculations against their results for atomic hydrogen, finding strong agreement.   Results in some inelastic tritium scattering channels have been also been obtained at magnetic field $B=0$ by Zygelman in Ref~\cite{zygelman2010electronic}. The formalism used therein is not amenable to applied magnetic fields, but the results are an additional point of comparison for our predictions in the zero field limit.  The results derived here are, to our knowledge, the first comprehensive set of spin-polarized tritium scattering cross sections spanning the full magnetic field and temperature ranges required to design and operate high-flux, cold atomic tritium experiments.  

The remainder of this paper will be structured as follows. Section~\ref{sec:Hamiltonian} introduces the spin basis and the effective Hamiltonian that will be used in our calculations. Section~\ref{sec:Potentials} discusses the triplet and singlet potentials that are used as input to the scattering calculation.  Section~\ref{sec:Elastic} presents the elastic scattering cross section calculations, and  Section~\ref{sec:Inelastic} presents the inelastic results.  Section~\ref{sec:Systematics} discusses the systematic uncertainties of our calculation. Finally, Section~\ref{sec:Conclusions}  summarizes our conclusions.  

This paper is accompanied by a comprehensive set of tables of both cross sections and rate constants for all relevant scattering channels, as well as an open source Python code for reproducing and extending these results.

\section{Spin Bases and Hamiltonian~\label{sec:Hamiltonian}}

The T-T system is formed from two spin-half tritons with individual masses of 3.016 amu and two spin-half electrons.  Under either Adiabatic or Born Oppenheimer approximations, the center-of-mass wave function is written in expanded form
\begin{equation}
    \Psi(\mathbf{r},\mathbf{R},t)=\sum_k\psi_k^{el}(\mathbf{r},\mathbf{R})\chi_k(\mathbf{R},t).
\end{equation}
Here, $\bf{r}$ is a composite variable encoding all of the electron coordinates, $\mathbf{R}$ is the separation vector between nuclei, and $\psi_k^{el}$ is the wave function of the $k$'th state of the electrons for nuclei at fixed separation $\mathbf{R}$.  $\chi_k(\mathbf{R})$ is an expansion coefficient that depends on $\mathbf{R}$, which can also be interpreted as the wave function for nuclear motion.   In the limit of large nuclear mass, the electronic and nuclear dynamics become separable~\cite{stanke2017adiabatic}.  With the two nuclei clamped at separation $\mathbf{R}=R\hat{z}$, the Schrodinger equation for electron-only motion can be solved to obtain a set of stationary states with energies $U_k(R)$, with distinct solutions for symmetric (spin-singlet) and antisymmetric (spin-triplet) spatial electron wave functions.  This $R$-dependent energy sources the potential for nuclear motion, per the Born approximation:
\begin{equation}
    \left[-\frac{\hbar^2}{2\mu}\nabla_R^2+U_k(R)\right] \chi(R,t)=i\hbar \frac{d}{dt}\chi(R,t),\label{eq:SchrodNucl}
\end{equation}
with $\mu$ the reduced mass of the two nuclei. The adiabatic approximation is similar, but allows for mass-dependent corrections to be added to $U_k(R)$ to account for first-order coupling between electron and nuclear motions,
\begin{equation}
    \left[-\frac{\hbar^2}{2\mu}\nabla_R^2+U_k(R)+\frac{1}{\mu}c_k(R)\right] \chi_k(R,t)=i\hbar \frac{d}{dt}\chi_k(R,t).\label{eq:SchrodNucl}
\end{equation}
These adiabatic corrections are functions of $R$, and  scale universally as $1/\mu$, allowing them to be applied by scaling a single mass-independent function $c_k(R)$ that depends on the shape of the electron wave function~\cite{pachucki2014accurate}.

For attractive potentials that support bound states,  Eq.~\ref{eq:SchrodNucl} can be solved to find stationary negative energy solutions corresponding to molecules.  It also admits positive-energy solutions that represent unbound systems, and Eq.~\ref{eq:SchrodNucl} is the starting point for scattering calculations for pairs of atoms in either singlet or triplet potentials.

\begin{figure}
    \centering
    \includegraphics[width=0.99\linewidth]{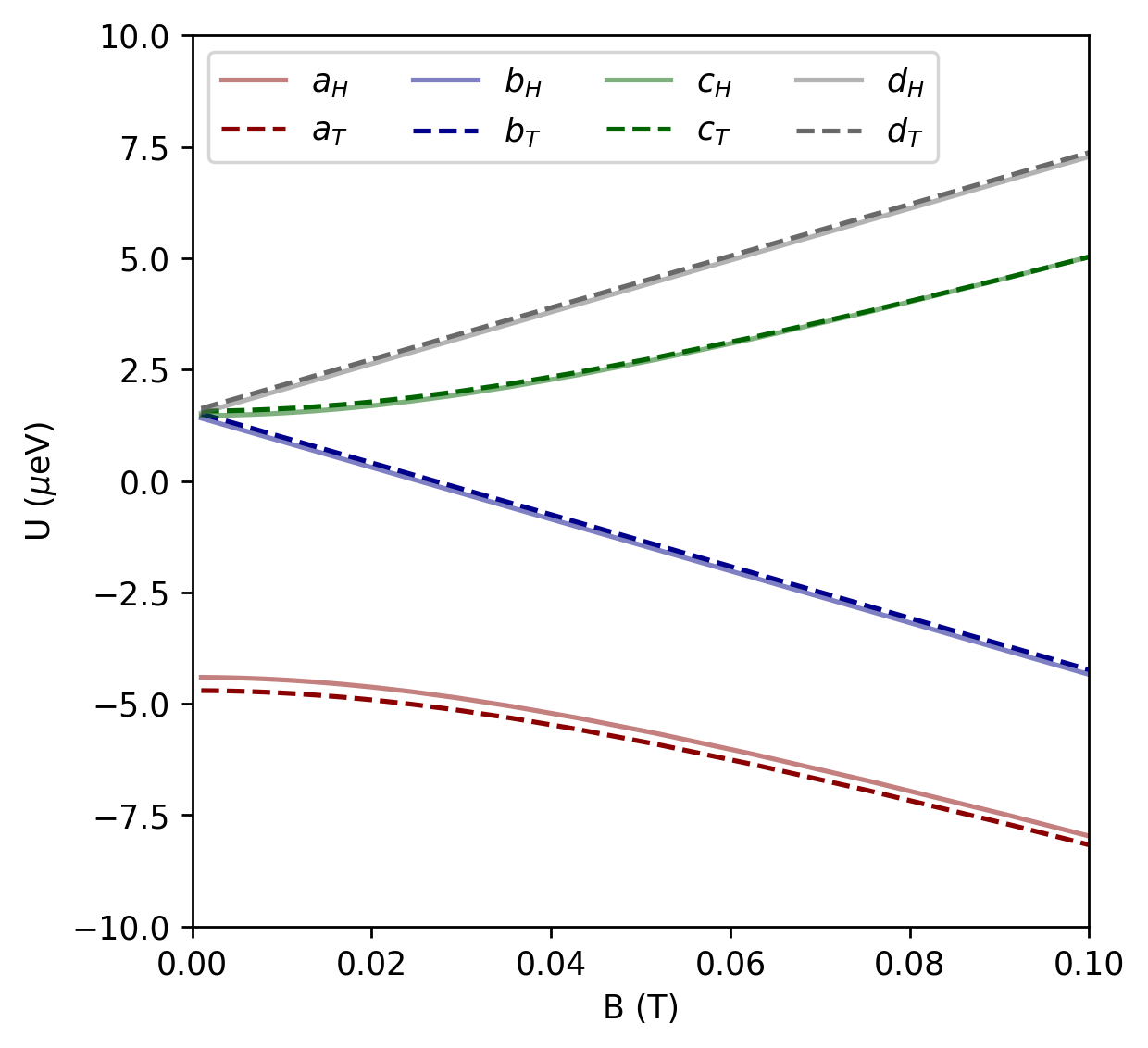}
    \caption{The Zeeman-hyperfine levels of atomic H and T, calculated using the Breit Rabi method.  The four single-atom eigenstates are labeled following the convention $a$,$b$,$c$,$d$ from lowest to highest energy in an applied B field. The $d$ states are the primary states of interest to atomic trapping, since they are low-field-seeking and stable against spin exchange collisions.}
    \label{Zeeman_Levels_of_T}
\end{figure}

Additional terms are required in the Hamiltonian to account for spin effects.  The effective masses of the four allowed [electron, nucleus] spin combinations within a given atom are displaced by the Zeeman and hyperfine effects, with the  four basis states of increasing energy conventionally labeled (in hydrogen) as $a$,$b$,$c$,$d$.   The relevant Hamiltonian for atom $(i)$ is given by

\begin{equation}
    \boldsymbol{H}_{int}^{(i)}=H^{(i)}_{hf}+H^{(i)}_{Zeeman}=\frac{a}{4}\boldsymbol{\sigma}^{(i)}_e\boldsymbol{\sigma}^{(i)}_p+\mu_e\boldsymbol{B}\boldsymbol{\sigma}^{(i)}_e-\mu_p\boldsymbol{B}\boldsymbol{\sigma}^{(i)}_p,
\end{equation}
where $a$= 1420 GHz (hydrogen) and 1517 GHz (tritium) and $a$ is the zero-field hyperfine constant. \(\mu_i\) represent either the electron or nuclear magnetic moments, and \(\sigma\) are the  Pauli spin matrices.  Diagonalizing this ``Breit–Rabi'' Hamiltonian yields the four eigenstates shown in Fig.~\ref{Zeeman_Levels_of_T}. These can be decomposed in terms of pure electron and nuclear basis states as
\begin{eqnarray}
\ket{a}  &=& \cos(\theta)\ket{\downarrow \cancel\uparrow}-\sin(\theta)\ket{\uparrow\cancel\downarrow},\nonumber\\
\ket{b}  &=& \ket{\downarrow \cancel\downarrow},\nonumber\\
\ket{c} &=& \cos(\theta)\ket{\uparrow \cancel\downarrow}+\sin(\theta)\ket{\downarrow\cancel\uparrow},\nonumber\\
\ket{d}  &=& \ket{\uparrow\cancel\uparrow},
\end{eqnarray}
where an un-slashed arrow represents an electron spin and a slashed one represents a nuclear spin.  The mixing angle $\theta$ is given by
\begin{equation}
    \tan(2\theta)=a/[2B(\mu_e+\mu_p)]\label{eq:Theta}.
\end{equation}
The set of ``good quantum numbers'' to describe this system are different in the high and low B-field limits. At $B=0$ the state energies are determined solely by the hyperfine interaction, and the three upper states are degenerate. As such, in this limit the hyperfine quantum numbers $|F,m_f\rangle$ are most appropriate, and these map to the $a$,$b$,$c$,$d$ basis as
\begin{eqnarray}
\lim_{B\rightarrow 0}\ket{a} &=& \ket{F=0,m_F=0} \nonumber\\
\lim_{B\rightarrow 0}\ket{b} &= & \ket{F=1,m_F=-1}\nonumber \\
\lim_{B\rightarrow 0}\ket{c} &= & \ket{F=1,m_F=0}\nonumber \\
\lim_{B\rightarrow 0}\ket{d} &= & \ket{F=1,m_F=1} 
\end{eqnarray}
At high magnetic fields (B $\gtrsim$ 0.5~T) the Zeeman interaction between the electron magnetic moment and the magnetic field dominates the Hamiltonian and the Breit Rabi states become pure spin states,
\begin{eqnarray}
\lim_{B\rightarrow \infty}\ket{a} &\rightarrow & \ket{\downarrow \cancel\uparrow}\nonumber \\
\lim_{B\rightarrow \infty}\ket{b} &\rightarrow & \ket{\downarrow \cancel\downarrow} \nonumber\\
\lim_{B\rightarrow \infty}\ket{c} &\rightarrow & \ket{\uparrow \cancel\downarrow}\nonumber \\
\lim_{B\rightarrow \infty}\ket{d} &\rightarrow & \ket{\uparrow\cancel\uparrow}
\end{eqnarray}
The magnetic field ranges of experimental interest are in the intermediate regime in between these two extremes. 
At any field strength, the four states can be divided into ``low-field-seeking'' states $c,d$ whose energy is reduced at lower magnetic field, and ``high-field-seeking'' ones which accelerate toward magnetic field maxima, $a$, and $b$.  The interactions among low-field-seeking states $c$ and $d$ are of most interest to experiments with magnetic traps, because the high-field-seeking atoms will quickly exit the experiment, leaving only low-field-seekers behind. 

Working now in the letter state $|a\rangle,|b\rangle,|c\rangle,|d\rangle$ basis, to evaluate the forces between the atoms the central part of the interaction described by Eq.~\ref{eq:SchrodNucl}  must be projected onto the relevant singlet and triplet sub-spaces of the electron wave functions.  We introduce the ``central potential'' operator,
\begin{eqnarray}
\boldsymbol{V}_{central}&=&V_S(R)\mathcal{P}_S+ V_T(R)\mathcal{P}_T,\label{eq:VCentral}\\ V_X(R)&=&\left[U_X(R)+\frac{1}{\mu}c_X(R)\right],
\end{eqnarray}
where $X=\{S,T\}$, with $U_T$ and $U_S$ the triplet and singlet potentials for clamped nuclei, $c_T$ and $c_S$ the triplet and singlet adiabatic corrections, and $\mathcal{P}_S$ and $\mathcal{P}_T$ are operators that project pairs of hyperfine levels onto the electron singlet and triplet basis states.  These operators can be written conveniently in terms of  the $i$'th electron spin $\boldsymbol{s}_i$  as
\begin{equation}
    \mathcal{P}_T=\frac{1}{4}(1+4\boldsymbol{s}_i\cdot\boldsymbol{s}_j), \quad  \quad \mathcal{P}_T=\frac{1}{4}(3-4\boldsymbol{s}_i\cdot\boldsymbol{s}_j).
\end{equation}
Arbitrary pairs of  hyperfine basis states are not eigenstates of the singlet and triplet projection operators, and so the central potential can lead to spin-changing as well as elastic scattering interactions.  This operator can be re-written as 
\begin{equation}
    V_{central}=\frac{1}{2} (V_T-V_S) (\boldsymbol{s_i\cdot s_j}),
\end{equation}
which is manifestly invariant under independent rotations of the spatial and spin co-ordinates. As such, it is clear that interactions mediated by the central potential do not change the total spin or orbital angular momentum, imposing selection rule $\Delta S=\Delta m_s=\Delta l=\Delta m=0$ (total spin, z-spin, total orbital angular momentum and z-orbital angular momentum, respectively).  

The remaining set of spin-changing transitions are driven by magnetic dipole interactions.   The dipole-dipole Hamiltonian couples the electron magnetic moments, taking the form
\begin{eqnarray}
    \boldsymbol{V}_{dipole}=-\frac{\mu_0\mu_e^2}{r^3}\sqrt{\frac{1}{20\pi }} 
 \sum_{\gamma=-2}^2(-1)^\gamma Y_{2,-\gamma}(\boldsymbol{\hat{R}})\Sigma^{ee}_{2\gamma} \label{eq:dipoleV}.
\end{eqnarray}
Here, $Y$ are spherical harmonics and $\Sigma^{ee}_{2\gamma}$ is a rank 2 tensor operator acting on electron spins that can be derived from the dipole-dipole interaction $3(\boldsymbol{s}_1\cdot\hat{\boldsymbol{R}})(\boldsymbol{s}_2\cdot\hat{\boldsymbol{R}})-(\boldsymbol{s}_1\cdot\boldsymbol{s}_2)$. Dipolar interactions between identical bosons are subject to the following selection rules:
\begin{eqnarray}
    \Delta l &=& 0, \pm 2,\nonumber \\
    \Delta m_F &=& 0,\pm 1,\pm 2.\label{eq:selection}
\end{eqnarray}
Each value of $\gamma$ in Eq.~\ref{eq:dipoleV} drives transitions with $\gamma=\Delta m_F$. The dipolar interaction allows $dd$ initial states to convert to high-field-seeking spin states that are no longer trapped, through converting electron spin angular momentum into orbital angular momentum.  Additional terms in the Hamiltonian coupling electron to nuclear and nuclear to nuclear dipole moments are suppressed by the ratio of the electron to the nuclear magneton and are neglected in our calculations due to being sufficiently small.  The total Hamiltonian is thus
\begin{equation}
\boldsymbol{H}=-\frac{\hbar^2}{2\mu}\nabla_R^2+\sum_{i=1}^2\boldsymbol{H}_{int}^{(i)} + \boldsymbol{V}_{central}+\boldsymbol{V}_{dipole}.
\end{equation}

When spin exchange and dipolar processes convert atom pairs between hyperfine levels of different effective masses, the energy released from the mass difference between initial and final states is converted into linear momentum in order to conserve energy. The final center-of-mass momentum $p'$ carried by hyperfine states $\alpha',\beta'$ is related to the initial momentum $p$ carried in  hyperfine states $\alpha,\beta$  via
\begin{equation}
     \frac{p'^2_{\alpha'\beta'}}{2\mu}+\epsilon_\alpha'+\epsilon_\beta' = \frac{p^2_{\alpha\beta}}{2\mu}+\epsilon_\alpha+\epsilon_\beta,
\end{equation}
with \(\epsilon_\zeta\) describing the magnetic-field dependent internal energy associated with the hyperfine level $\zeta$.

\begin{figure}[t]
    \centering
    \includegraphics[width=0.99\linewidth]{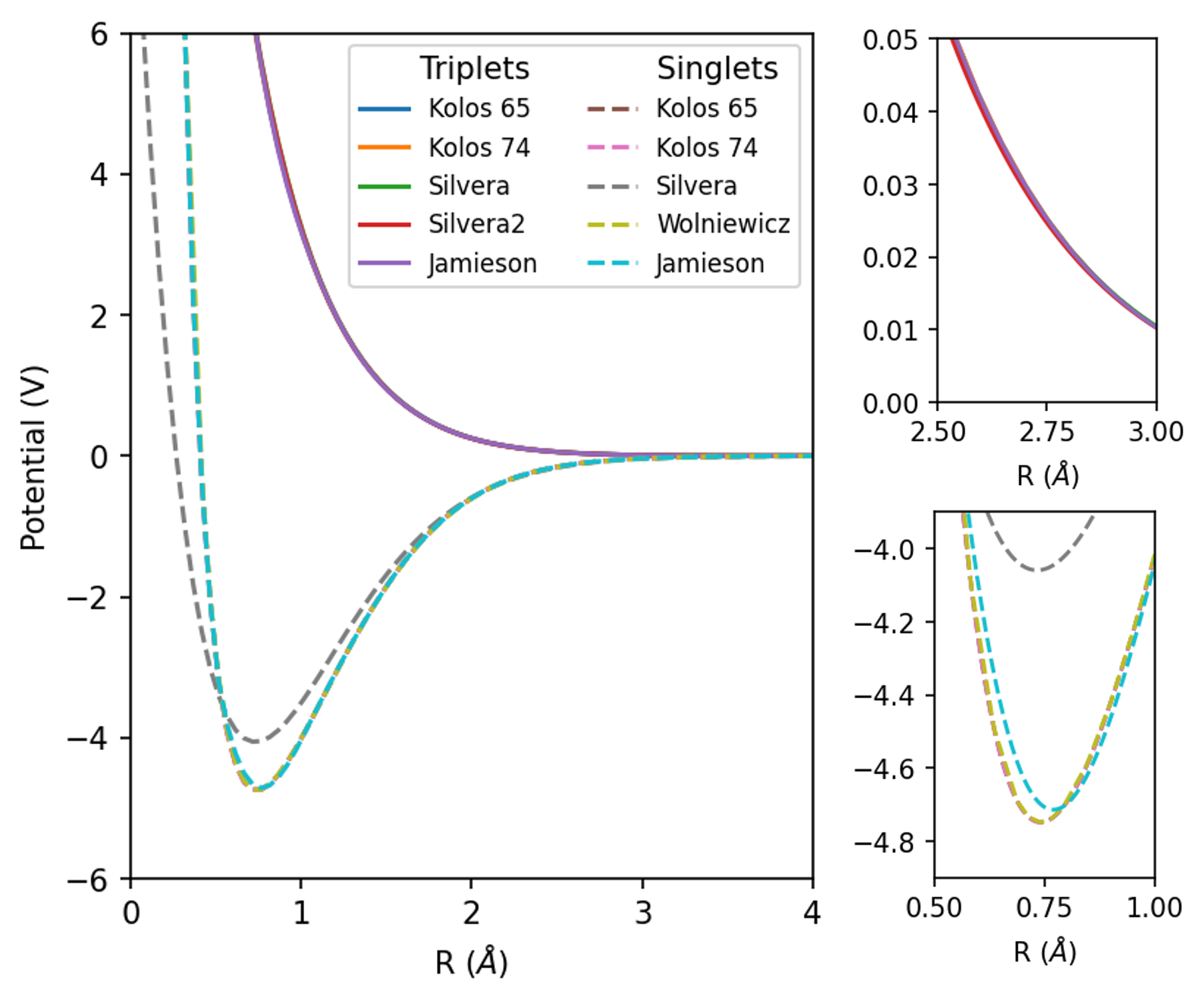}
    \caption{The singlet and triplet potentials used in this work, drawn from Refs.~ ~\cite{kol1965potential,kolos1974variational,Silvera1986,Jamieson1999,wolniewicz1993relativistic}. The right panels shown zoomed-in segments of the left panel to better illustrate the level of agreement between the various calculations.}
    \label{fig:Potentials}
\end{figure}

\section{Potentials\label{sec:Potentials}}

To obtain the quantities of interest,  we must specify the triplet and singlet interaction potentials $V_T$ and $V_S$.  Both singlet and triplet potentials for atomic hydrogen were calculated in a series of important early papers by Kolos and collaborators~\cite{kolos1963nonadiabatic,kol1965potential,kolos1974variational,Kolos1975}. Adiabatic corrections to the singlet and triplet states were obtained in Refs~\cite{kol1964accurate} and~\cite{kolos1990adiabatic}, respectively.    Wolniewicz calculated the corrections to the energy from relativistic effects of the electron motion, as reported in Ref.~\cite{wolniewicz1993relativistic} for the singlet state.  These calculatons were improved and extended by Jamieson, Dalgarno and Wolniewicz in 2000~\cite{jamieson2000calculation}, including both adiabatic and relativistic corrections for both singlet and triplet states for $R$ from 1 to 20 Bohr.  The Jamieson {\em et. al.} potentials can be considered to supersede the earlier Kolos and Wolniewicz versions.  We follow the convention used by others in reporting results derived from this suite of calculations as ``Born Oppenheimer'' potentials, though it is worthy of mention that they are adiabatically corrected, so do not strictly follow the Born Oppenheimer approximation. For distances below 1 Bohr where the most modern papers do not provide coverage, we patch the Jamieson potential onto the Kolos 1974 potential, which report values over somewhat shorter distances.

Silvera \cite{Silvera1980} developed a parameterized form of the Kolos triplet potential which has been a workhorse for atomic triplet scattering calculations in hydrogen and tritium~\cite{haugen1989spin,Al-Maaitah2012,friend1980dilute}.   Silvera adds to the fitted repulsive short-distance potential a term that accounts for long distance attractive behaviour using an approach introduced by Ahlrichs {\em et. al.}~\cite{ahlrichs1977intermolecular} via a Hartree-Fock-Dispersion (HFD) functional form,
\begin{eqnarray}
    V_{silvera} =& A&\exp\left(a-b r-c r^2\right) \nonumber \\
    &+&f(R)\left(-\frac{C_6}{R^6}-\frac{C_8}{R^8}-\frac{C_{10}}{R^{10}}\right).\label{eq:Silvera}
\end{eqnarray}
Constants $C_n$ are obtained by Deal in Ref~\cite{deal1972long} and the remaining constants are fitted to the Kolos potential.  The function $f(R)$ smoothly switches the long range part of the potential on at long distances. Two forms reported by Silvera in different publications differ in their prescription for $f(R)$, with Ref~\cite{silvera1978isotropic} introducing
\begin{equation}
f(R)=\left\{ \begin{array}{cc}
\exp\left[-\left(1.28\frac{R_{min}}{R}-1\right)^{2}\right] & \quad R<1.28R_{min}\\
1 & \quad R>1.28R_{min}\label{eq:SilveraSwitch}
\end{array}\right.
\end{equation}with $R_{min}$ being the position of the triplet potential minimum, $R_{min}=4.16\AA$; and Ref~\cite{Silvera1980} presenting
\begin{equation}
f_1(R)=\exp \left[-\left(1.28 \frac{R_{min}}{R}-1\right)^2\right]    
\end{equation}
Most works that use the Silvera potential have adopted the former switching function~\cite{friend1980dilute,Al-Maaitah2012}, including subsequent publications from Silvera {\em et. al}~\cite{Silvera1986}.  As such, for the Silvera triplet potential we have used Eq.~\ref{eq:Silvera}  with switching function Eq.~\ref{eq:SilveraSwitch} and constants as given in Ref.~\cite{Silvera1986}.  In Ref.~\cite{Silvera1980} Silvera also provides an approximate analytic form for the singlet potential.  It is cautioned that this potential is not sufficiently accurate for quantitative calculations, and it notably differs significantly from the data points of Kolos~{\em et. al.} that it is based on. As such, we have not considered the Silvera singlet as a reliable singlet potential choice in this work.

For all of the Born Oppenheimer potentials, we have followed Silvera by extrapolating all of the tabulated data to $R\rightarrow\infty$ using a HFD form, with constants from Deal.  To test for sensitivity to this assumption,  comparative calculations were also made assuming a pure Van Der Waals long-distance potential, proportional to $R^{-6}$. This leads to changes in the zero-energy triplet scattering length of up to 0.6\% with gradually reducing effect as the energy increases, and is included as a source of systematic uncertainty in our uncertainty budget.  We conclude that our results are not strongly influenced by the choice of potential extrapolation function at long distances.

The suite of potentials discussed above  are shown in Figure~\ref{fig:Potentials}. The two smaller panels on the right show zoomed-in regions of the left panel to indicate the spread of the datasets.   For many of the results in this paper we will report only Silvera and / or Jamieson / Born Oppenheimer results, with the other older potentials serving as a tool to assess the uncertainty in the calculations.

\begin{figure}[t]
    \centering
    \includegraphics[width=0.99\linewidth]{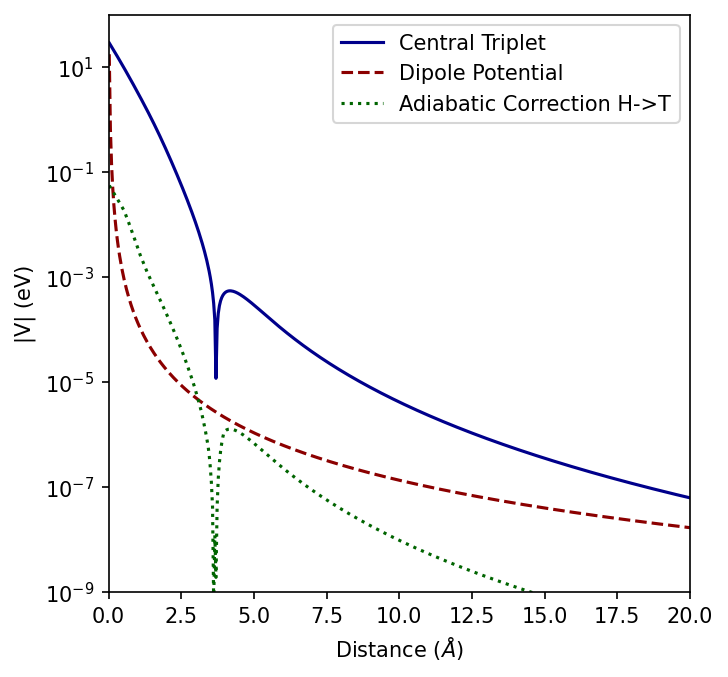}
    \caption{Comparison of the triplet potential, its adiabatic correction and the spatial part of the dipole potential. The dipolar effects are treated as a first-order perturbation in scattering calculations, whereas the adiabatic term is added as a mass-dependent diagonal term to the triplet potential.}
    \label{fig:AdiabaticCorrection}
\end{figure}

Several of the potentials  above are supplied with mass-dependent adiabatic corrections already applied for hydrogen, and the relevant publications do not independently report these corrections.  To adiabatically correct those potentials for T-T scattering we therefore subtract the hydrogen correction $c(R)$ that was reported in Ref.~\cite{kolos1990adiabatic}, and re-add the mass-scaled correction for tritium. Since the adiabatic correction affects the scattering cross sections at only the few-percent level, this protocol appears sufficiently accurate for present purposes.  The $H\rightarrow T$ adiabatic correction is shown for comparison with the triplet central potential and dipole potential in Fig~\ref{fig:AdiabaticCorrection}.

\section{Elastic Cross Section\label{sec:Elastic}}
A system of two tritium $d$ states is in a pure electron triplet $|\uparrow\uparrow \rangle $, and as such the calculation in the  important elastic channel $dd\rightarrow dd$ reduces to  potential scattering in the spherical triplet potential.  Because the internal energy in the initial and final states is always equal in elastic scattering, the cross sections are independent of applied magnetic field, so a calculation at $B=0$ suffices for all fields. 
\begin{figure}[t]
    \centering
    \includegraphics[width=0.99\linewidth]{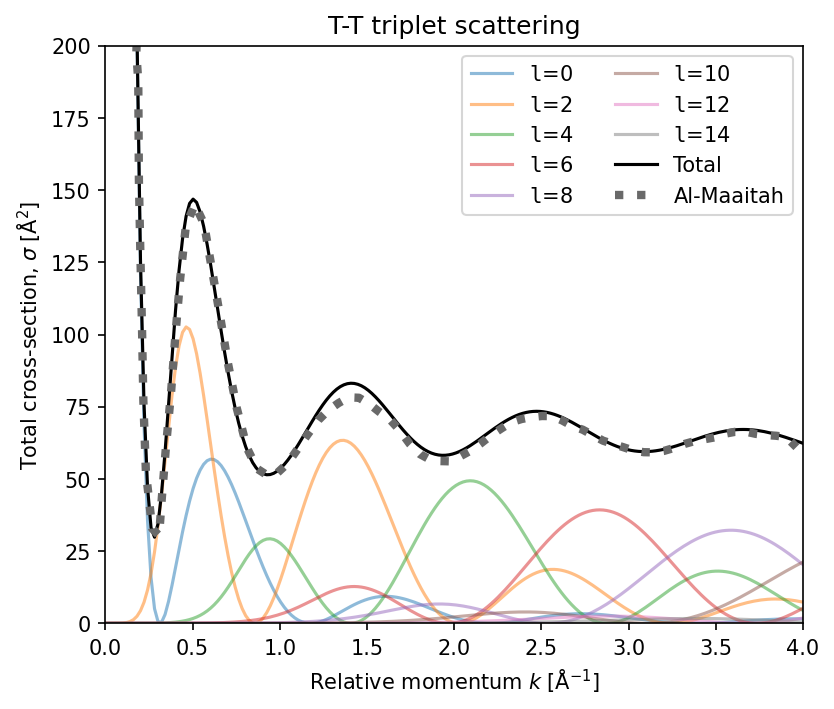}\\
    \includegraphics[width=0.99\linewidth]{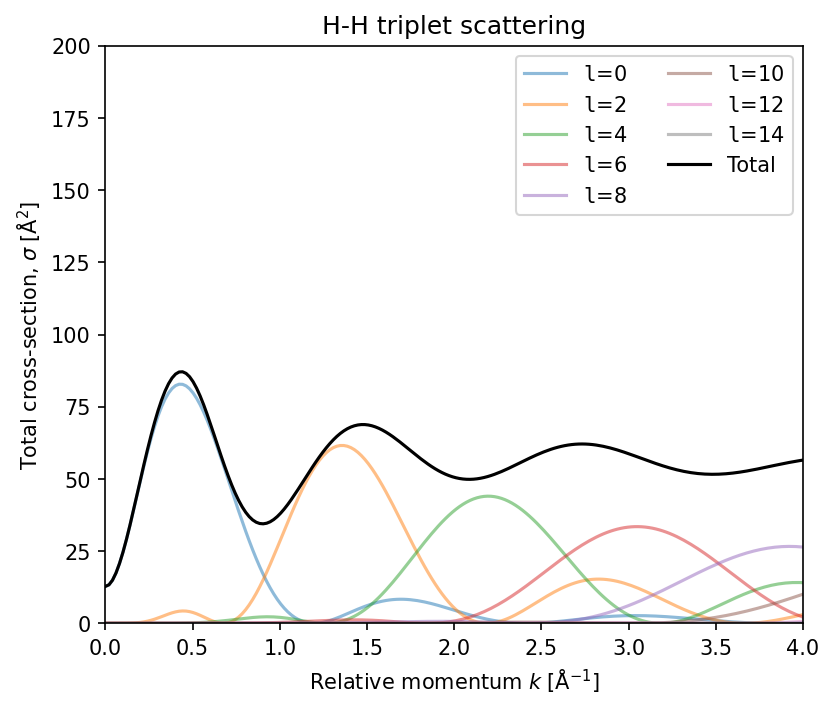}
    \caption{Calculation of the elastic scattering cross sections for T-T (top) and H-H (bottom) collisions as a function of energy. The total scattering cross section is shown, along with its decomposition into various partial waves.  These cross sections are independent of magnetic field.  Also shown for comparison is the calculation of ~\cite{Al-Maaitah2012}, with agreement at the few percent level. Triplet scattering  in T-T becomes very large at low momentum transfer, eventually approaching the scattering length shown in Table~\ref{tab:ScatLengths}.}
    \label{fig:Elastic}
\end{figure}
Defining the $z$ axis based on the separation between the two particles, only $m=0$ terms appear in the wave function, and we write
\begin{equation}
    \chi(\theta,\phi,R)=Y_l^0(\theta,\phi)\frac{u_{El}(R)}{R}.
\end{equation}
Following standard methods~\cite{weinberg2015lectures}, we solve the radial Schrodinger equation with the centripetal term for finite $l$ to find $u_{El}$,
\begin{equation}
    \left\{\frac{d}{dR^2}+\frac{2\mu}{\hbar}\left[E-V(R)-\frac{l(l+1)\hbar^2}{2\mu R^2}\right]\right\}u_{El}(R)=0.\label{eq:SchrodingerRadial}
\end{equation}
This equation is solved numerically from small $R$ to a large radius $\rho$ where both the potential and centripetal terms become negligible~\cite{o2000optical}.  Beyond $\rho$, the wave functions match accurately to free-particle solutions with energy $E=\hbar^2k^2/2m$ for wave vector $k$, and the large distance behavior is described entirely in terms of phase shifts $\delta_l$ of the relevant spherical waves, as
\begin{eqnarray}
    \lim_{R\to \infty} \psi(\boldsymbol{R})&=& e^{i\boldsymbol{kz}}+ \frac{e^{ikR}}{R}f(\theta,\phi),\\
        f(\theta) &=& \sum_{l=0} (2l+1)\frac{e^{(2i\delta_l-1)}}{2ik}P_l(\cos(\theta)).
\end{eqnarray}
We will use $u_{El}$ later in the calculation in spin-changing rates, but for present purposes we require only the scattering phase shift of the $l$'th partial wave $\delta_l(k)$, which can be extracted from $u_{El}$ as
\begin{eqnarray}
\delta_l(k)=\arctan\left[\frac{kj_l'(k\rho)-\Delta_l(k)j_l(k\rho)}{kn_l'(k\rho)-\Delta_l(k)n_l(k\rho)}\right]\nonumber\\
\quad \mathrm{with}\quad \Delta_l(k) = \frac{\rho \,u_{El}'(\rho )-u_{El}(\rho)}{\rho\, u_{El}(\rho)},
\end{eqnarray}
where $n_l$ and $j_l$ are the spherical Bessel functions.  Finally, the scattering cross section can be calculated as a sum over independent partial waves as
\begin{equation}
    \sigma=\int d\Omega|f(\theta,\phi)|^2  = \frac{8\pi}{k^2}\sum_{l(even)}(2l+1)\sin^2(\delta_l).\label{eq:DifferentialECrossSection}
\end{equation}

Symmetrization of the wave function for identical bosons requires that only even $l$ partial waves contribute, and at low energy (below 100~mK) scattering is dominated by the s-wave $l=0$ channel.  The zero-energy limit of the phase shift can be used to extract the s-wave scattering length $a_0$,  \begin{equation}
    a_0 = -\lim_{k\to0}\frac{\tan(\delta_0(k))}{k},
\end{equation} 
which determines the low energy limit of the elastic scattering cross section,
\begin{equation}
    \lim_{k\rightarrow0}\sigma=8\pi a_0^2.\label{eq:lowECrossSection}
\end{equation} 
The factor of 8 in Eq.~\ref{eq:lowECrossSection} and Eq. ~\ref{eq:DifferentialECrossSection} reflects that the scattering process is for two identical bosons.  

The triplet s-wave scattering lengths that we obtain for H and T are presented for comparison against past works in Table~\ref{tab:ScatLengths}, and the energy-dependent scattering lengths are shown in Fig.~\ref{fig:Elastic}.  The adiabatic correction to the potential has around a 1\% impact on the H scattering lengths and 2-3\% on the T scattering lengths.  

\begin{figure}[t]
    \centering
    \includegraphics[width=0.99\linewidth]{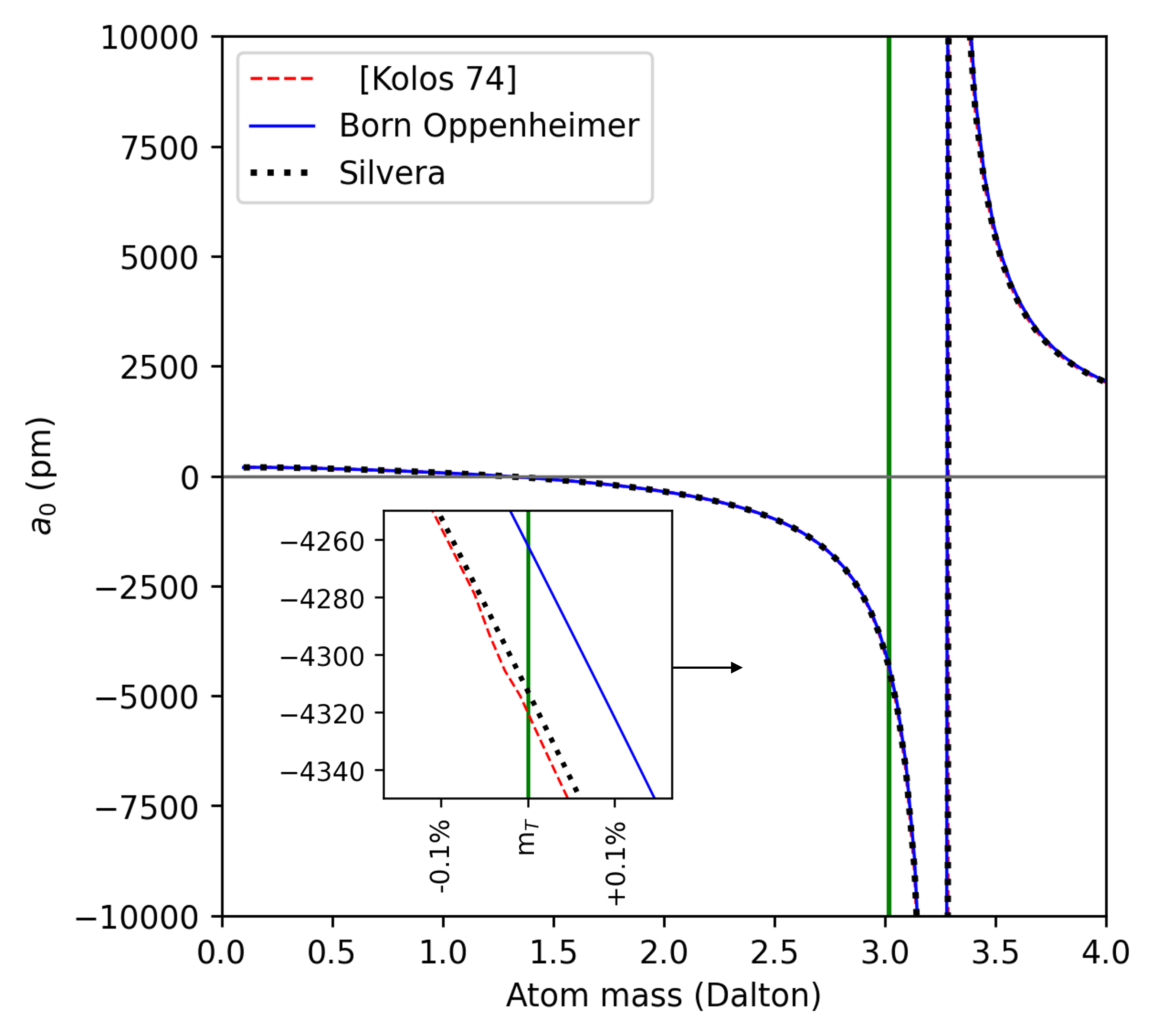}
    \caption{The mass dependence of the calculated triplet scattering length, accounting for both the effect of reduced mass in the radial Schrodinger equation and the mass-dependent adiabatic correction.  The pole near m=3.3 Dalton indicates the presence of a bound state that is just above threshold for T-T scattering, leading to a large negative triplet scattering length for T-T collisions.  The inset shows a zoom into the region around the tritium mass $m_T$ to illustrate the strong sensitivity of the cross section to the atom mass (and similarly the potential shape and other inputs to the calculation) in this region.}
    \label{fig:MassDep}
\end{figure}

As has been noted by others~\cite{Al-Maaitah2012}, the low energy cross section for T-T scattering is vastly larger than that for H-H scattering.  This is because of a slightly above-threshold triplet bound state that happens to appear near atom mass $m\sim3$. This can be seen in Figure~\ref{fig:MassDep}, where the calculation is repeated for various hypothetical values of the atom mass.  The pole at $~\sim$3.3 indicates that for this specific mass, a new triplet bound state would be on resonance at zero energy.  This near-threshold state is responsible for the large negative scattering length for T-T elastic interactions.  The resultant steepness of the scattering length near the tritium mass $m=m_T$ also makes the low energy spin-polarized tritium cross sections rather more sensitive to details of the potentials than the equivalent processes in hydrogen.  As such we have taken care to account for theoretical uncertainties to the extent possible, in Section~\ref{sec:Systematics}.  

\begin{table}[h]
\begin{tabular}{|l|c|c|}
\hline
Reference & $a_0^T$ (pm) & $a_0^H$ (pm) \\
\hline
{\bf This Work (Born Oppenheimer)}    & {\bf -4310} & {\bf  70.35}  \\ {\bf This Work (Silvera)}     & {\bf -4263} & {\bf 71.33}  \\
\hspace{0.15cm}     { \em This Work (Kolos '74)}     & -4321 & 71.00  \\

Sandouqa (Born Oppenheimer) ~\cite{sandouqa2018thermophysical}& -4370 &         \\
Sandouqa (Silvera) ~\cite{sandouqa2018thermophysical}         & -3820 &         \\
Joudeh (Born Oppenheimer)~\cite{Joudeh2013}    &       & 70.5    \\
Joudeh (Silvera)~\cite{Joudeh2013}             &       & 72.1    \\
Stwalley ~\cite{Stwalley2004}                   & -4314 & 70.4    \\
Markic~\cite{markic2007quantum}                      &       & 69.7    \\
Blume~\cite{blume2002formation}                       & -4344 & 70.4    \\
Al Maaitah~\cite{Al-Maaitah2012}                  & -3635  &         \\
Friend and Etters~\cite{friend1980dilute}           &       & 72    \\
\hline
\end{tabular}
\caption{Calculated scattering length from this work and from other references, for both T-T (left) and H-H (right) triplet elastic scattering.\label{tab:ScatLengths}}
\end{table}

\section{Inelastic cross sections\label{sec:Inelastic}}
Due to the difference in magnitude between the dipolar and central potentials evident in Fig.~\ref{fig:AdiabaticCorrection}, it is appropriate to treat the dipolar scattering contributions as perturbations to the Hamiltonian and apply first-order perturbation theory to calculate dipolar scattering rates. Stoof {\em et. al.} compare this perturbative approach to a non-perturbative coupled channels calculation in Ref.~\cite{Stoof1987} and validate the accuracy of the method for hydrogen at the 5\% level.

The rate constant $G$ for scattering process $\alpha\beta\rightarrow\alpha'\beta'$ is given by
\begin{equation}
    G_{\alpha'\beta'\to \alpha\beta} = (4\pi^3\hbar^2)\mu p_{\alpha'\beta'}|T_{\alpha\beta\to \alpha'\beta'}|^2,\label{eq:G}
\end{equation}
where $T$ is the transition operator, which can be obtained via the distorted-wave Born approximation~\cite{weinberg2015lectures}, as 
\begin{equation}
    T_{\alpha\beta\to \alpha'\beta'} = \braket{\chi_{DW;\alpha'\beta'}^-|V_{dipole}|\chi_{DW;\alpha\beta}^+}+\mathcal{O}(V_{dipole}^2).\label{eq:DWBA}
\end{equation}
The wave functions $\chi^\pm_{DW,\alpha\beta}$ are the incoming and outgoing distorted wave that are solutions to the Schrodinger equation with the central potential, Eq.~\ref{eq:SchrodingerRadial}.  We have suppressed additional indices for economy of notation, but it is noted that $T_{\alpha\beta\to \alpha'\beta'}$ also depends on the linear momentum and orbital angular momenta of the initial and final states.

After substituting the dipole potential and the distorted spherical wave solutions into Eq.~\ref{eq:G} the following expression for the rate constant $G$ is obtained,
\begin{eqnarray}
G_{\alpha\beta\to\alpha'\beta'}=\frac{\hbar^2\mu\mu_0^2\mu_e^4}{5\pi p_{\alpha'\beta'}p^2_{\alpha\beta}}\left[\int dr \frac{u_{E',l'}^*(r)u_{E,l}(r)}{r^3}\right]^2\\
\times 2\sqrt\pi {\mathcal{G}\left(\begin{array}{ccc}
l & l' & 2\\
m & -m' & \gamma
\end{array}\right)}\nonumber
\\
    \times\,\left|\braket{\{\alpha'\beta'\}|\mathcal{P}_T\Sigma_{2,\gamma}\mathcal{P}_T|\{\alpha\beta\}}\right|^2.\nonumber~\label{eq:GFactor}
\end{eqnarray}
The sequence of $\mu$'s above are the reduced mass ($\mu$), the magnetic permeability of free space ($\mu_0$) and the electron magnetic dipole moment ($\mu_e$). The expression is factorized into a spin part and a spatial part.  The spatial part includes a one-dimensional integral that can be evaluated numerically, and the normalization convention used for the radial wave functions $u_{El}$ is such that they match to real sine waves of unit amplitude as $R\rightarrow\infty$. $\mathcal{G}$ is the Gaunt coefficient,
\begin{equation}
\mathcal{G}\left(\begin{array}{ccc}
l & l' & 2\\
m & -m' & \gamma
\end{array}\right)=\int d\Omega \,Y_{lm}(\theta,\phi)Y^*_{l'm'}(\theta,\phi)Y_{2\gamma }(\theta,\phi),
\end{equation}
which gives rise to the selection rules of Eq.~\ref{eq:selection}. 

\begin{figure}
    \centering
    \includegraphics[width=0.99\linewidth]{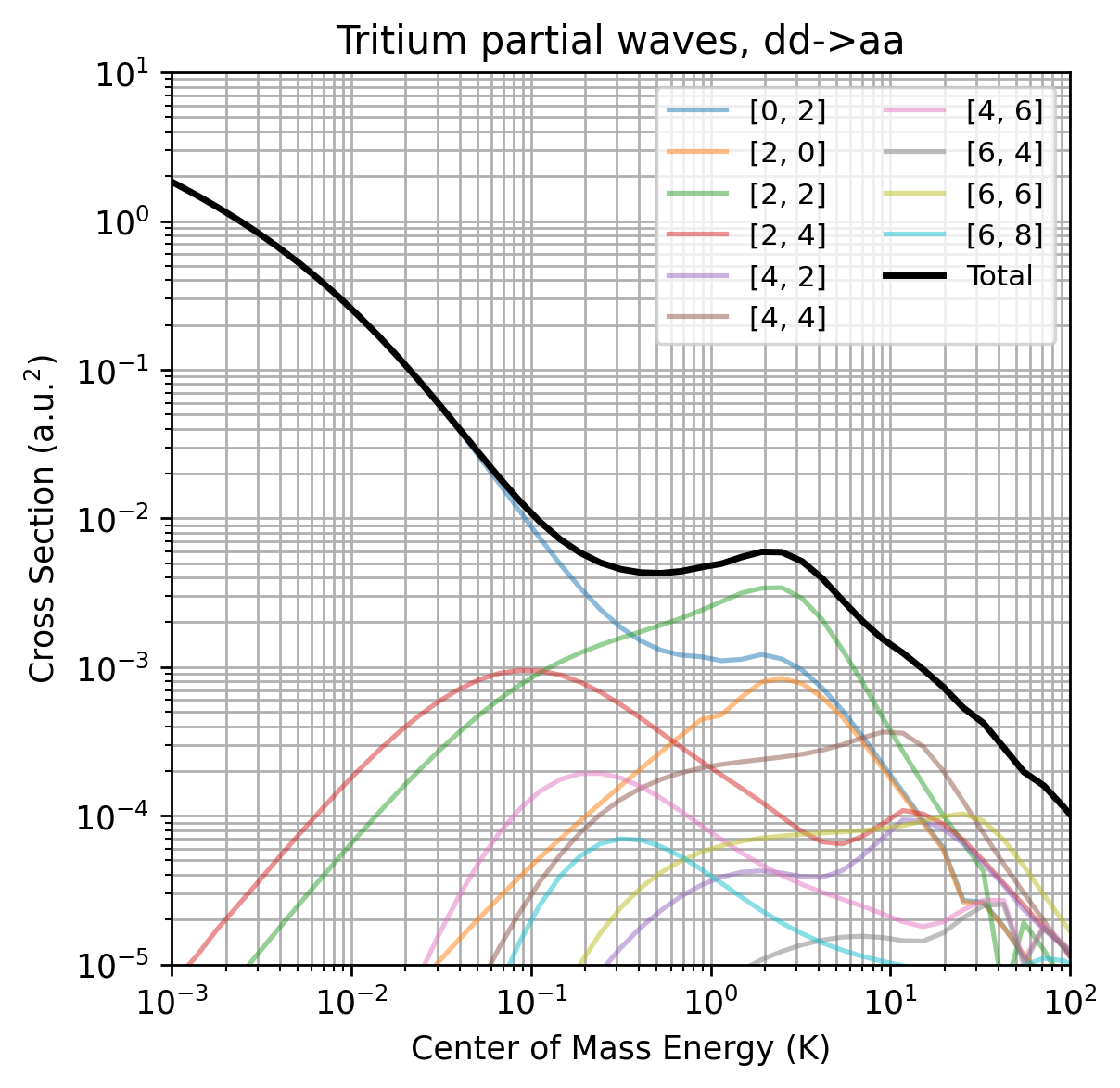}
    \caption{Illustration of the energy-dependent dipolar scattering cross section as a function of temperature, decomposed into its partial wave contributions. The various lines are labeled via initial and final state orbital angular momentum $l$ for each contribution.  Only even $l$ partial waves are included, due to the symmetry requirements of the initial state. }
    \label{TTPartialDipole}
\end{figure}

\begin{figure}[t]
    \centering
    \includegraphics[width=0.99\linewidth]{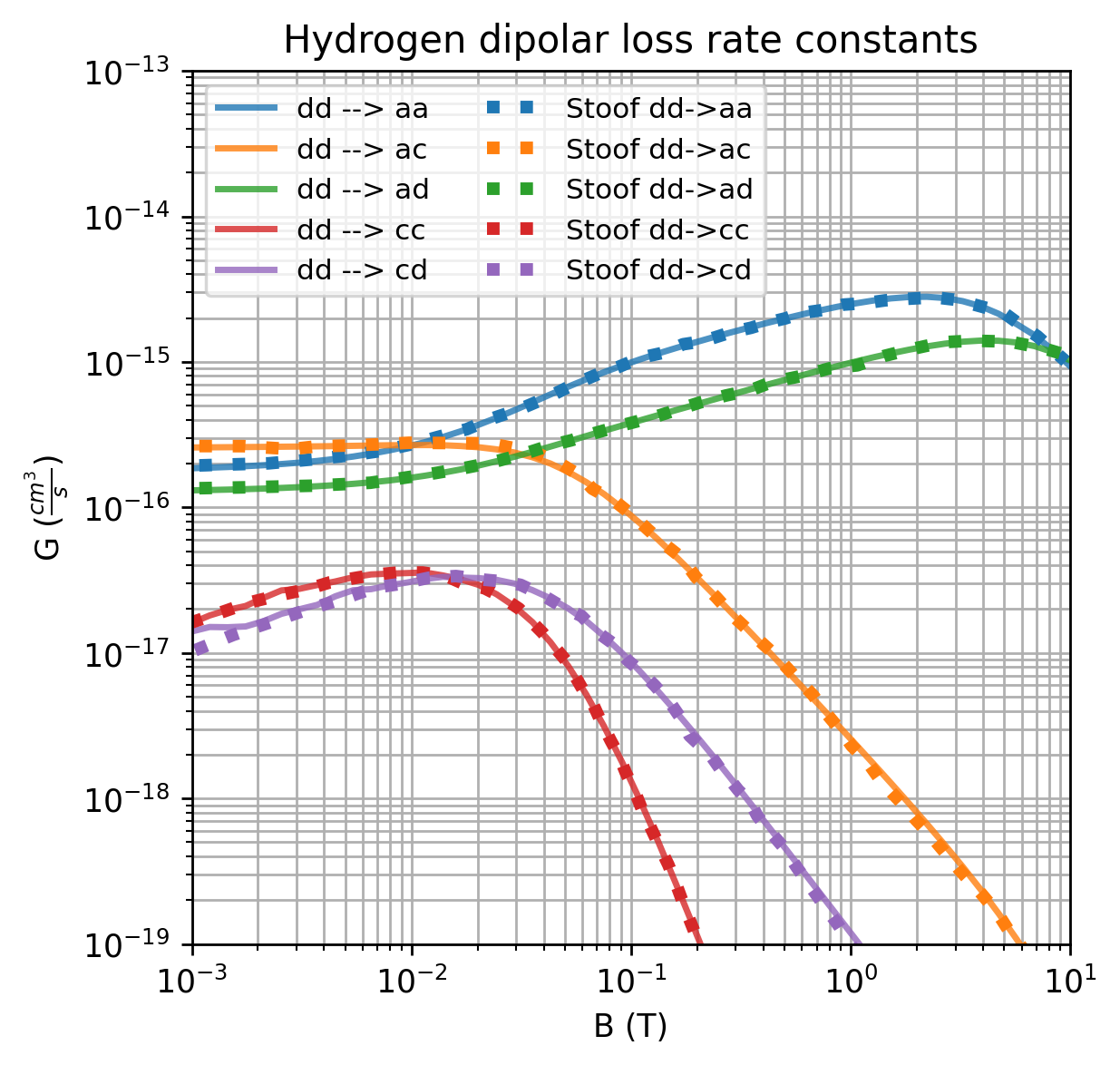}\\
    \includegraphics[width=0.99\linewidth]{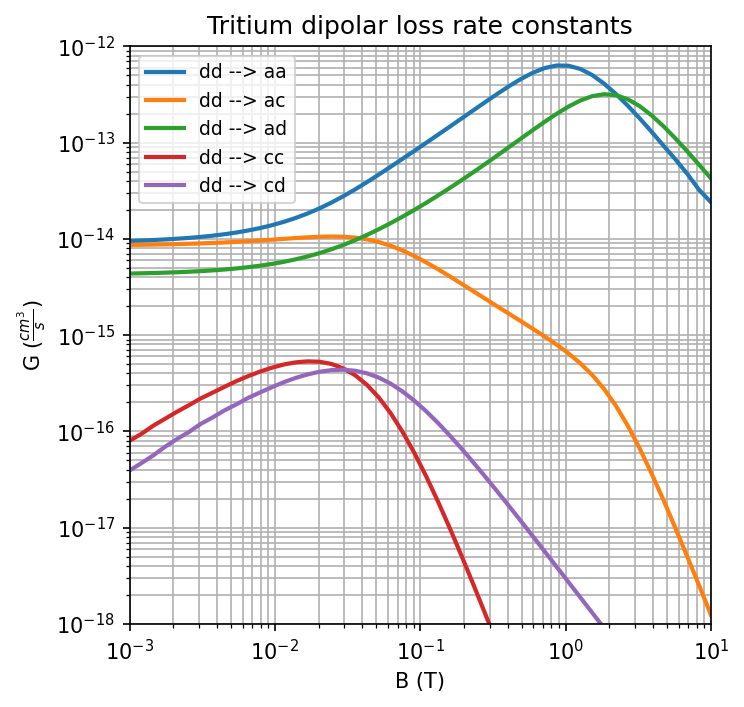}
    \caption{Calculated rate constants for dipolar interactions in H-H (top) and T-T (bottom) collisions. This calculation is performed using the Silvera triplet potential. The H-H rates are bench-marked against results reported in Ref.~\cite{Stoof1987}, showing strong agreement. The T-T rates are enhanced relative to the H-H rates due to a larger distortion of the incoming waves by the central potential.}
    \label{fig:DipolesRateVsB}
\end{figure}
The spin part involves the triplet projection operators $\mathcal{P}_T$ and the dipole tensor operator $\Sigma$ for the relevant hyperfine channel. We evaluate these terms by first expressing the $\Sigma$ operator in the $|\tilde a\rangle,|\tilde b\rangle,|\tilde c\rangle,|\tilde d\rangle$ basis, where the tilde indicates that the states are defined at the $\theta=0$ limit (see Eq.~\ref{eq:Theta}).  In this basis, the spin eigenstates and Zeeman-hyperfine eigenstates are equivalent, so the $\Sigma$ and $\mathcal{P}_{S,T}$ operators have a simple form. A basis rotation is then applied to the operator to take it to the $|a\rangle,|b\rangle,| c\rangle,|d\rangle$ basis in order to evaluate the spin matrix elements of Eq.~\ref{eq:GFactor}. The curly brackets indicate the use of symmetrized wave functions, defined via
\begin{equation}
    \ket{\{\alpha\beta\}}=\frac{\ket{\alpha\beta}+(-1)^l\ket{\beta\alpha}}{\sqrt{2(1+\delta_{\alpha\beta})}}.
\end{equation}

A related expression to Eq.~\ref{eq:GFactor} appears as Eq.~40 of Ref.~\cite{Stoof1987}, which notably includes terms with spin singlet projectors as well as the triplet-triplet contribution.  There it is reported that the spin singlet terms are of vanishing significance compared to the triplet contributions in H-H dipolar scattering.  This hierachy is even  stronger in T-T scattering where there are additional large triplet enhancements, and as such we can safely neglect the S=0 contributions to Eq.~\ref{eq:GFactor}.

The predicted rates depend on magnetic field, temperature, isotope and hyperfine channel. The important dipolar spin-changing channels for trapped H and T experiments are $dd\rightarrow aa$, $dd\rightarrow ac$, $dd\rightarrow ad$, $dd\rightarrow cc$ and $dd\rightarrow cd$.  Ref.~\cite{Stoof1987} presented $T=0$ rates for H-H scattering as a function of magnetic field from 10$^{-3}$ to 10~T. Our results compare favorably to these predictions, which presents an important point of validation, as shown in Fig.~\ref{fig:DipolesRateVsB}, top.  The second panel of Fig.~\ref{fig:DipolesRateVsB} shows the corresponding rates for T-T scattering. We note that the T-T rates are universally higher than the H-H rates,  by a factor of approximately 50.  This increase is due largely to the enhancement caused by triplet-state wave distortion, which enters the formalism via Eq.~\ref{eq:DWBA}.

\begin{figure}
    \centering
    \includegraphics[width=0.99\linewidth]{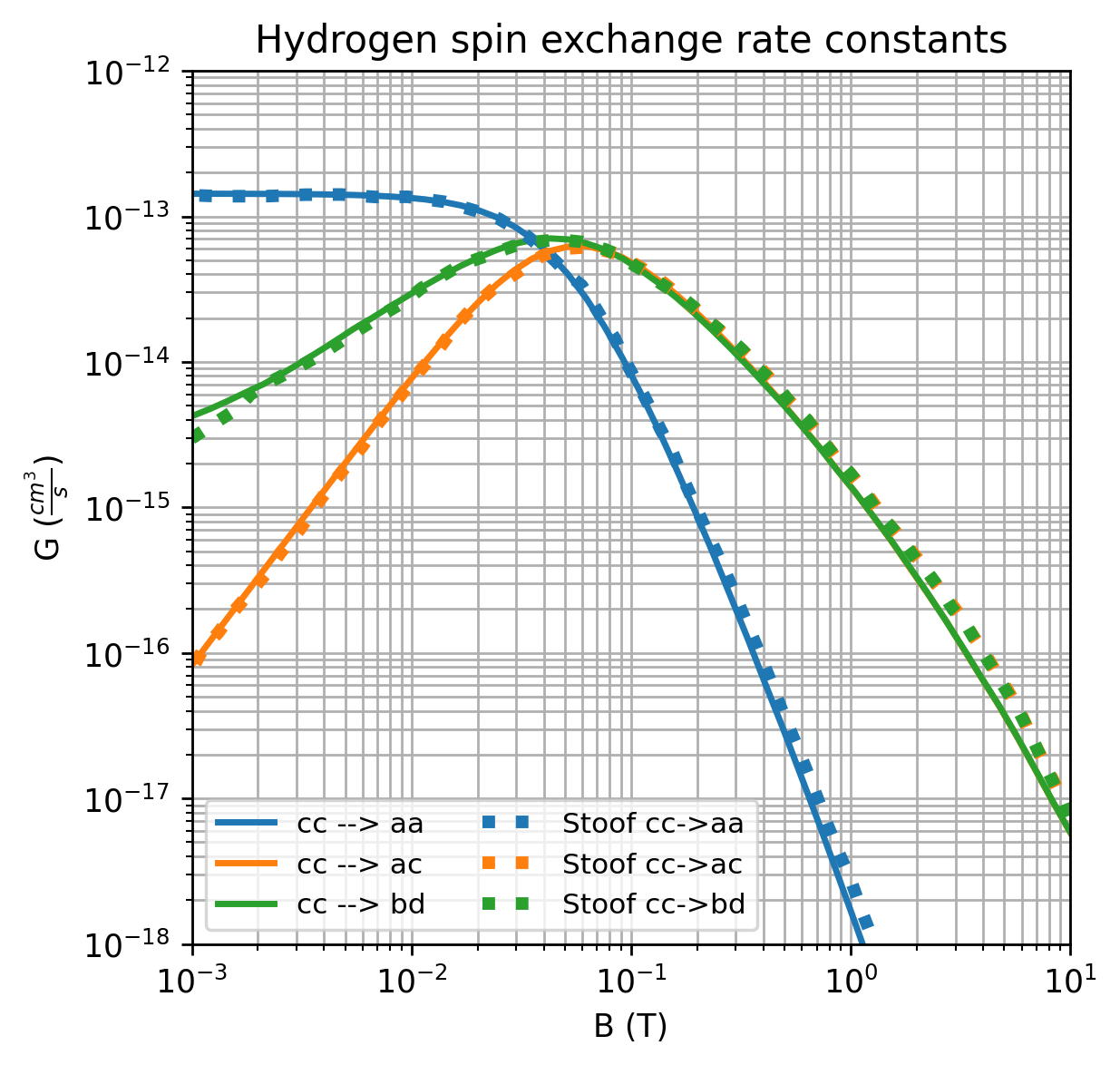}\\
    \includegraphics[width=0.99\linewidth]{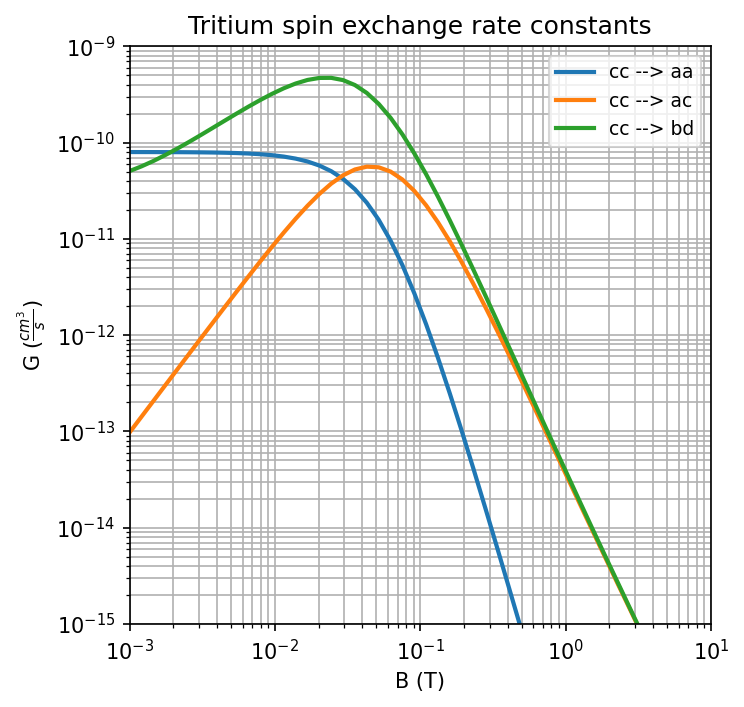}
    \caption{Calculated rate constants for spin exchange interactions in H-H (top) and T-T (bottom) collisions. This calculation is performed using the Silvera triplet potential and the Jamieson singlet. The H-H rates are bench-marked against results reported in Ref.~\cite{Stoof1987}, showing strong agreement. The T-T rates are enhanced since spin-exchange interactions are directly mediated by the hyperfine non-diagonal terms of the central potential.}
    \label{fig:SpinExchange}
\end{figure}

Spin exchange processes are mediated by non-diagonal hyperfine matrix elements of the central potential, rather than the dipole operator.  The relevant transition matrix element is therefore 
\begin{equation}
    T_{\alpha\beta\to \alpha'\beta'} = \braket{\chi_{DW;\alpha'\beta'}^-|V_{central}|\chi_{DW;\alpha\beta}^+}+\mathcal{O}(V_{dipole}^2).\label{eq:DWBA2}
\end{equation}
The spatial parts of this matrix element are greatly simplified by their expression in terms of the singlet and triplet phase shifts $\delta^{S}_l$ and $\delta^{T}_l$  
\begin{eqnarray}
\braket{\chi_{DW;\alpha'\beta'}^-|V_T|\chi_{DW;\alpha\beta}^+}=e^{2i\delta^{T}_l},\\ \braket{\chi_{DW;\alpha'\beta'}^-|V_S|\chi_{DW;\alpha\beta}^+}=e^{2i\delta^{S}_l}.\nonumber
\end{eqnarray}
After some simple manipulation, the following rate constant is obtained, as also reported in Refs.~\cite{koelman1987spin,Stoof1987},
\begin{eqnarray}    G_{\alpha\beta\to\alpha'\beta'}&=&\frac{\pi\hbar^2}{\mu p_{\alpha\beta}}\sum_l(2l+1)\left[\frac{p_{\alpha'\beta'}p_{\alpha\beta}}{p^2}\right]^{2l+1}\\ &\times&\sin^2\left(\delta_l^{T}(p)-\delta_l^{S}(p)\right)\nonumber\\ 
&\times& \left|\braket{\{\alpha'\beta'\}|\mathcal{P}_T-\mathcal{P}_S|\{\alpha\beta\}}\right|^2.\nonumber
\end{eqnarray}
This expression makes use of the degenerate-internal-states approximation, which ascribes a single representative momentum scale $p$ for evaluation of the phase shifts. Ref.~\cite{Stoof1987} prescribes using the average momentum in the channel for highly accurate results,
\begin{equation}
    p^2=p_i^2+\mu (\epsilon_\alpha+\epsilon_\beta-\epsilon'_\alpha-\epsilon_\beta').
\end{equation}

\begin{figure}[t]
    \centering
    \includegraphics[width=0.99\linewidth]{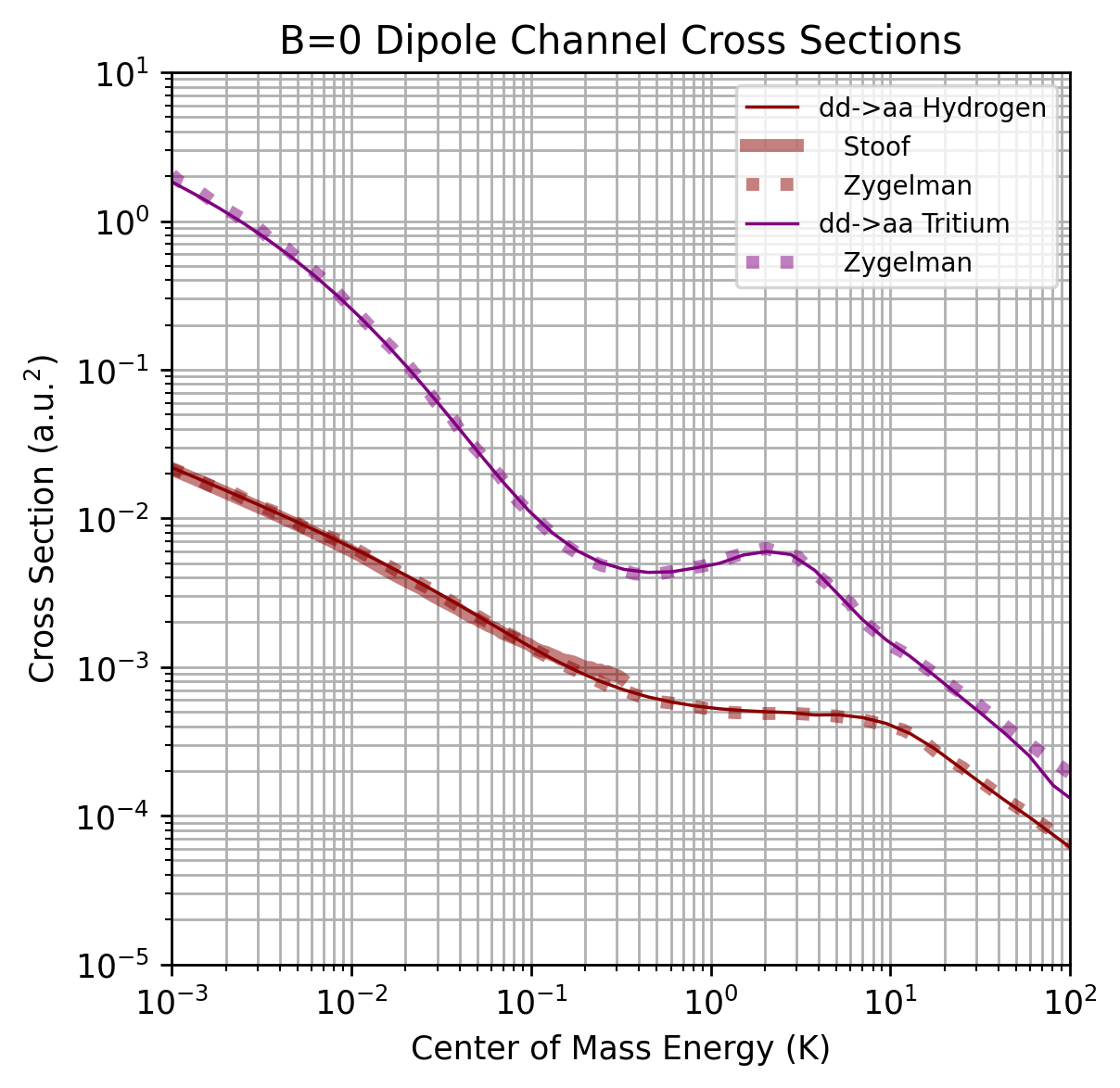}\\
    \includegraphics[width=0.99\linewidth]{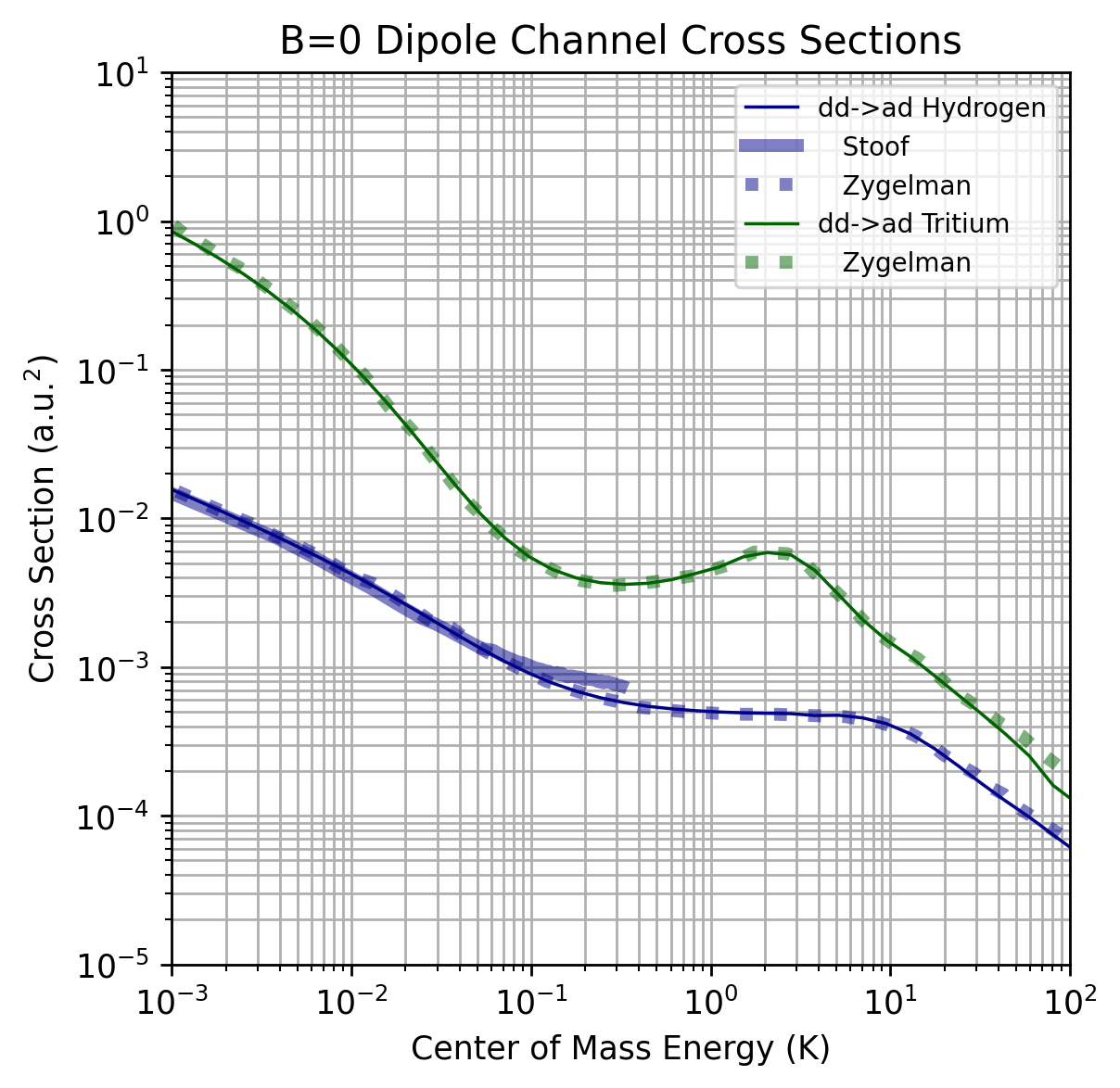}
    \caption{Comparison of energy-dependent cross section predictions to those reported in the literature for channels $dd\rightarrow aa$ and $dd\rightarrow ad $.  The thin lines show our predictions, the solid thick lines are the calculations of Ref.~\cite{Stoof1987}, and the dotted lines of Ref.~\cite{zygelman2010electronic}.  We find excellent agreement with Ref ~\cite{Stoof1987} in all cases.  Both our results and those of  Ref ~\cite{Stoof1987} are consistent with the predictions of Ref.~\cite{zygelman2010electronic} for the $dd\rightarrow ad$ channel, but discrepant by a constant factor of almost exactly two for $dd\rightarrow aa$.
    \label{fig:TempDependent}}
\end{figure}

The most relevant spin exchange channels for tritium experiments are those of the relatively short lived low-field seeking $c$ states, of which the dominant channels are $cc\rightarrow aa$, $cc\rightarrow ac$ and $cc\rightarrow bd$. Fig.~\ref{fig:SpinExchange}, top shows the predictions from our work compared to those of Ref.~\cite{Stoof1987} for hydrogen in the zero temperature limit, over magnetic fields ranging from 10$^{-3}$ to 10 T.  Accurate agreement is observed.  Fig~\ref{fig:SpinExchange}, bottom shows our predictions for tritium spin exchange processes. As with the dipolar loss rates of $d$ atoms, the spin exchange loss rate of $c$ atoms is also significantly larger in T-T scattering than in H-H scattering, driven substantially by the enhanced triplet s-wave phase shift.  The enhancement in the spin exchange channels is as large as 10$^3$ for some parameter points, suggesting that $c$ states are expected to have an especially short lifetime in the magnetically trapped tritium system.

To calculate rates at higher temperatures, we must evaluate the spatial matrix elements at finite incident momentum, and also incorporate higher partial wave contributions.  An example calculation including the higher partial waves is provided for the largest channel, $dd\rightarrow aa$ at zero magnetic field in Fig.~\ref{TTPartialDipole}, with the rates evaluated at the mean collision energy  in the center of mass frame, which is related to the temperature via $E_{com}=\frac{2}{\pi}k_BT$.  The work of Ref.~\cite{zygelman2010electronic} provides a point of comparison for this channel at $B=0$ in both H-H and T-T. To facilitate comparison to the results of Ref.~\cite{zygelman2010electronic} we convert the calculated rate constant $G$ into an effective cross section $\sigma$ via
\begin{equation}
\sigma=2G\sqrt{\frac{\mu}{2E_{com}}}.
\end{equation}
Fig~\ref{fig:TempDependent} shows these comparisons for $dd\rightarrow aa$ and $dd\rightarrow ad$ channels. When making these comparisons it is important to note that in the notation of Ref.~\cite{zygelman2010electronic}, two contributions must be added to the cross section for $dd\rightarrow ad$ and $dd\rightarrow da$, whereas in the approach of Ref.~\cite{Stoof1987} these are combined into a single symmetrized channel~\cite{zygelman2005hyperfine,zygelmancomm}.  We use the latter convention in this work. In both channels, we find strong agreement with the H-H calculations of  both Ref.~\cite{Stoof1987} and Ref.~\cite{zygelman2010electronic}. For T-T scattering, good agreement is also found with the results of Ref~\cite{zygelman2010electronic}.  These calculations follow different methodologies to one another, so their comparison provides two independent and meaningful validation of our results.

\begin{figure}[t]
    \centering
    \includegraphics[width=0.99\linewidth]{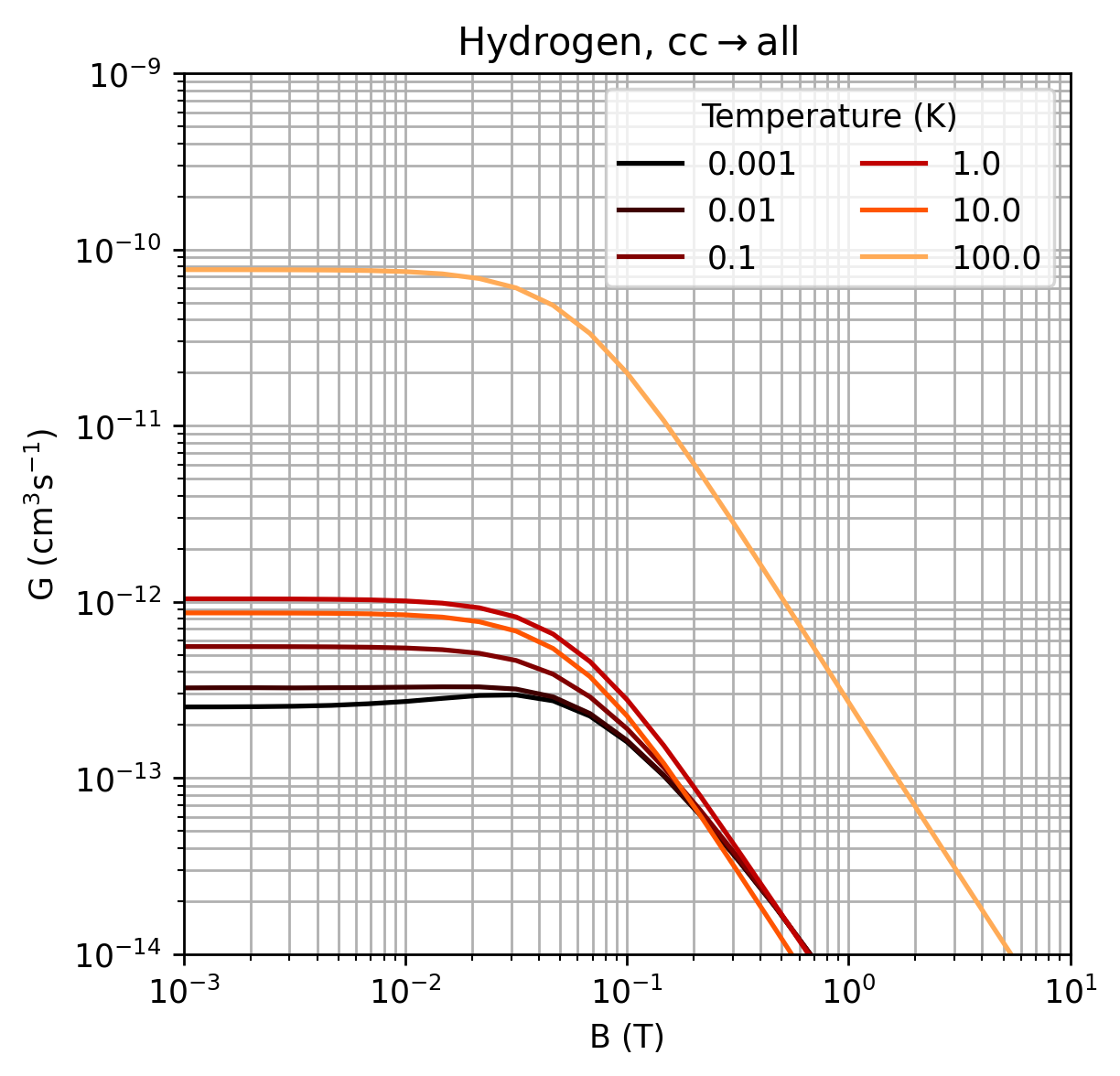}\\
    \includegraphics[width=0.99\linewidth]{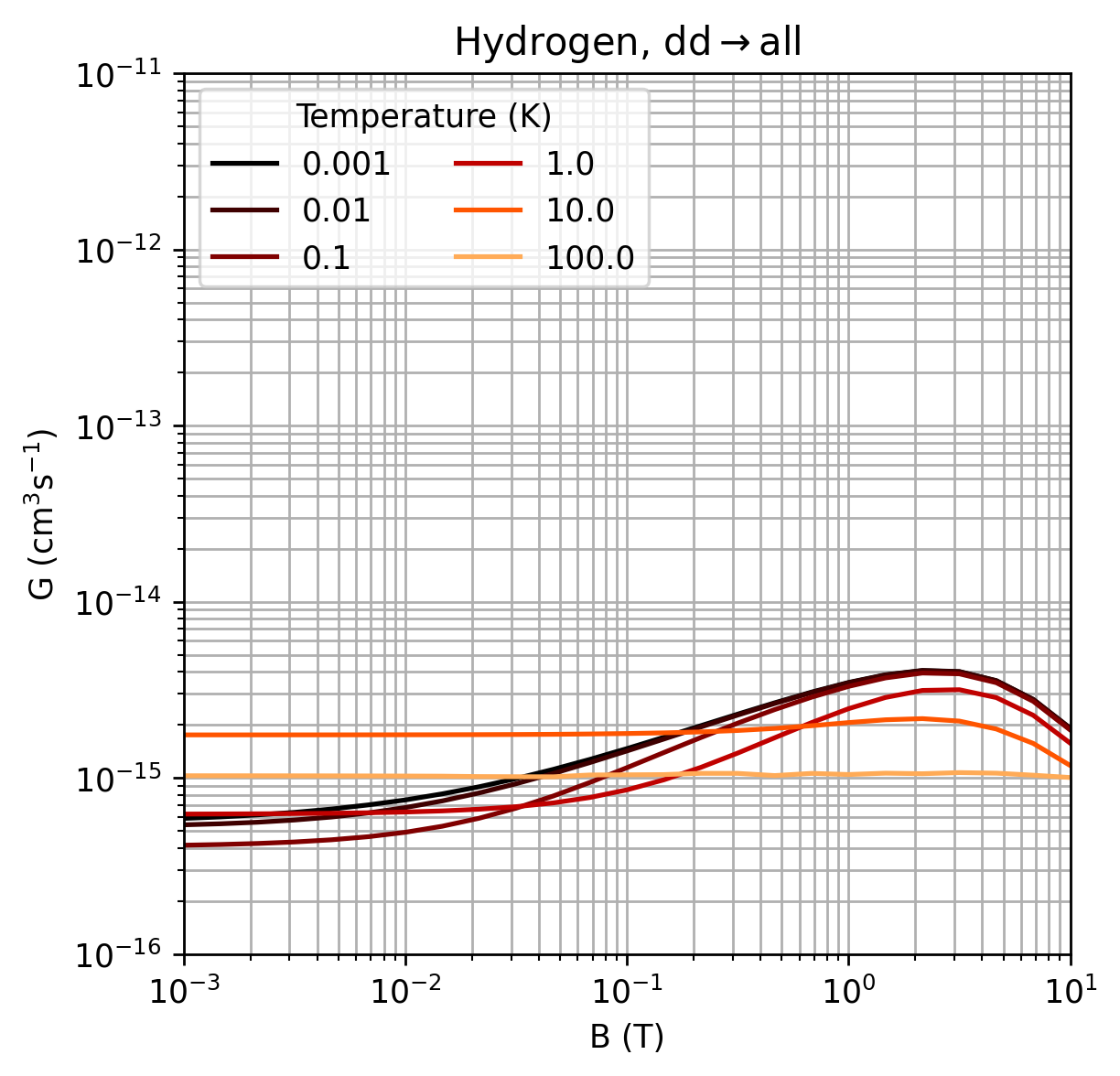}
    \caption{The summed rate constants for losses of high-field-seeking states relevant to magnetic hydrogen trapping experiments, shown as a function of temperature and magnetic field.  These results are calculated using the Silvera triplet potential and Jamieson singlet.   A channel-by-channel breakdown of these rates is provided in supplementary Figures~\ref{fig:AllDipoleRates} and ~\ref{fig:AllSpinExRates}.}
    \label{fig:ScanBAndT_H}
\end{figure}

\begin{figure}[t]
    \centering
    \includegraphics[width=0.99\linewidth]{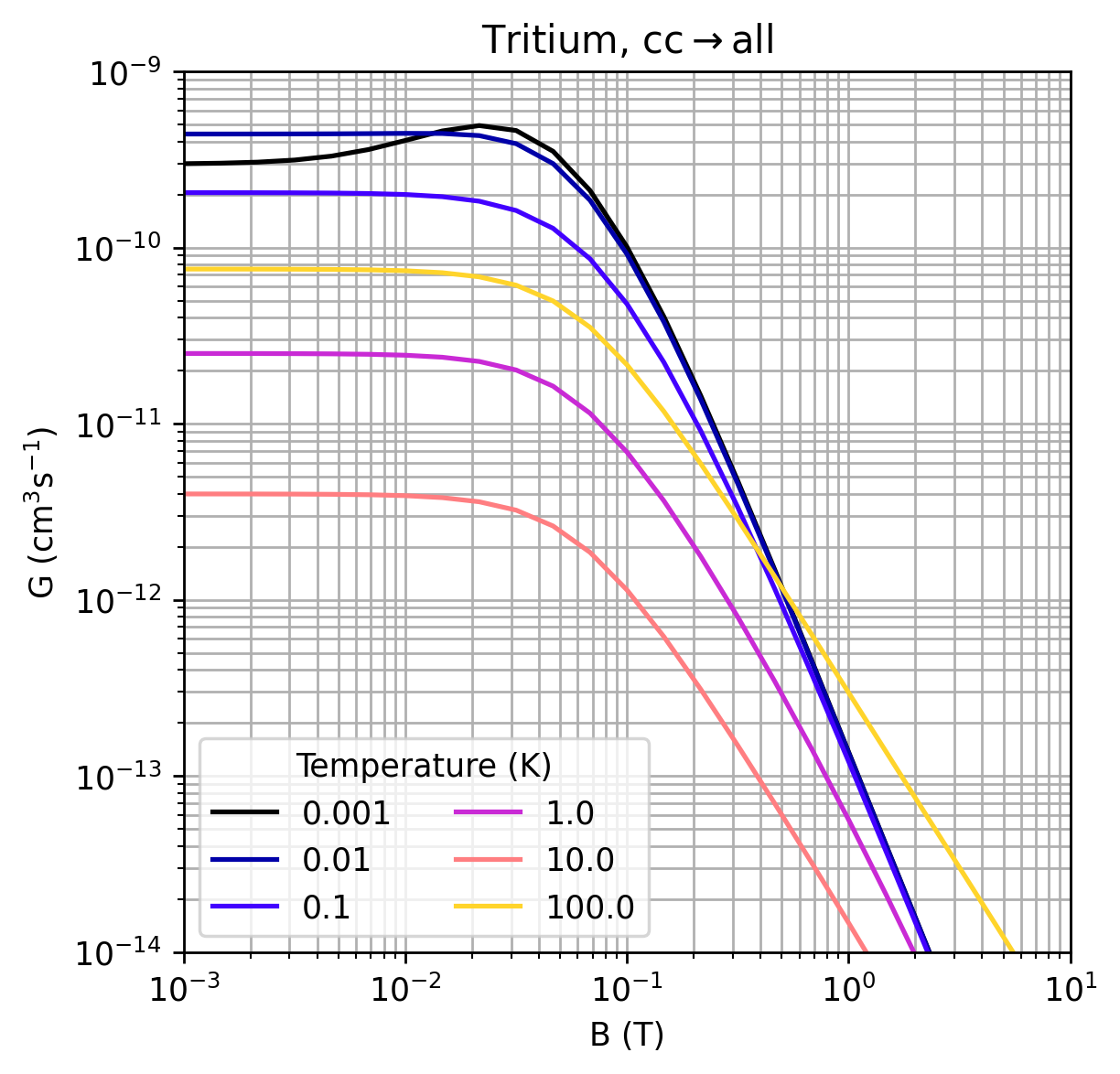}\\
    \includegraphics[width=0.99\linewidth]{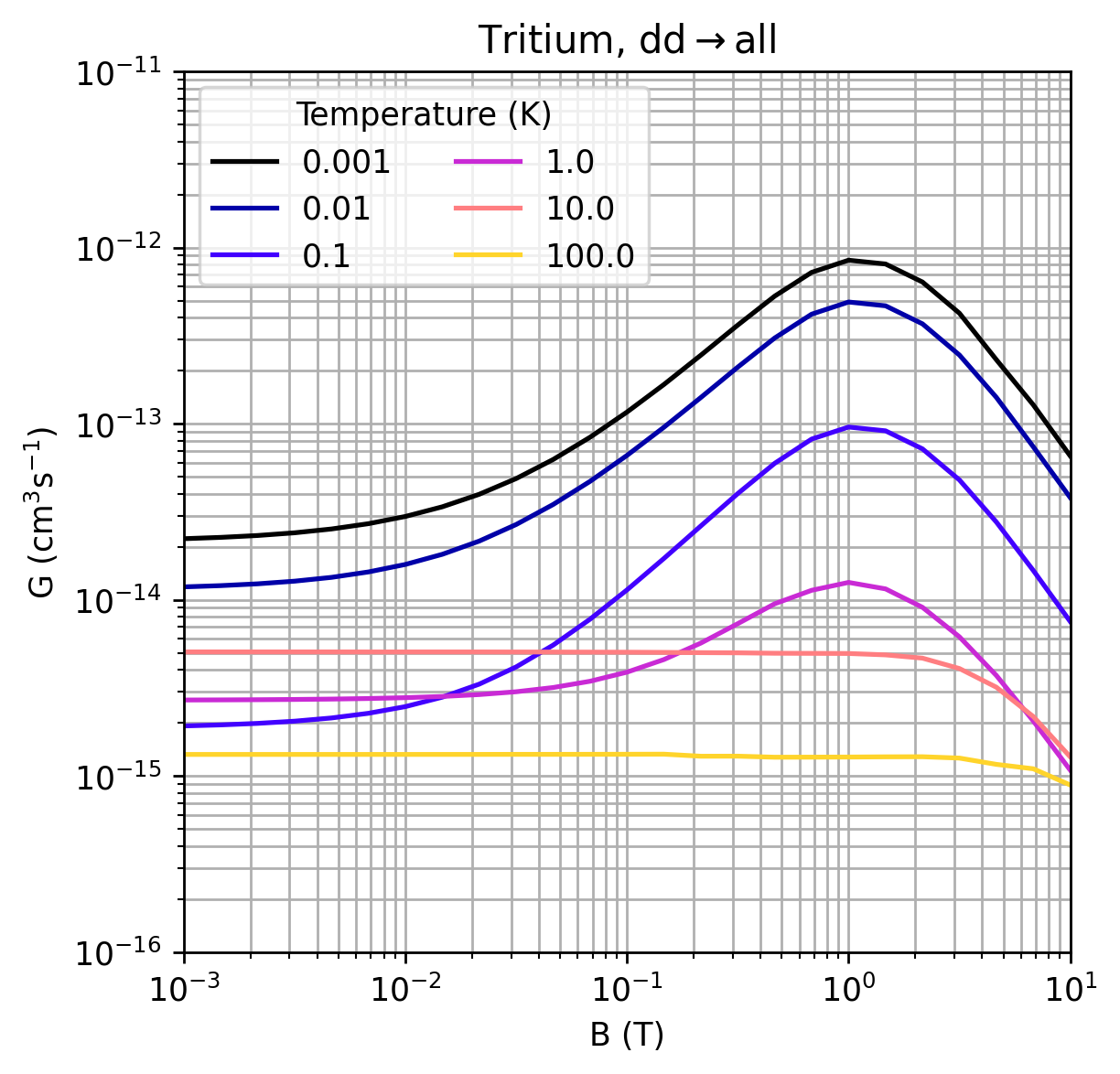}
    \caption{The summed rate constants for losses of high-field-seeking states relevant to magnetic tritium trapping experiments, shown as a function of temperature and magnetic field.  These results are calculated using the Silvera triplet potential and Jamieson singlet.   A channel-by-channel breakdown of these rates is provided in supplementary Figures~\ref{fig:AllDipoleRates} and ~\ref{fig:AllSpinExRates}.}
    \label{fig:ScanBAndT_T}
\end{figure}

\begin{figure*}[h]
    \centering
    \includegraphics[width=0.2\linewidth]{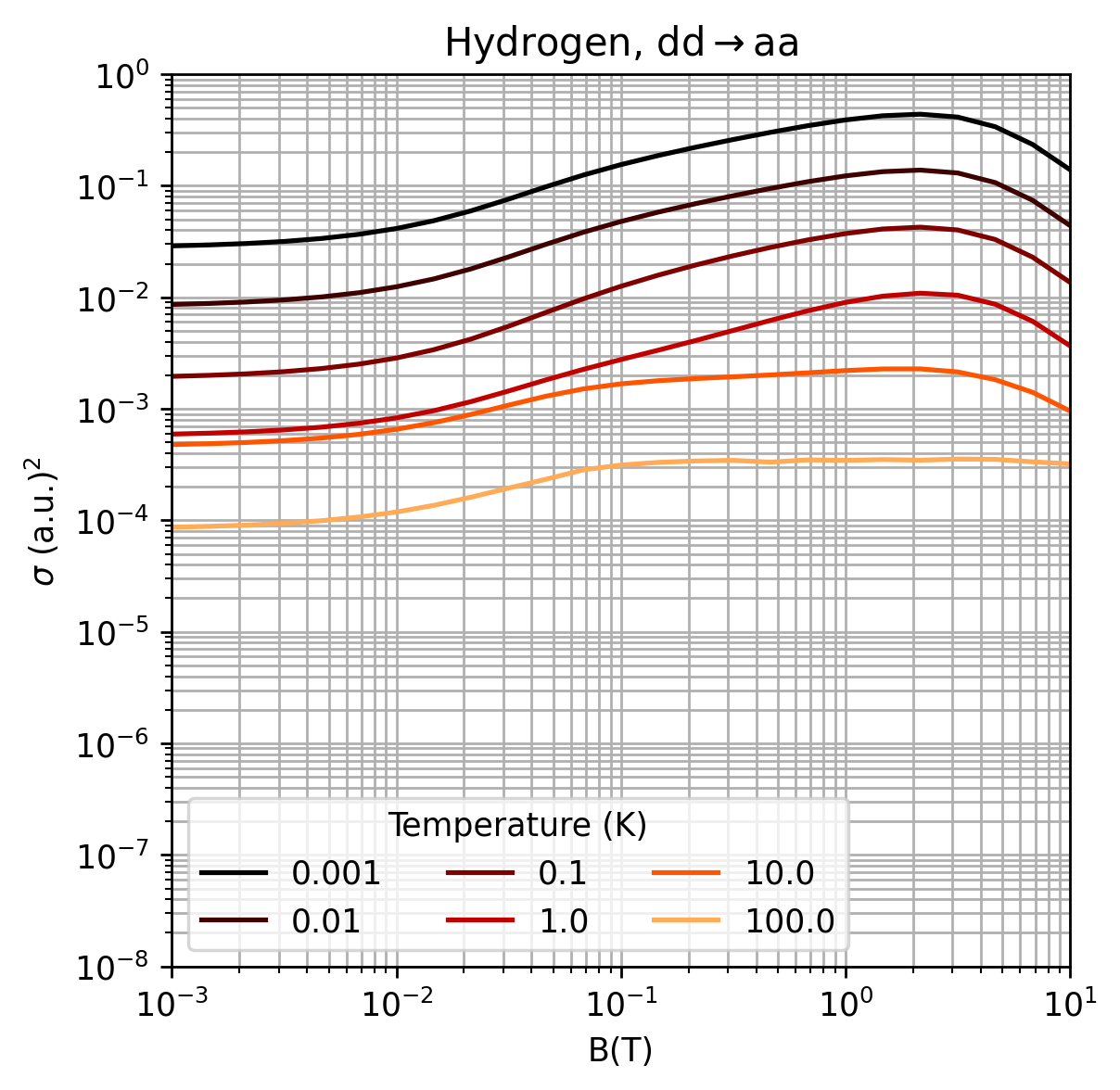}\includegraphics[width=0.2\linewidth]{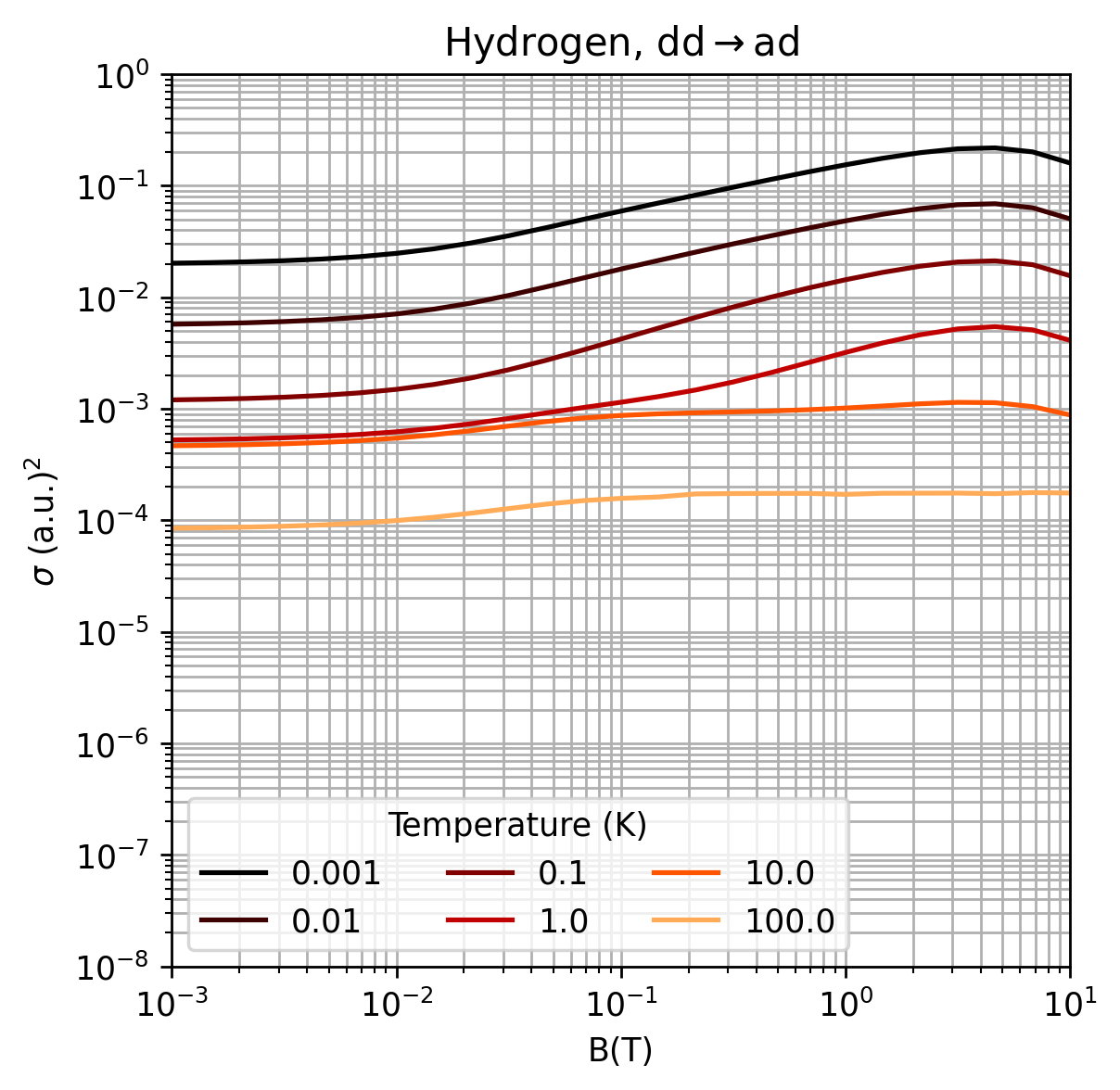}\includegraphics[width=0.2\linewidth]{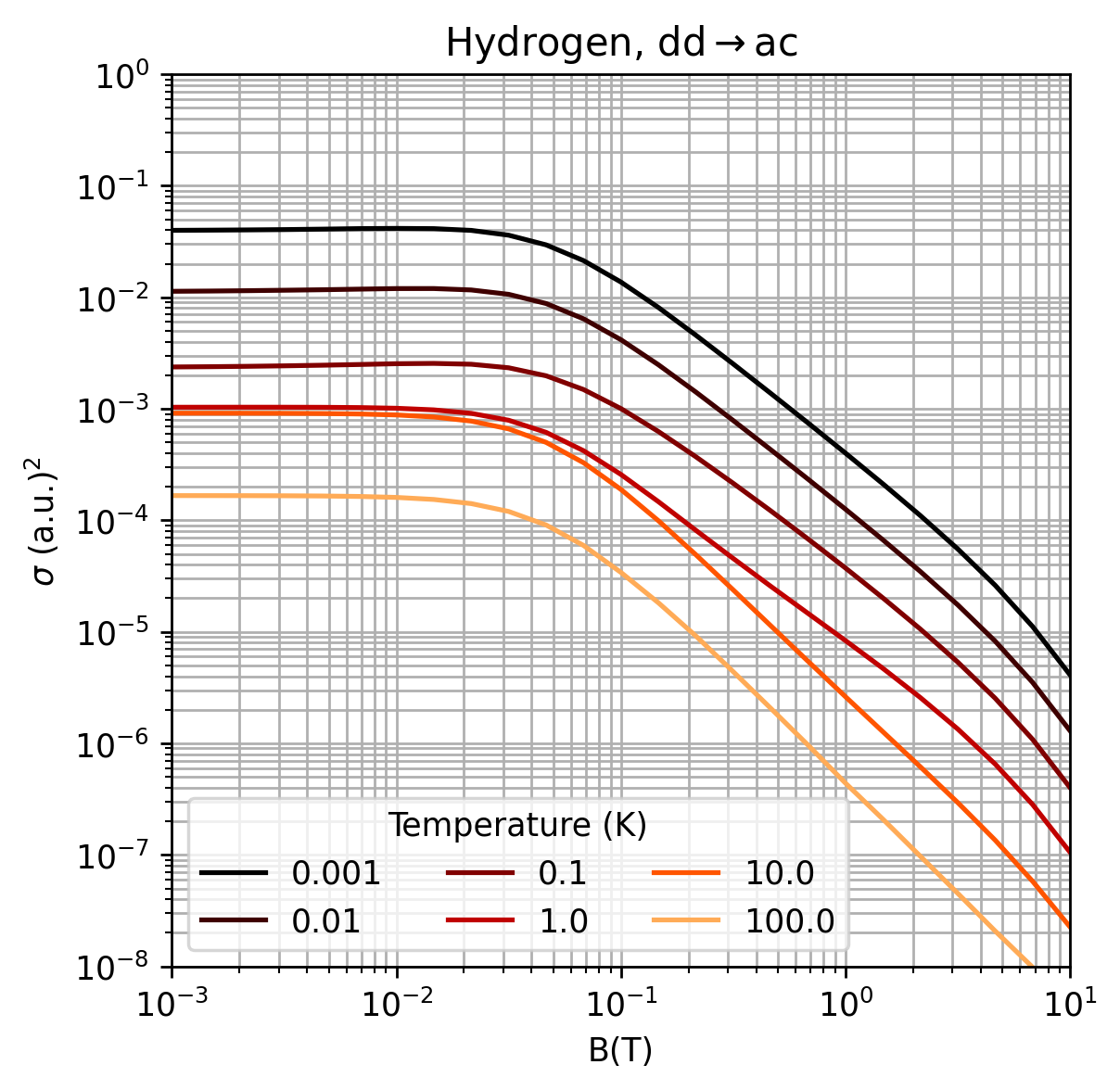}\includegraphics[width=0.2\linewidth]{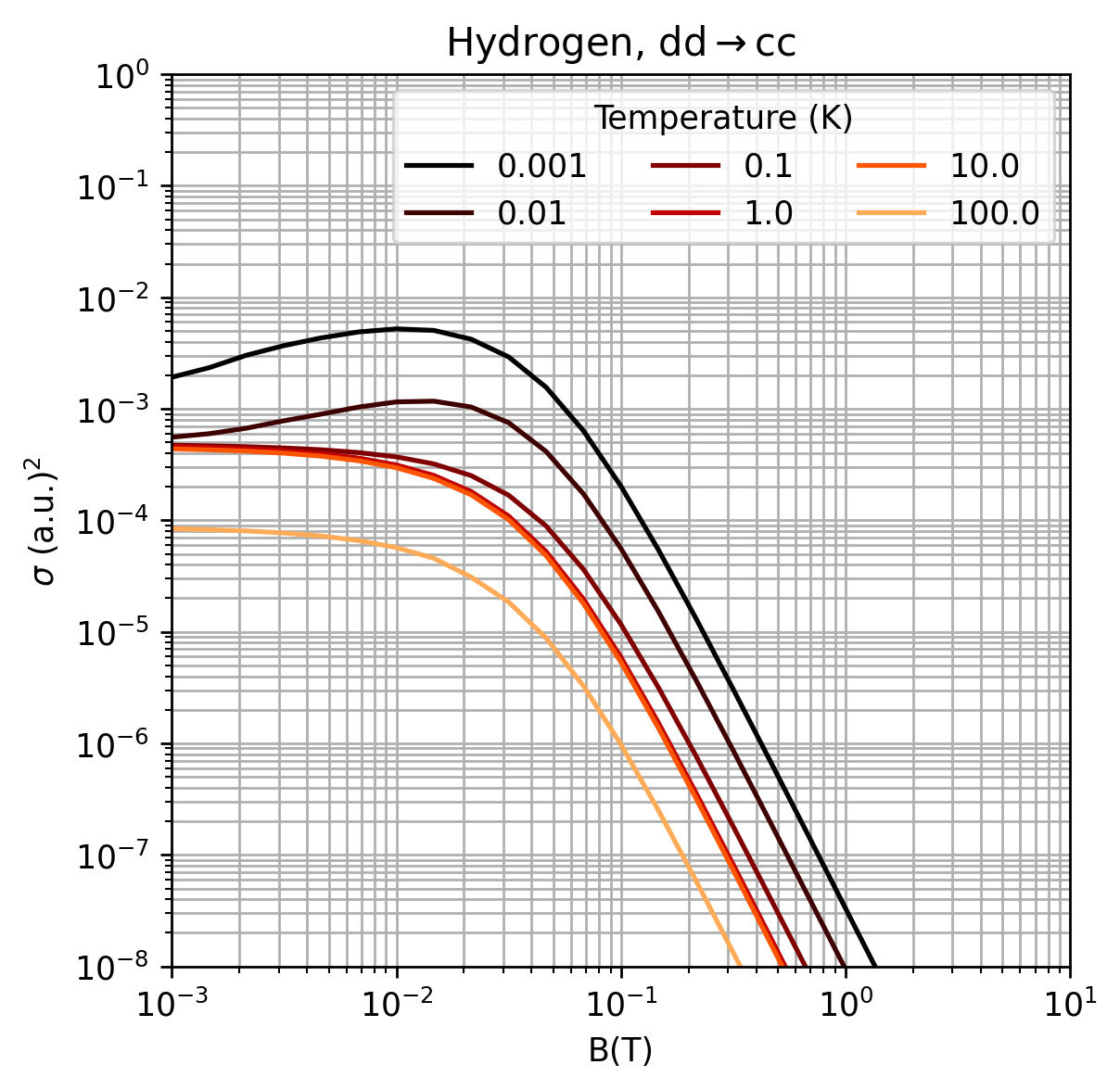}\includegraphics[width=0.2\linewidth]{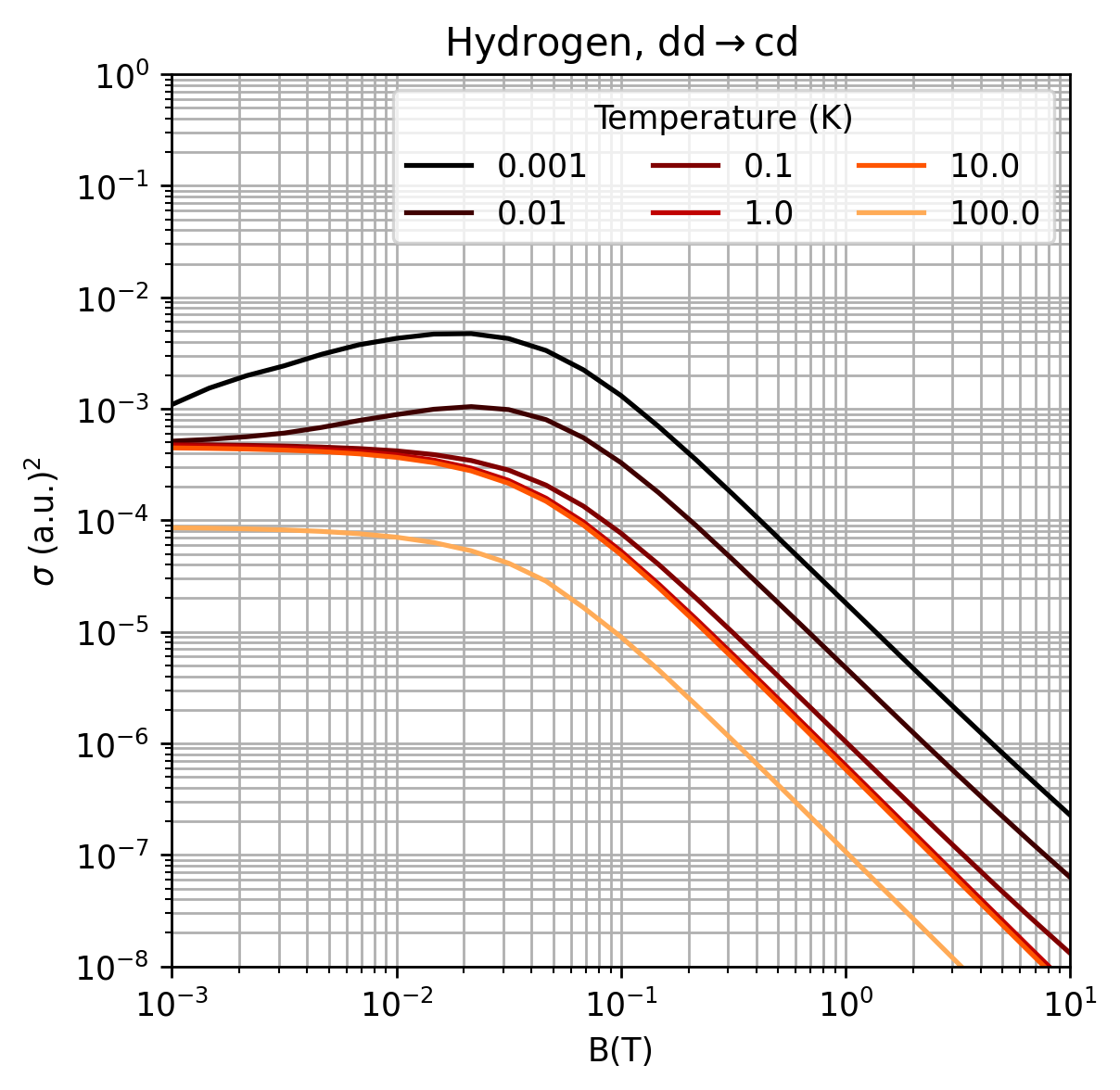}
    
    \includegraphics[width=0.2\linewidth]{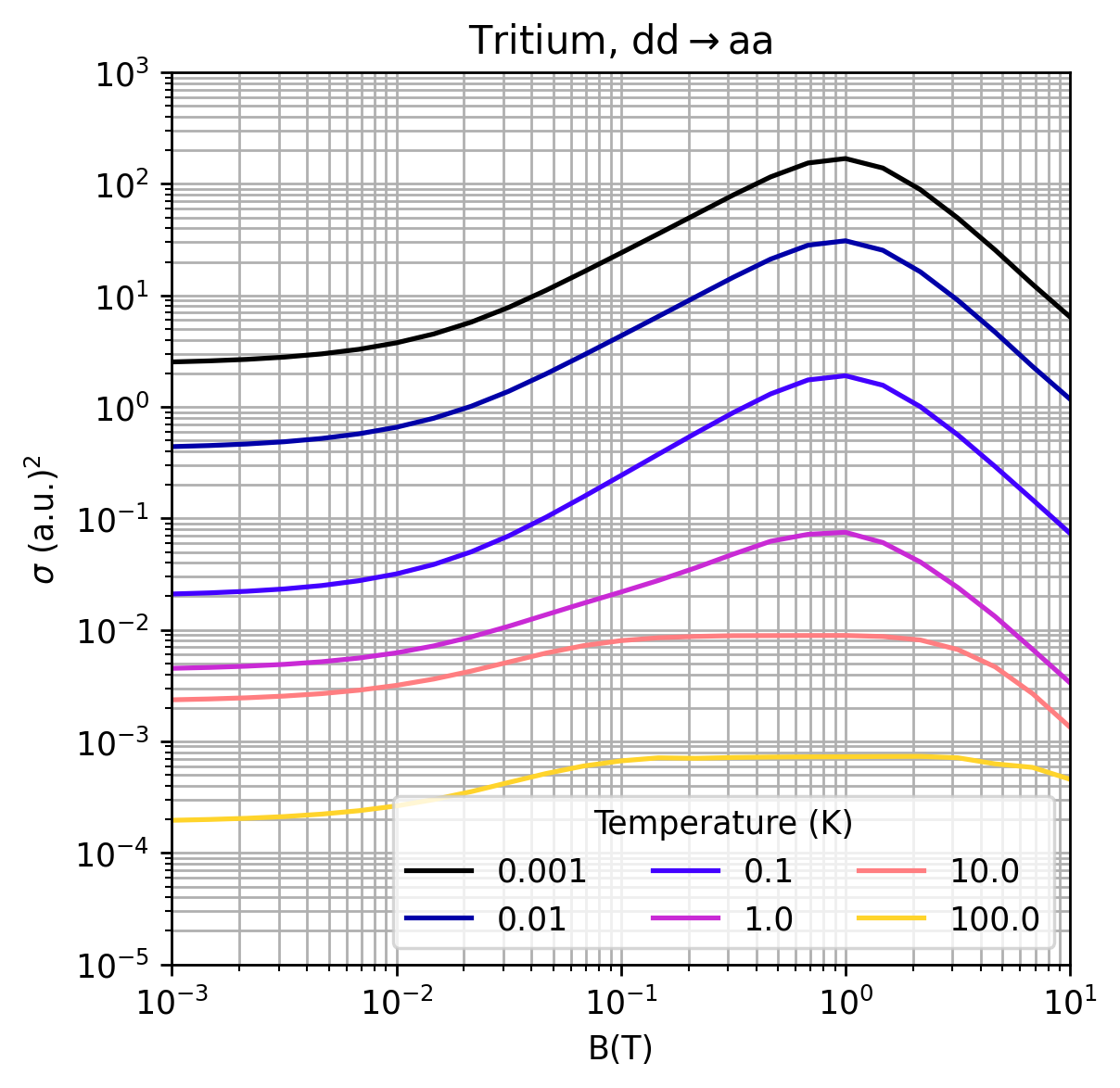}\includegraphics[width=0.2\linewidth]{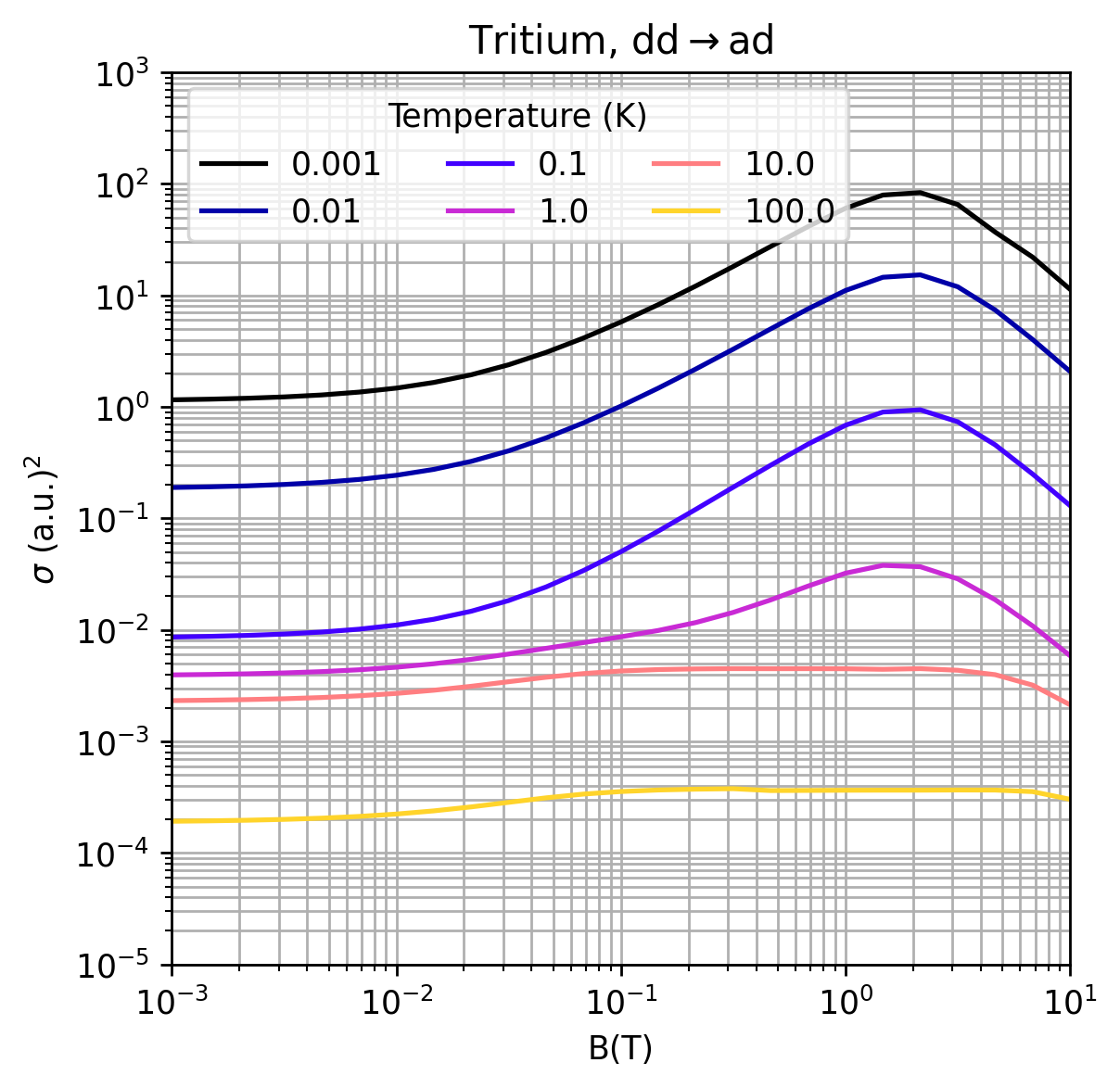}\includegraphics[width=0.2\linewidth]{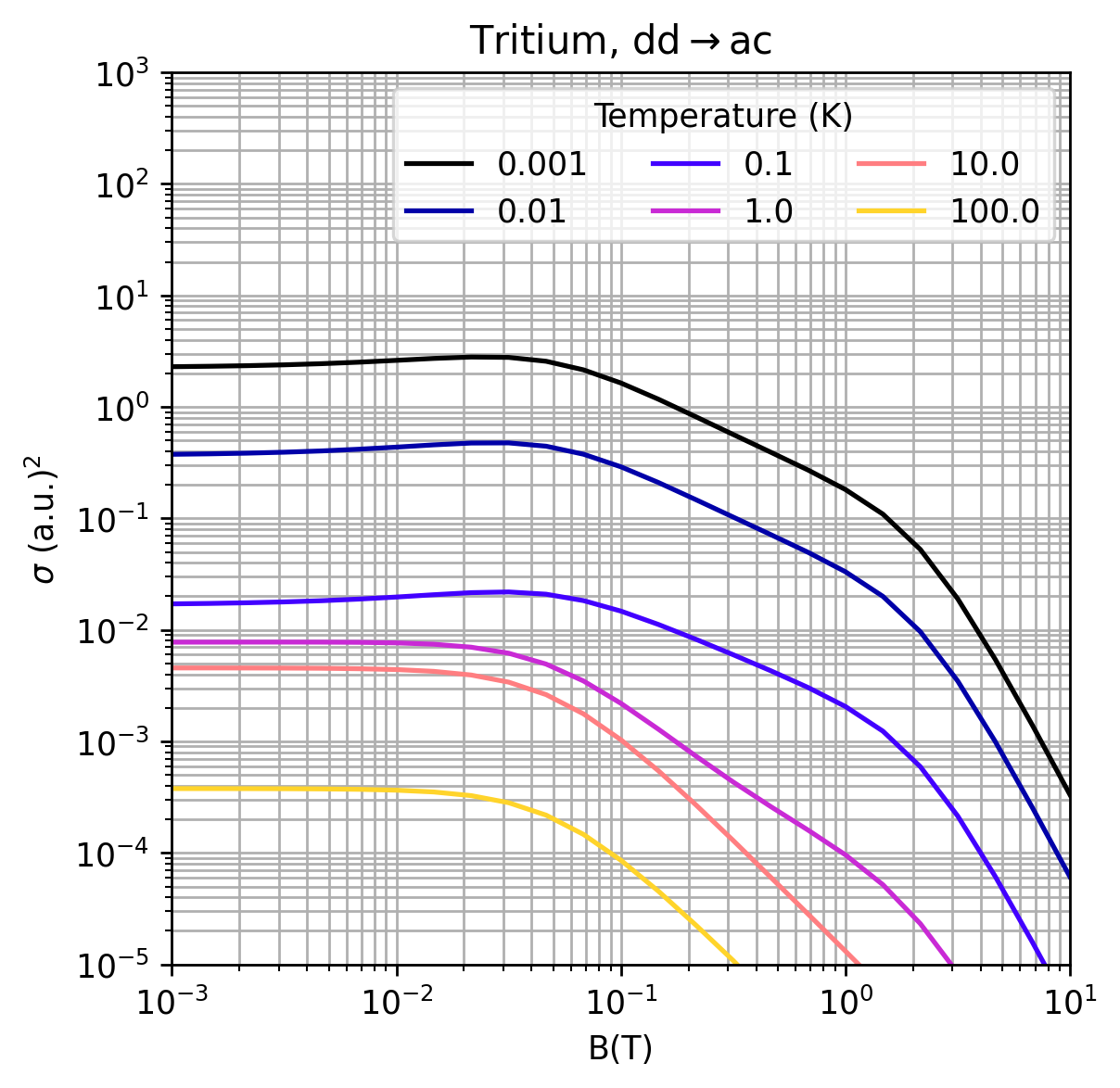}\includegraphics[width=0.2\linewidth]{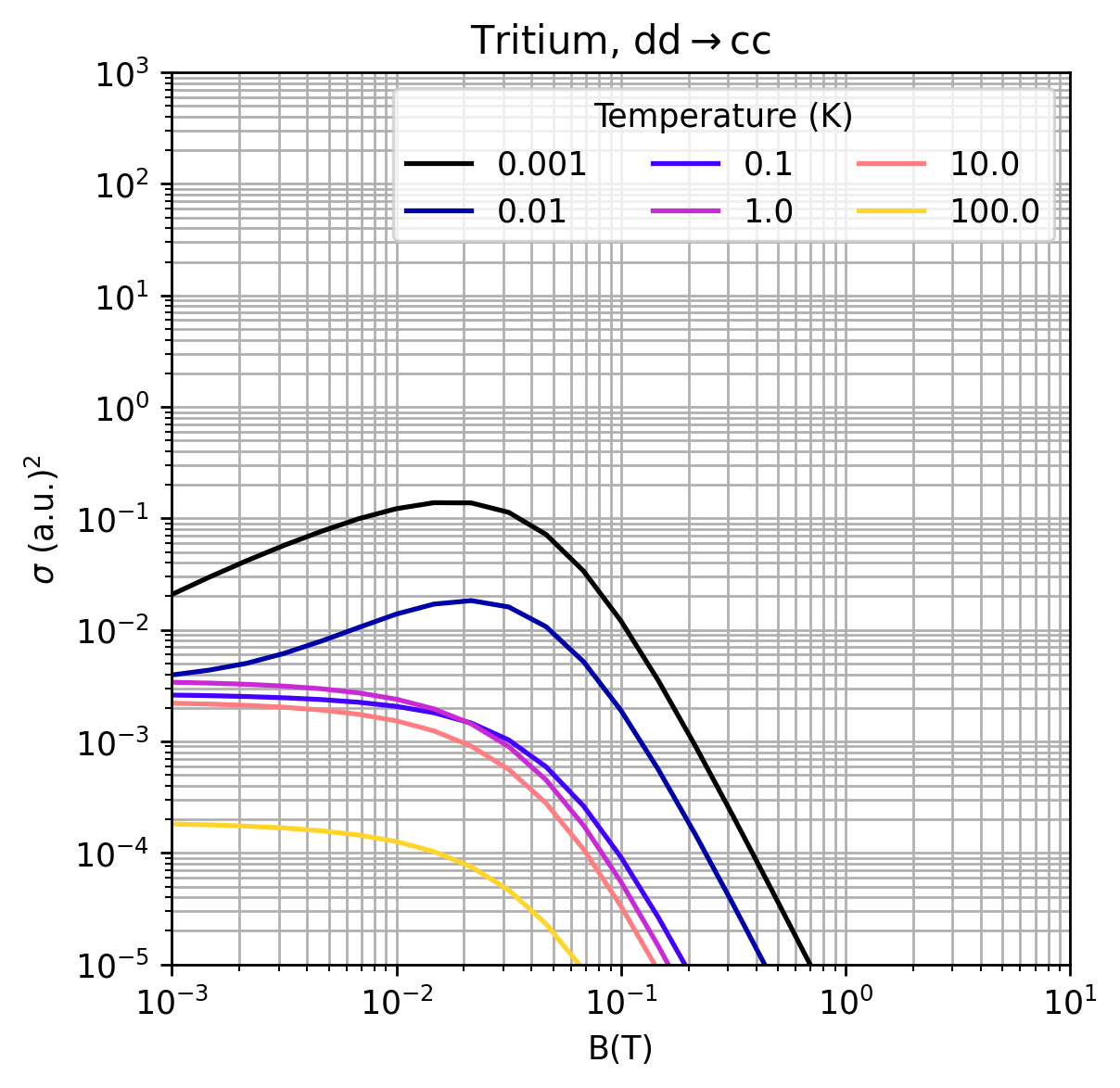}\includegraphics[width=0.2\linewidth]{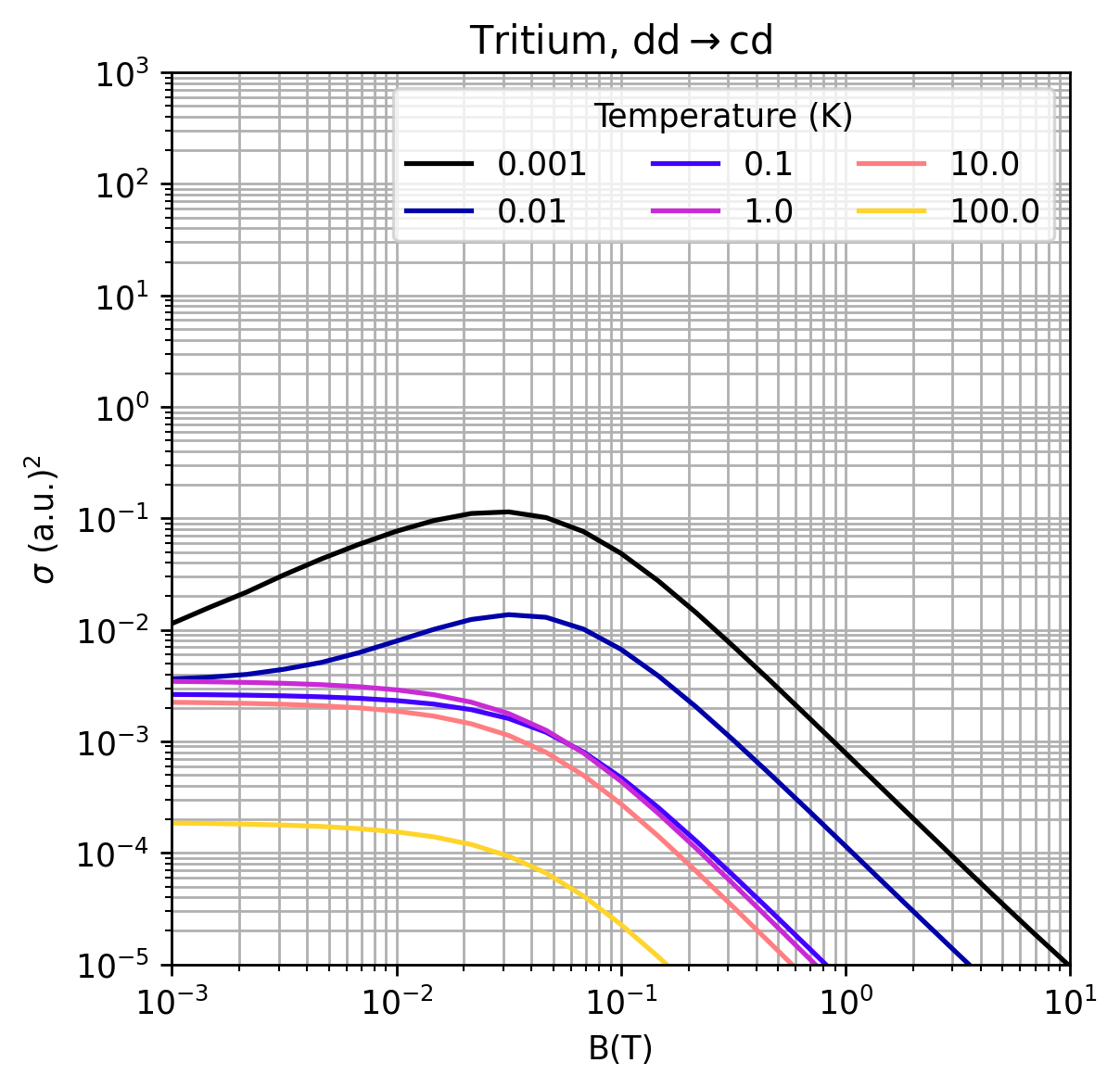}
        \caption{Dipole loss cross sections, all channels, all magnetic fields, all temperatures.}
    \label{fig:AllDipoleSigmas}
\end{figure*}

\begin{figure*}[h]
    \centering
    \includegraphics[width=0.3\linewidth]{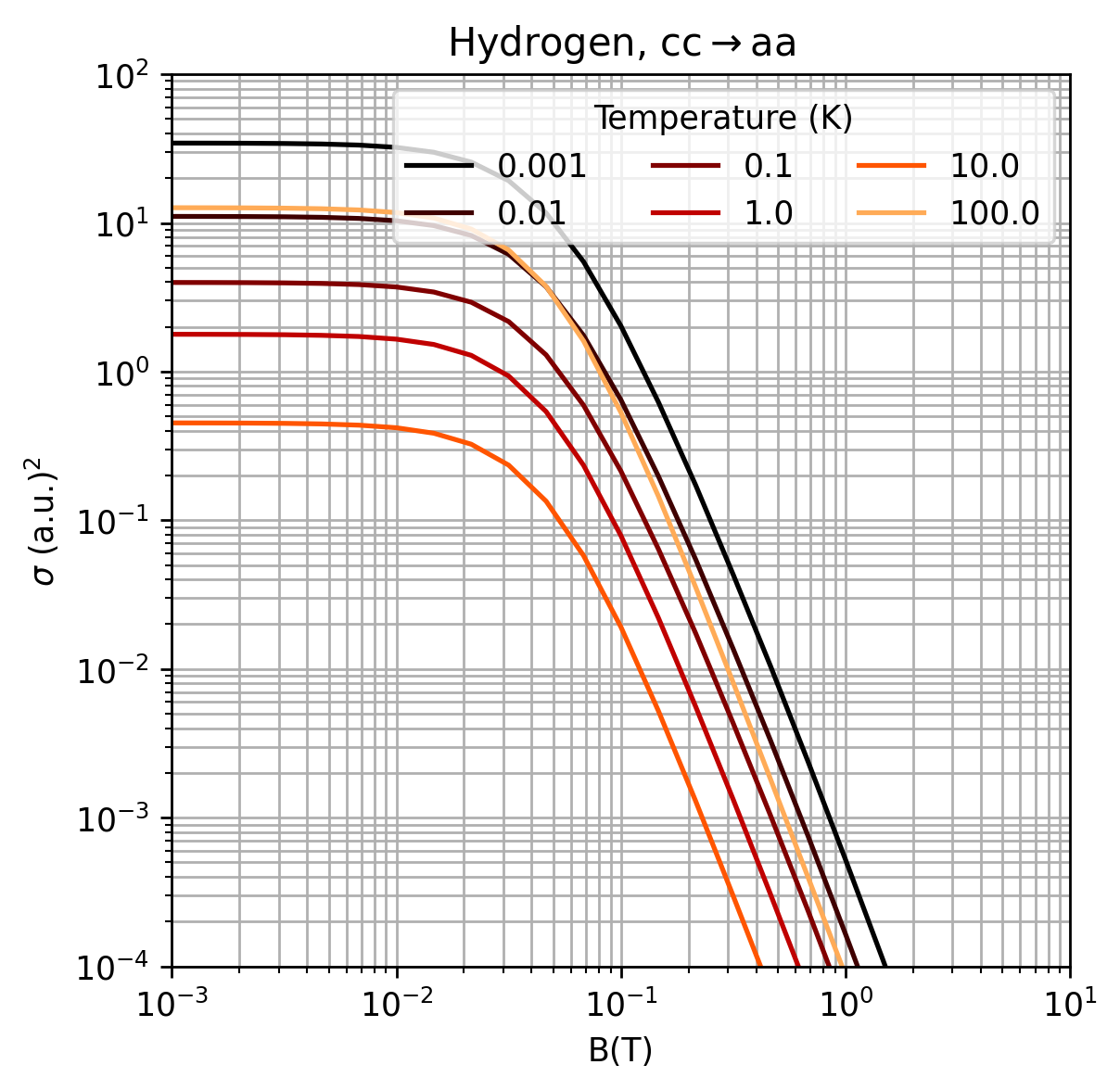}\includegraphics[width=0.3\linewidth]{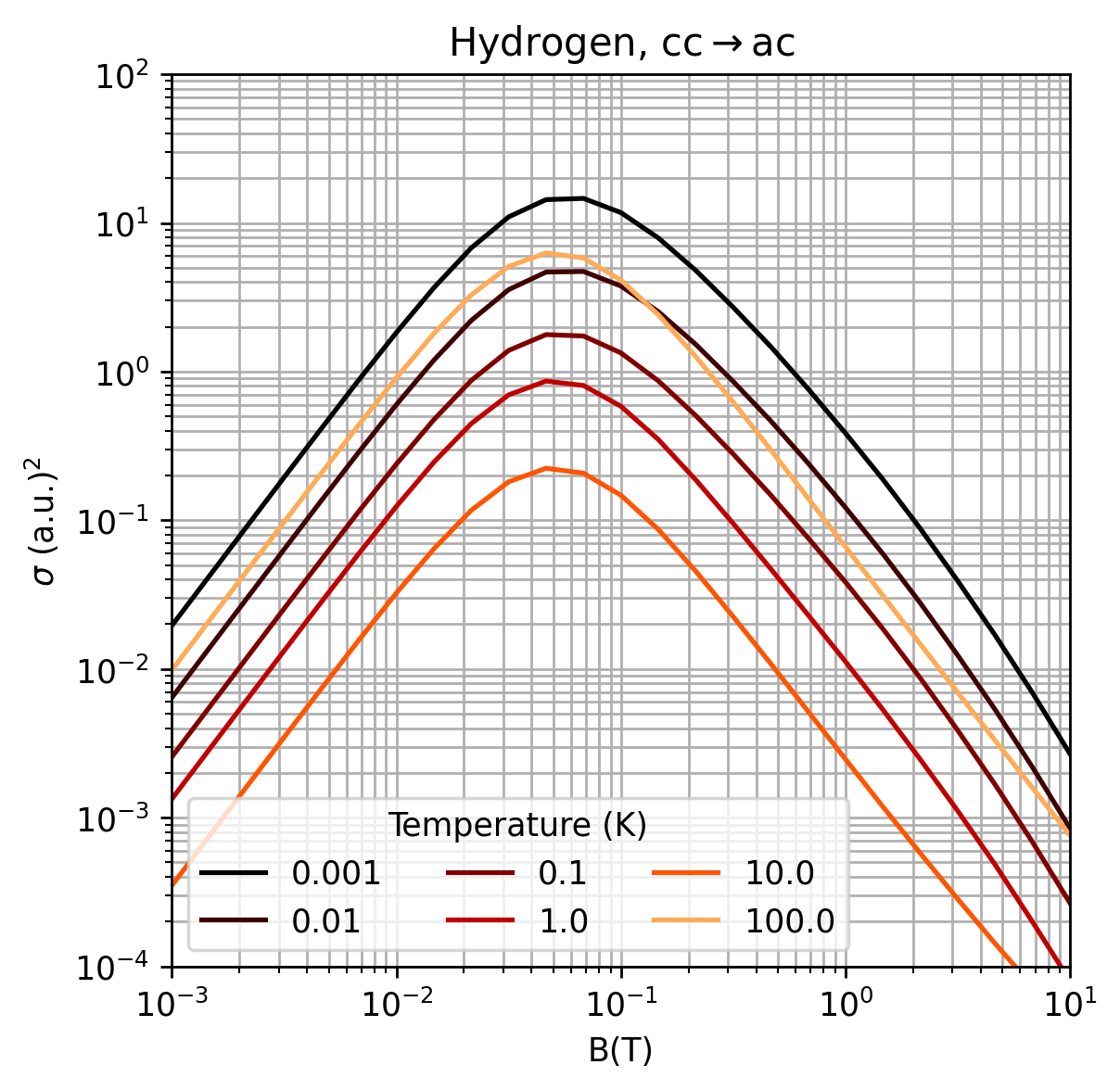}\includegraphics[width=0.3\linewidth]{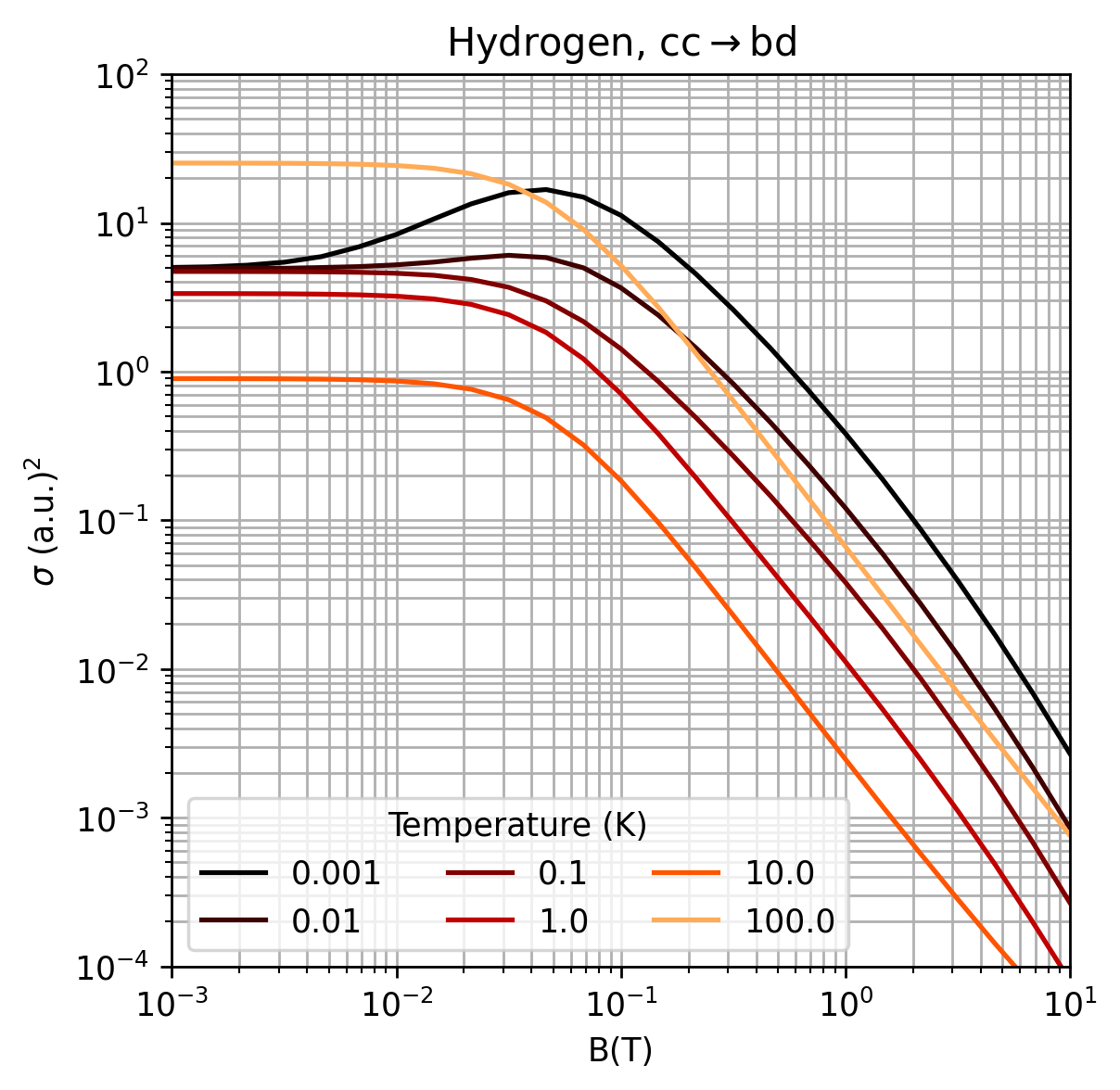}
    
    \includegraphics[width=0.3\linewidth]{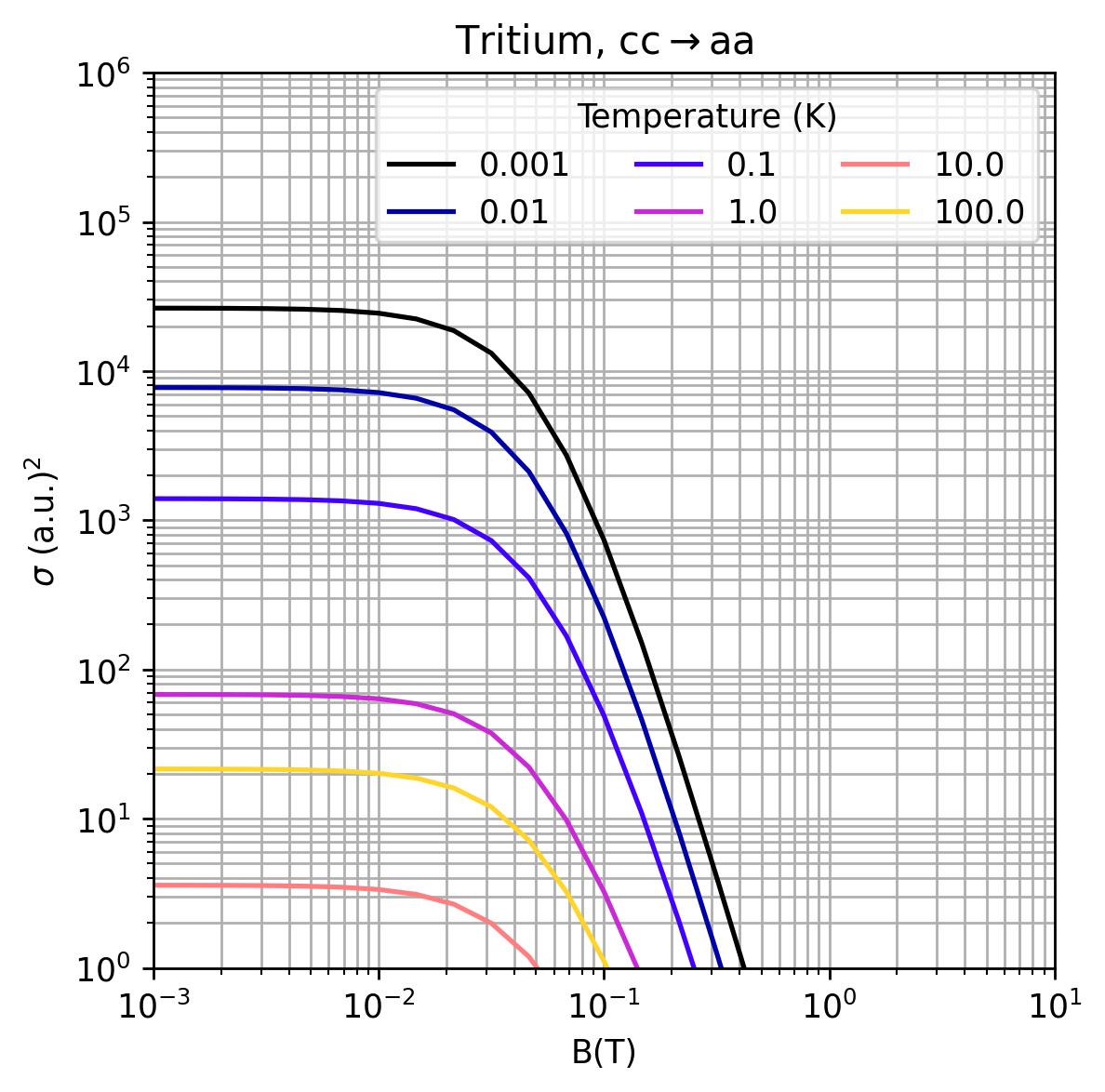}\includegraphics[width=0.3\linewidth]{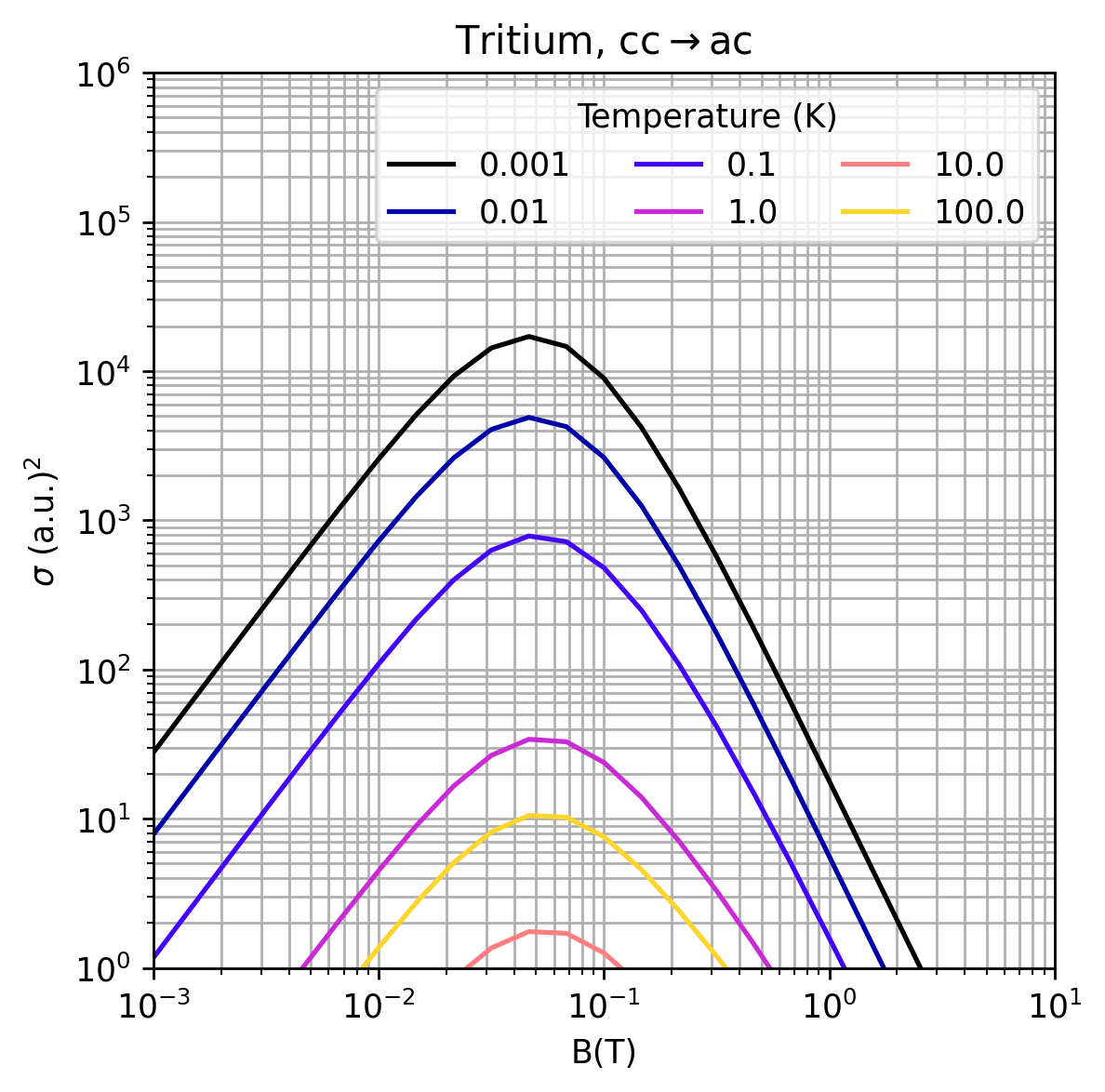}\includegraphics[width=0.3\linewidth]{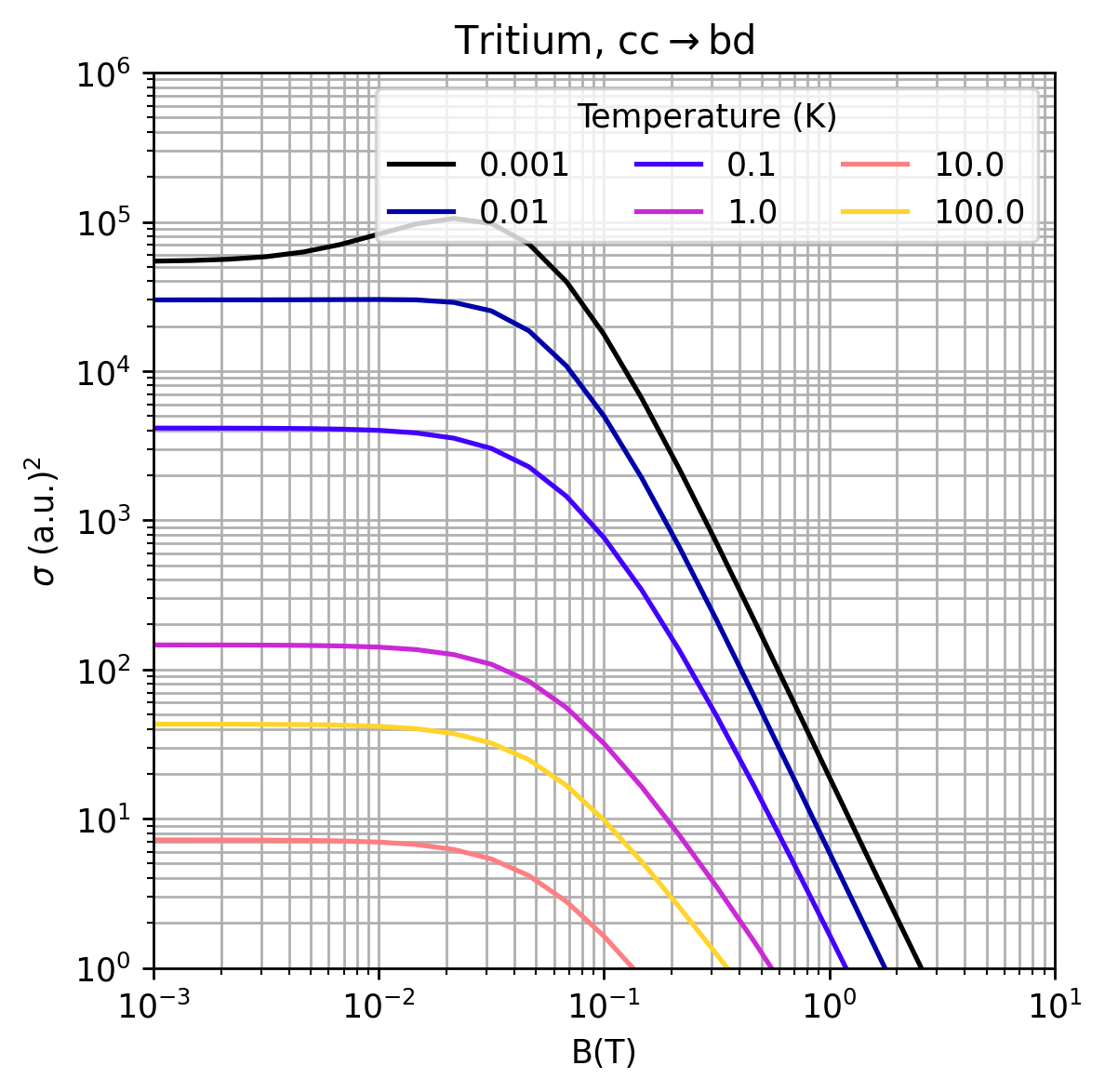}
        \caption{Spin exchange cross sections, all channels, all B field, all temperatures.}
    \label{fig:AllSpinExSigmas}
\end{figure*}

Figures ~\ref{fig:ScanBAndT_H} and ~\ref{fig:ScanBAndT_T}  present the primary new results derived from this work,  the summed rate constant for losses of high-field-seeking states as a function of magnetic field and temperature.  As expected, the $c$ states have a short trap lifetime  due to their spin-exchange interactions, whereas the $d$ states have a longer residence time in magnetic traps since they are only lost through dipole interactions. For tritium, there is a significant temperature dependence to the rates, driven largely by the rapidly falling cross section as the average momentum transfer increases.  For hydrogen, the dependence of the loss rates on temperature are significantly smaller at low temperatures, though the spin exchange processes rapidly climb in strength above 10~K as higher partial waves begin to contribute strongly to the cross section, which is also observed in the high temperature tritium predictions.  This effect is illustrated in the partial wave decomposition for H-H spin exchange scattering, shown in  Fig.~\ref{fig:HHPartialSpinEx}.   A channel-by-channel decomposition of these rates are provided in the supplementary Figures~\ref{fig:AllDipoleRates} and ~\ref{fig:AllSpinExRates}, alongside the derived scattering cross sections in Figures~\ref{fig:AllDipoleSigmas} and ~\ref{fig:AllSpinExSigmas}.

\section{Systematic Uncertainties\label{sec:Systematics}}

We assess the systematic uncertainties on the cross sections calculations based on the following sources of uncertainty:

{\bf Discrete triplet potentials:} The degree of uncertainty on the triplet potential is estimated by comparison of the Silvera, Jamieson (Born Oppenheimer) and Kolos (Born Oppenheimer) potentials. This source of uncertainty impacts all three cross section types.

{\bf Discrete singlet potentials:} Similarly, the singlet potential can be modeled as either the  Jamieson (Born Oppenheimer) and Wolniewicz (Born Oppenheimer) or Kolos (Born Oppenheimer) form.  The Silvera singlet is not considered as sufficiently accurate to inform our singlet uncertainty budget. Only the spin exchange cross sections are impacted by the singlet potential.  The spin exchange processes are highly sensitive to the singlet shape, and as such, this term is one of the larger uncertainties for these processes.

\begin{figure}[t]
    \centering
    \includegraphics[width=0.99\linewidth]{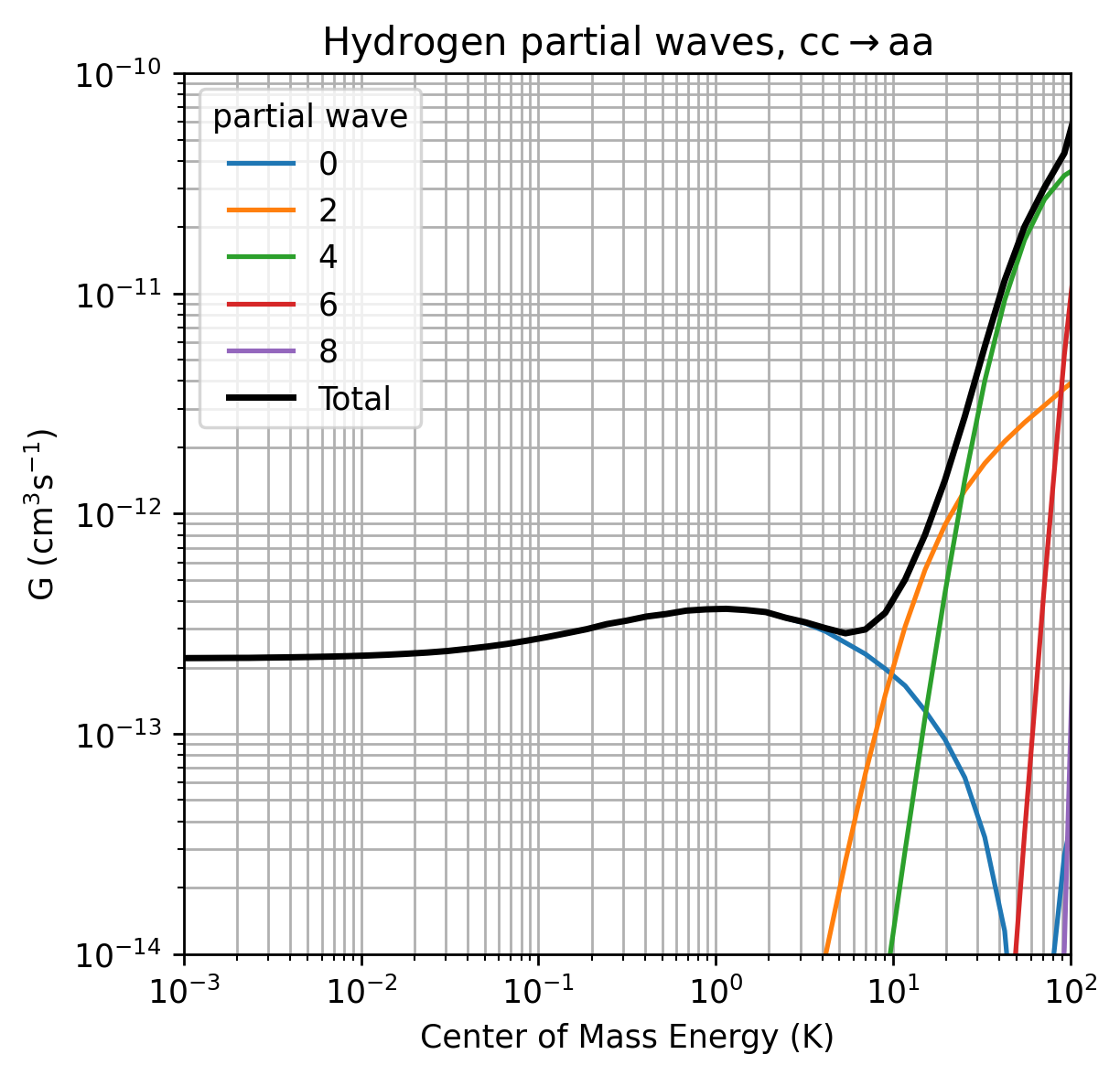}\\
    \caption{Partial wave contributions to the hydrogen spin exchange process $cc\rightarrow aa$.  We see the higher partial waves rapidly enhancing the cross section around 10~K, explaining the sharp increase in the H-H spin exchange cross section that is apparent between 10~K and 100~K on Fig~\ref{fig:ScanBAndT_H}.}
    \label{fig:HHPartialSpinEx}
\end{figure}

{\bf Potential shape:} To account for shape uncertainty in the potentials that is not covered by the above discrete perturbations, we consider a linear distortion of the form
\begin{equation}
    V_{perturbed} = V(r) (1+c \frac{r}{a_0})
\end{equation}
where the constant $c$ is picked based on the spread in the available potentials, to be of magnitude 0.1$\%$ per Bohr.

{\bf Adiabatic correction}: The triplet adiabatic correction is drawn from the older work~\cite{kolos1990adiabatic} where it is provided in tabular form. Since the correction was not calculated explicitly for each potential, we assign a 10\% systematic uncertainty to the magnitude of the applied correction.

{\bf Potential extrapolations:} At long distances where no calculations are available, the triplet potentials are extrapolated using an analytic form. To assess uncertainty introduced by this extrapolation we compare the HFD form of Eq.~\cite{Silvera1980} with a simple $r^{-6}$ Van Der Waals form. 

{\bf Non-adiabatic terms:} As has been reported in past works~\cite{jamieson2010dependence,williams1993mass}, the effects of non-adiabatic corrections are approximately captured by considering the difference between nuclear and atomic reduced masses as inputs to the scattering calculations.  To this end we compare predictions using these two cases as a proxy for the magnitude of un-modeled non-adiabatic effects.

{\bf Methodology:} The formulae used in our calculations of the dipolar and spin exchange rates involve approximations, in the former case the use of first order perturbation theory and the distorted wave Born approximation; in the second case, the use of the degenerate internal states approximation.  These were compared against coupled channel approaches in Ref.~\cite{Stoof1987} and the reported precisions were 4\% and 5\% respectively. We include these precision quantifications as contributions to the theoretical uncertainty. The formalism for the elastic scattering calculations are exact, and so no additional uncertainty is required in this case.

{\bf Fundamental constants and external data:}
The values of the tritium and hydrogen reduced masses, hyperfine constants, nuclear g-factors, electron magnetic moment and mass were all varied using their bounds from the CODATA tables~\cite{mohr2025codata}. This leads to a vanishingly small uncertainty contribution $<0.01\%$ in all cases and is thus determined to be negligible for all channels.

The approximate magnitudes of the uncertainties introduced by each of these effects are tabulated in Table~\ref{tab:Systematic} for three representative cross section channels.  The numerical value of the uncertainty is final state, temperature, magnetic field and isotope dependent. The numbers shown are intended to be representative for the conditions of interest to atom trapping experiments, so are evaluated in the low temperature limit and averaged over magnetic fields from 0 to 100~mT.  The code provided in the supplementary material can be used to obtain a specific uncertainty quantification at any given parameter point and channel, as needed.  The approximate uncertainties on our predicted cross sections are 7.6\% for the elastic channel, 8.5\% for the dipole loss channels, and 26\% for the spin exchange channels, with the dominant contributions arising from potential shape effects and the non-adiabatic uncertainty estimate that derives from comparing calculations using atomic and nuclear masses for tritium.

\begin{table}[t]
\centering
\begin{tabular}{|l|ccc|}
\hline
                                     & \textbf{Elastic}        & \textbf{Dipolar}        & \textbf{SpinEx.}  \\
                                     & {[}dd$\rightarrow$dd{]} & {[}dd$\rightarrow$aa{]} & {[}cc$\rightarrow$aa{]} \\
\hline
Discrete triplets &  6.4\%   &  4.7\% &  1.1\% \\
Discrete singlets & --  & --  &  12\% \\
Triplet shape & 4.0\%    &    4.2\%  &  22\%   \\
Singlet shape &   --   &   --      &    3.3\% \\             
Adiabatic correction & 0.52\% &  0.26\%  &   0.13\%    \\
Triplet extrapolation &   1.1\%  &    0.67\%  &  0.087\% \\
Non-adiabatic terms &  0.54\%   &  5.9\%  &  4.0\%  \\
Methodology& -- & 4\%& 5\% \\
Input constants  &  $<0.01\%$  &  $<0.01\%$   &  $<0.01\%$   \\  
\hline
Total   &  7.6\%        &        8.5\%                 &    26\%                          \\  

\hline
\end{tabular}
\caption{Table of approximate systematic uncertainties for elastic, dipolar and spin exchange channels. One representative channel is chosen to quantify the impact of each contribution.  To provide a compact presentation, we show here the fractional uncertainties averaged over $B$ field from 1 to 100~mT in the low temperature limit.   \label{tab:Systematic}}

\end{table}

\section{Conclusion\label{sec:Conclusions}}

We have presented new calculations of the cross sections and rate constants for atomic hydrogen and tritium scattering, for all channels relevant to experiments that aim to magnetically trap low-field-seeking states.  The calculations are reported as a function of both temperature and magnetic field over ranges spanning 1~mK-100~K, and 10$^{-3}$-10~T.  Among other applications, our results provide crucial numerical inputs to the design of an atomic tritium source for the Project~8 direct neutrino mass search, which will perform precision cyclotron radiation emission spectroscopy on the beta electrons produced by cooled and magnetically trapped atomic tritium.  

Relative to H-H scattering, the T-T system exhibits an increased triplet  scattering length due to a closely-above-threshold bound state. This produces a dramatic elastic scattering rate enhancement for the spin-polarized T-T system, with significant implications for its fluid properties. This enhancement is also manifest in the dipolar loss and spin exchange channels that determine the trap lifetime of the $d$ and $c$ low-field seeking states respectively, through distortion of the incoming s-wave initial state.    The dipolar loss rates in trapped atomic tritium are found to be enhanced over those for a similar system of trapped hydrogen by a factor of 50-100, whereas the spin exchange loss rates are enhanced by a factor of up to 1000, with dependencies on both magnetic field and temperature.

Our calculations have been bench-marked against past works where subsets of similar results have been reported~\cite{Al-Maaitah2012,Stoof1987,zygelman2010electronic}, with satisfactory consistency  in all cases.  Examination of the known  systematic uncertainties suggest that our results should be considered as accurate at the few \% level in elastic and dipolar channels, or to order-1 in spin exchange, with the leading source of imprecision in the latter case being the singlet potential shape.  This uncertainty could in principle be reduced using more modern approaches to calculation of the hydrogen singlet potential as, for example, in Ref.~\cite{lai2025precision}.

We accompany this paper with  comprehensive  tables of cross sections and rate constants, as well as open source Python code that can be used to reproduce or extend the presented results. This supporting material can all be found at~\cite{code}.

\section*{acknowledgments}
MGE and BJPJ are supported by the DOE Office of Nuclear Physics under awards DE-SC0024434 and DE-SC0019223.  We express our sincere thanks to Bernard Zygelman for guidance on the interpretation of his spin-exchange cross-section calculations.   We also thank Hamish Robertson, Rene Reimann, Stephen Kuenstner and Talia Weiss, Krishan Mistry for valuable comments on and corrections to this manuscript.  Finally, we thank the Project~8 collaboration for providing the inspiration for this work, and for  feedback during its development.

\bibliographystyle{unsrt}
\bibliography{PaperBib}
\newpage

\begin{figure*}[h]
    \centering
    \includegraphics[width=0.2\linewidth]{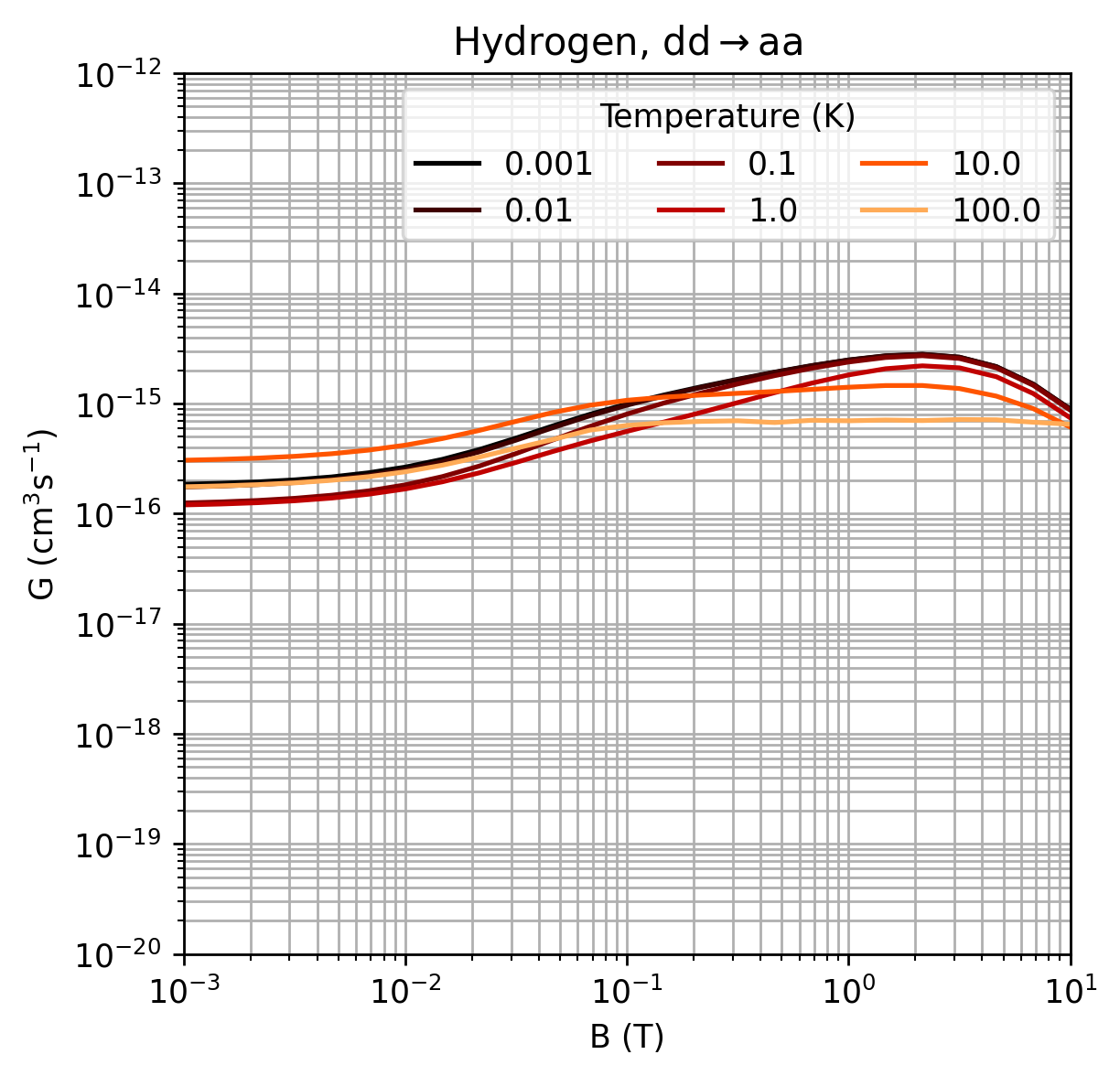}\includegraphics[width=0.2\linewidth]{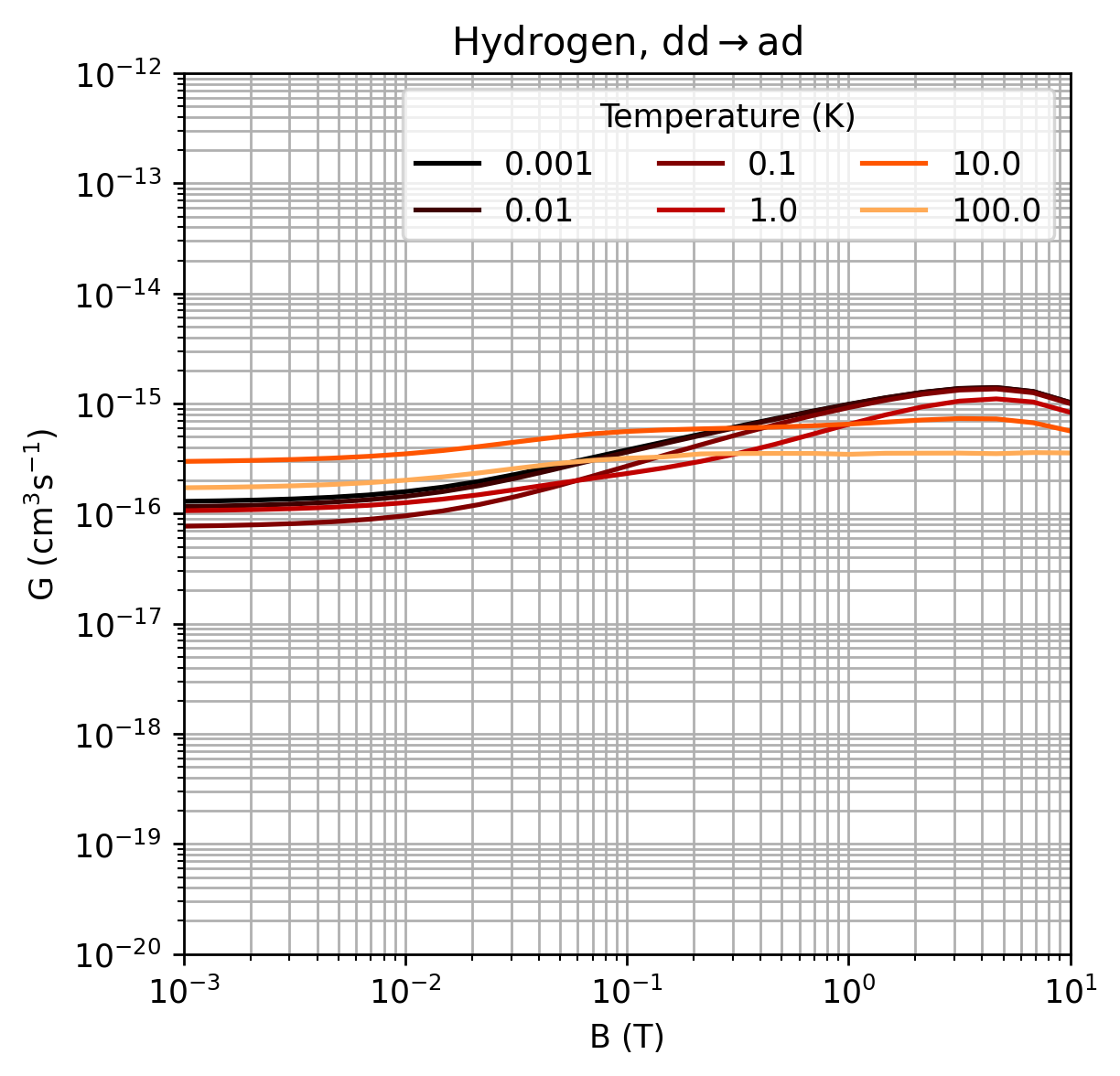}\includegraphics[width=0.2\linewidth]{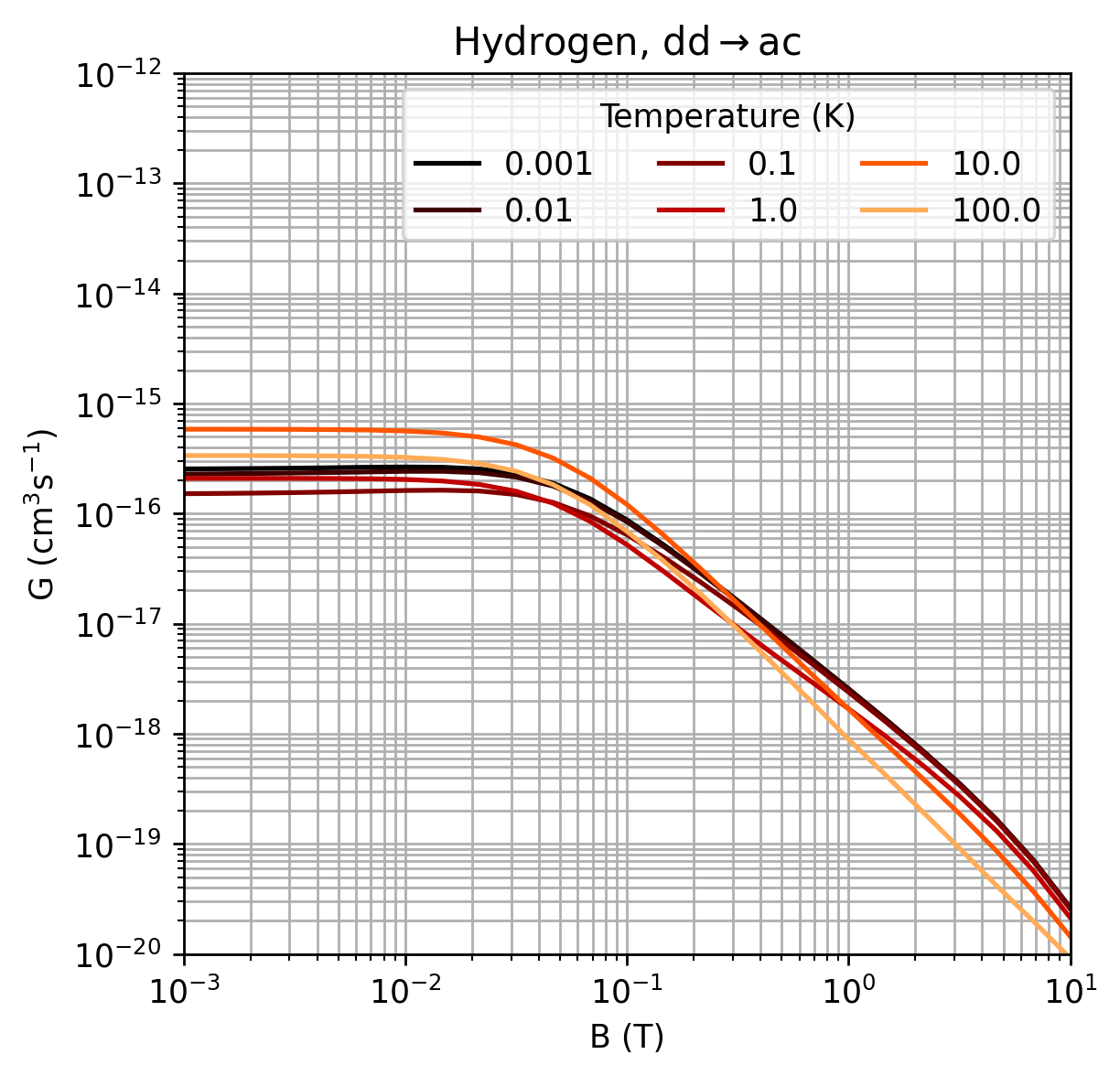}\includegraphics[width=0.2\linewidth]{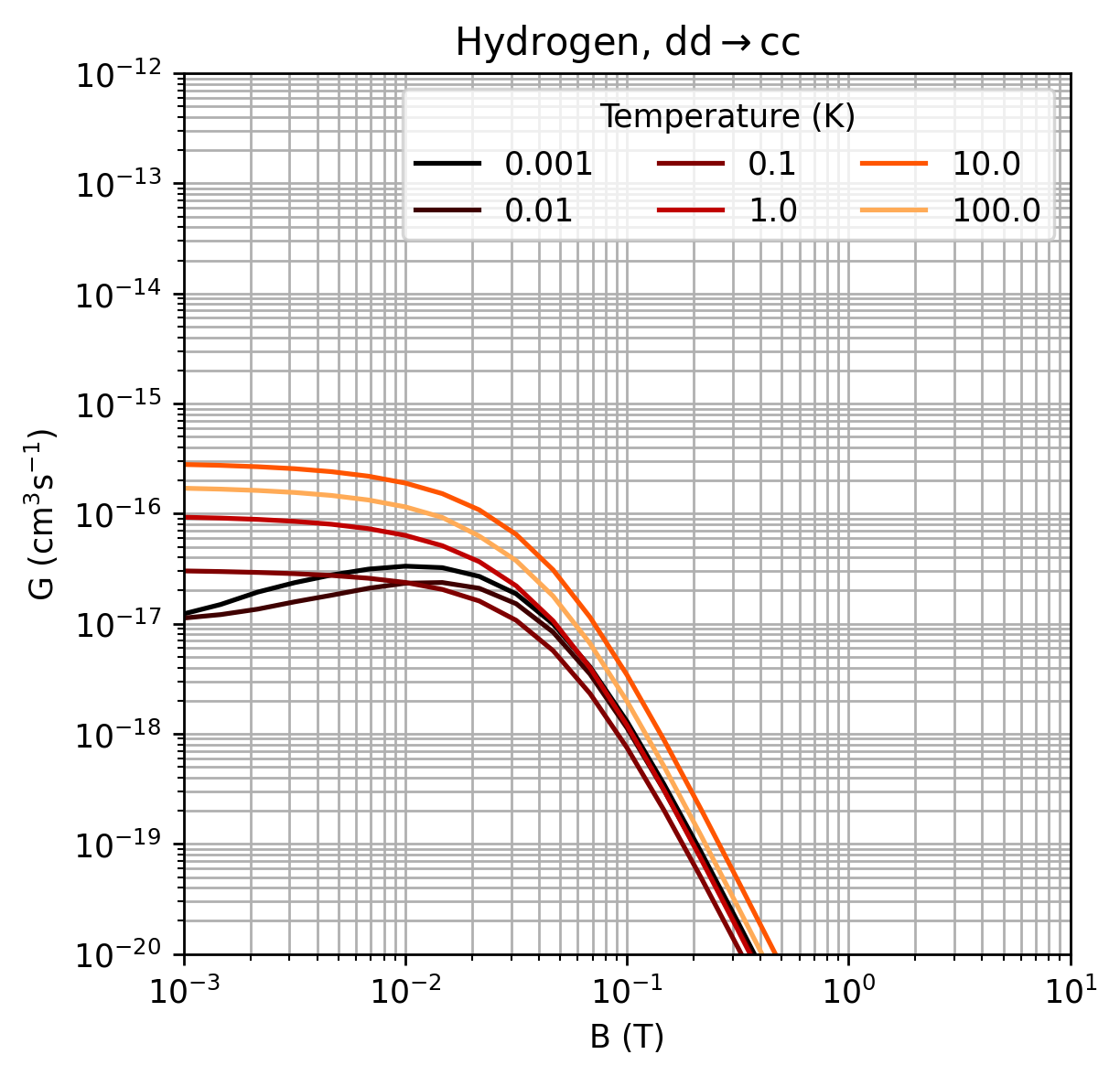}\includegraphics[width=0.2\linewidth]{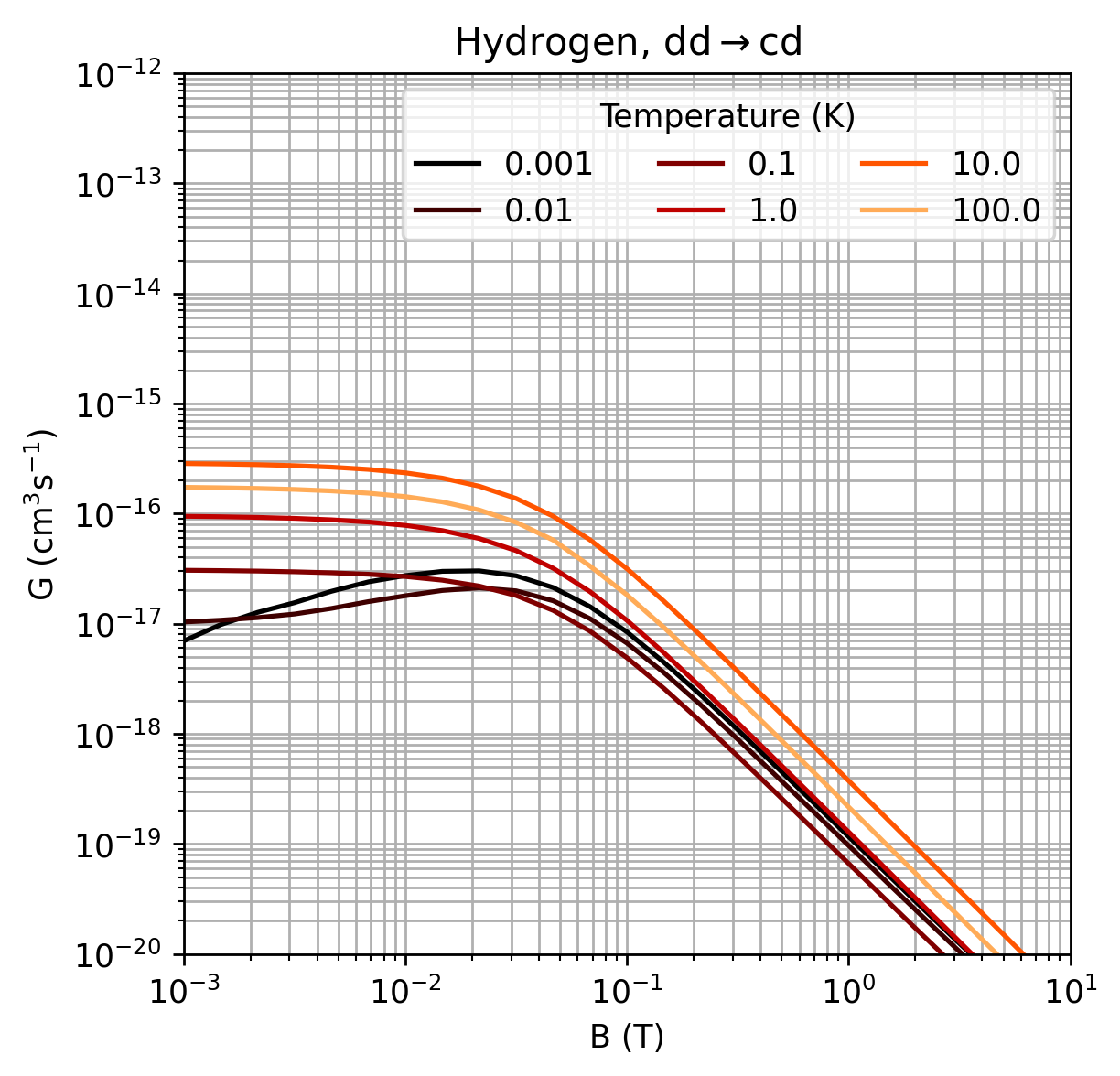}
    
    \includegraphics[width=0.2\linewidth]{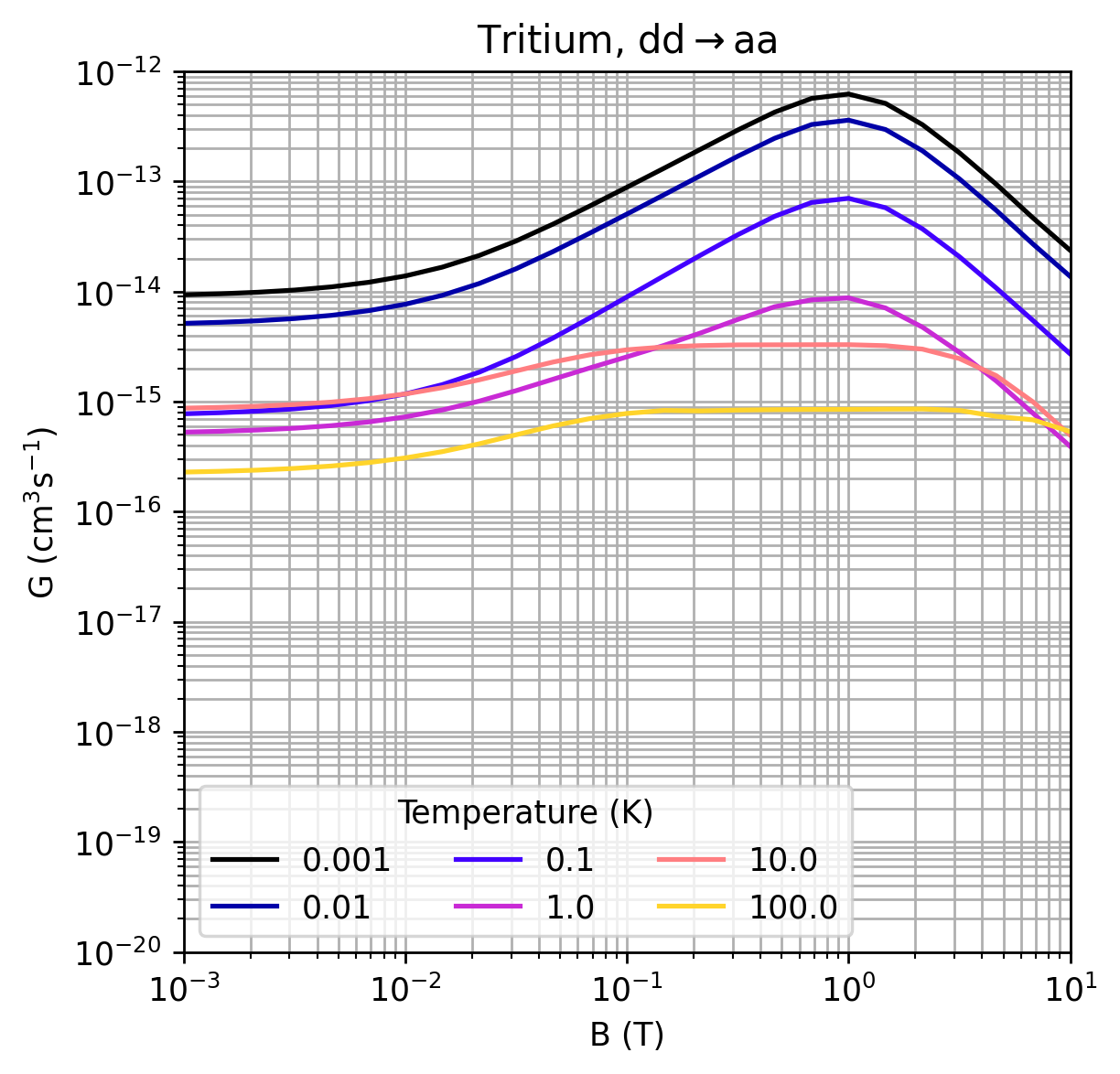}\includegraphics[width=0.2\linewidth]{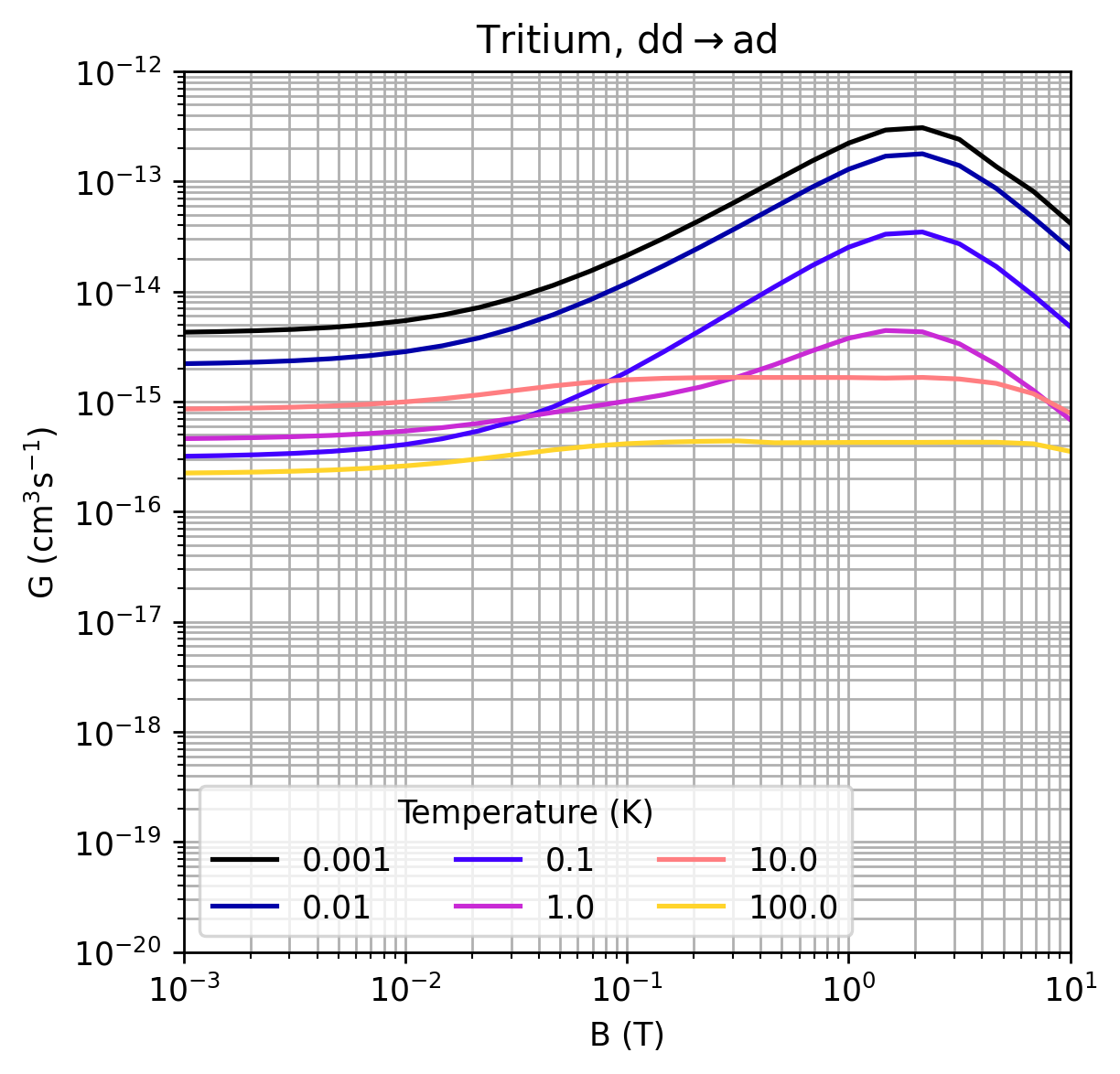}\includegraphics[width=0.2\linewidth]{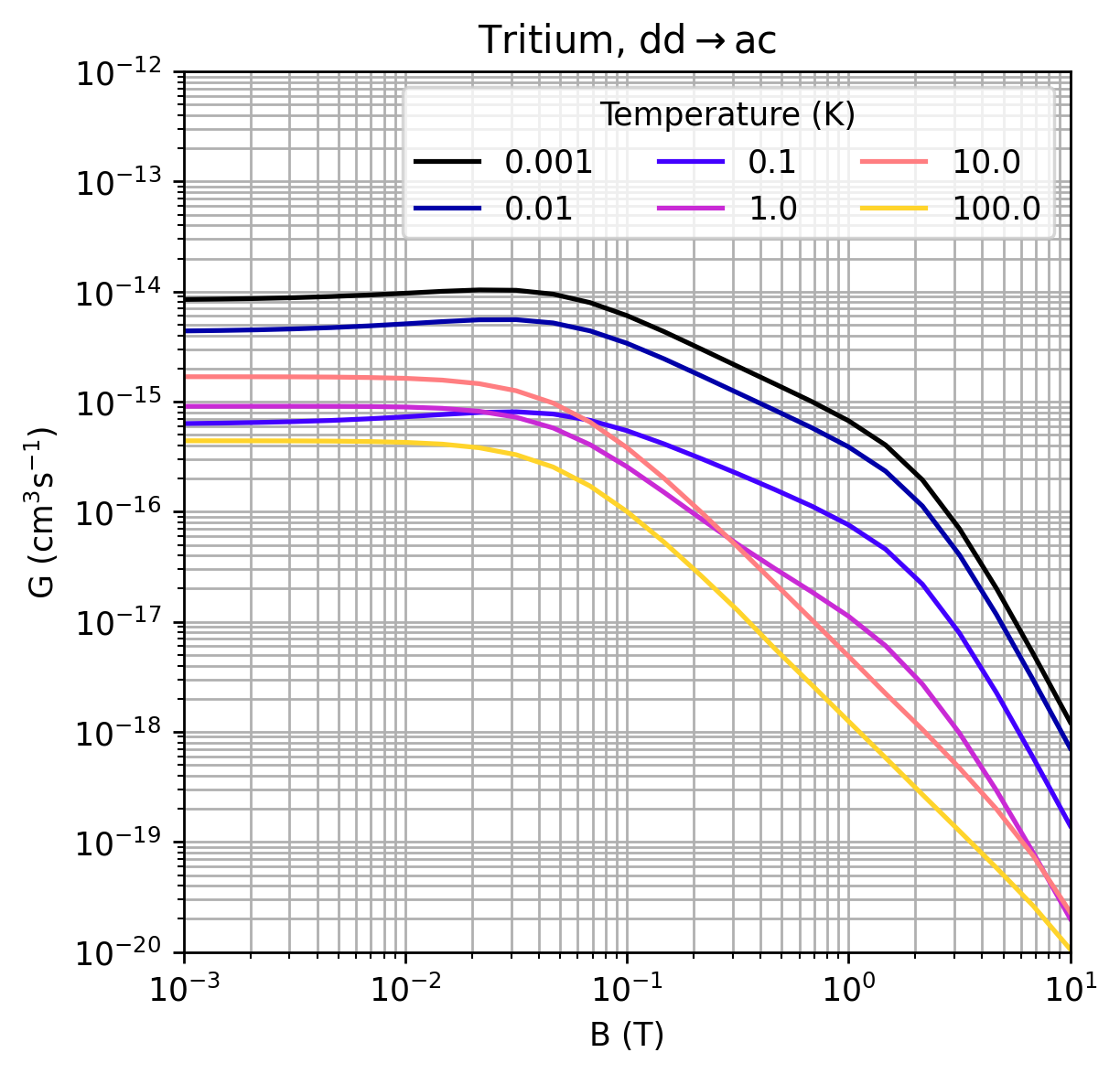}\includegraphics[width=0.2\linewidth]{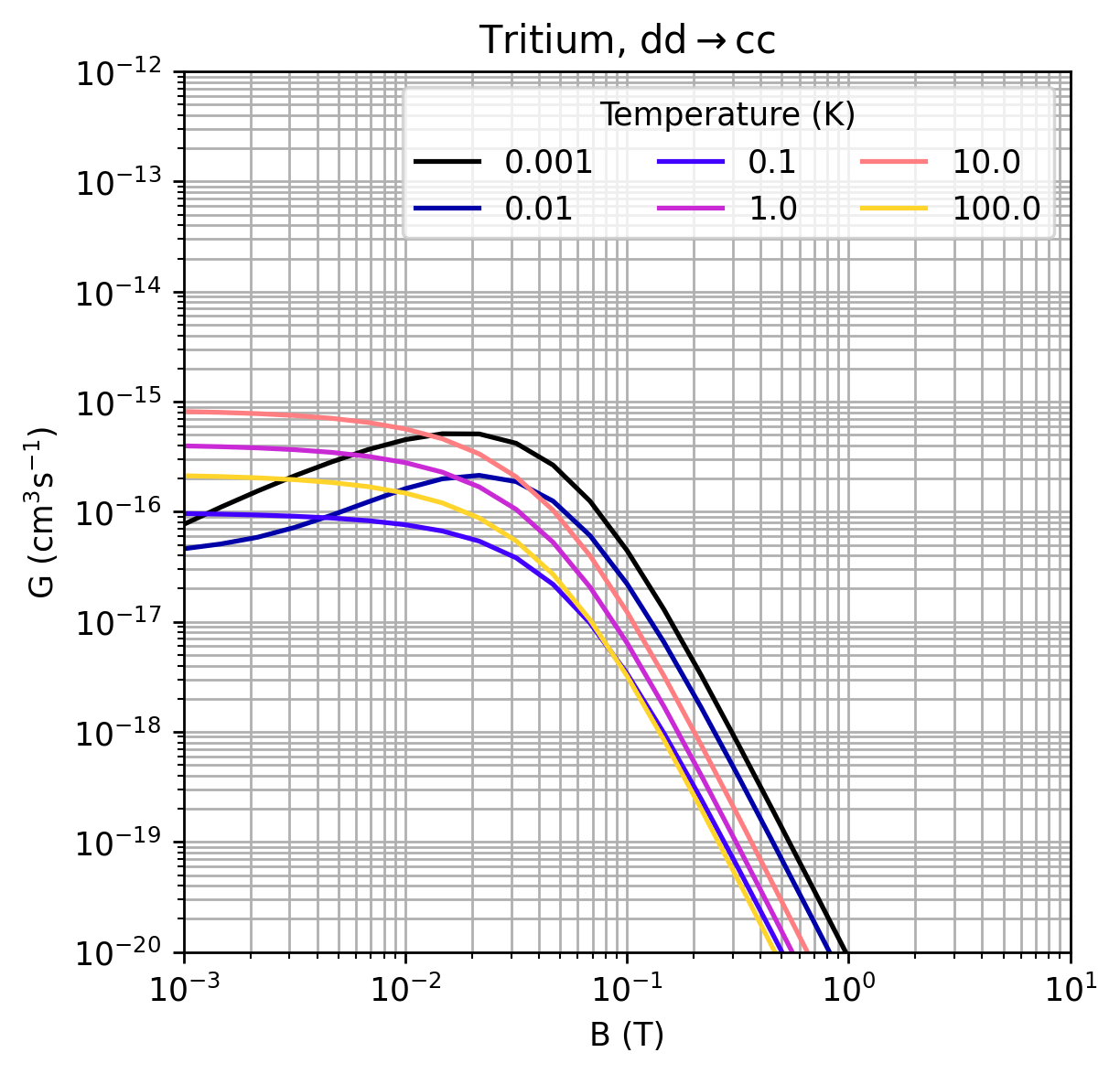}\includegraphics[width=0.2\linewidth]{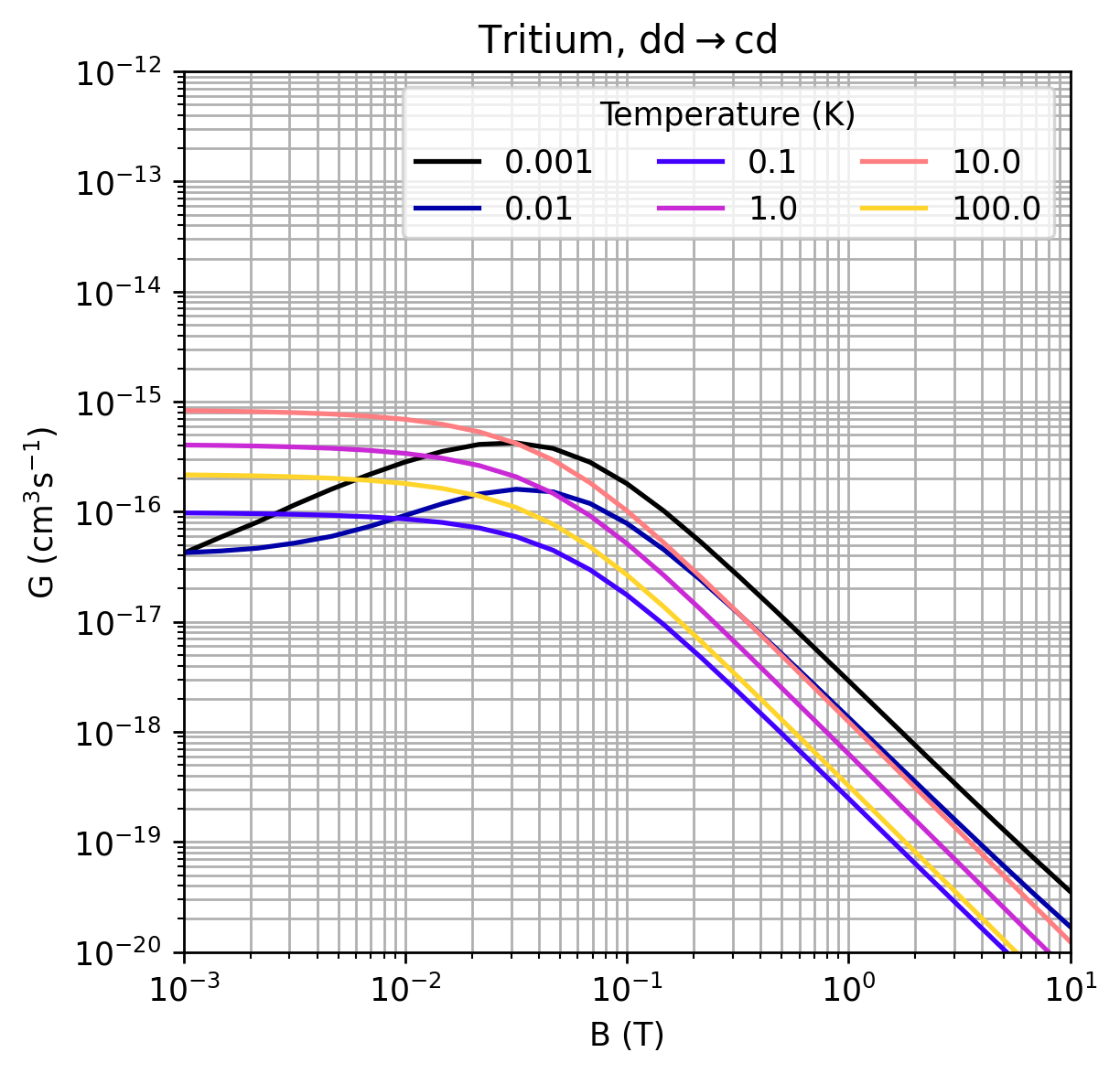}
        \caption{Dipole loss rate constants, all channels, all magnetic fields, all temperatures.}
    \label{fig:AllDipoleRates}
\end{figure*}

\begin{figure*}[h]
    \centering
    \includegraphics[width=0.3\linewidth]{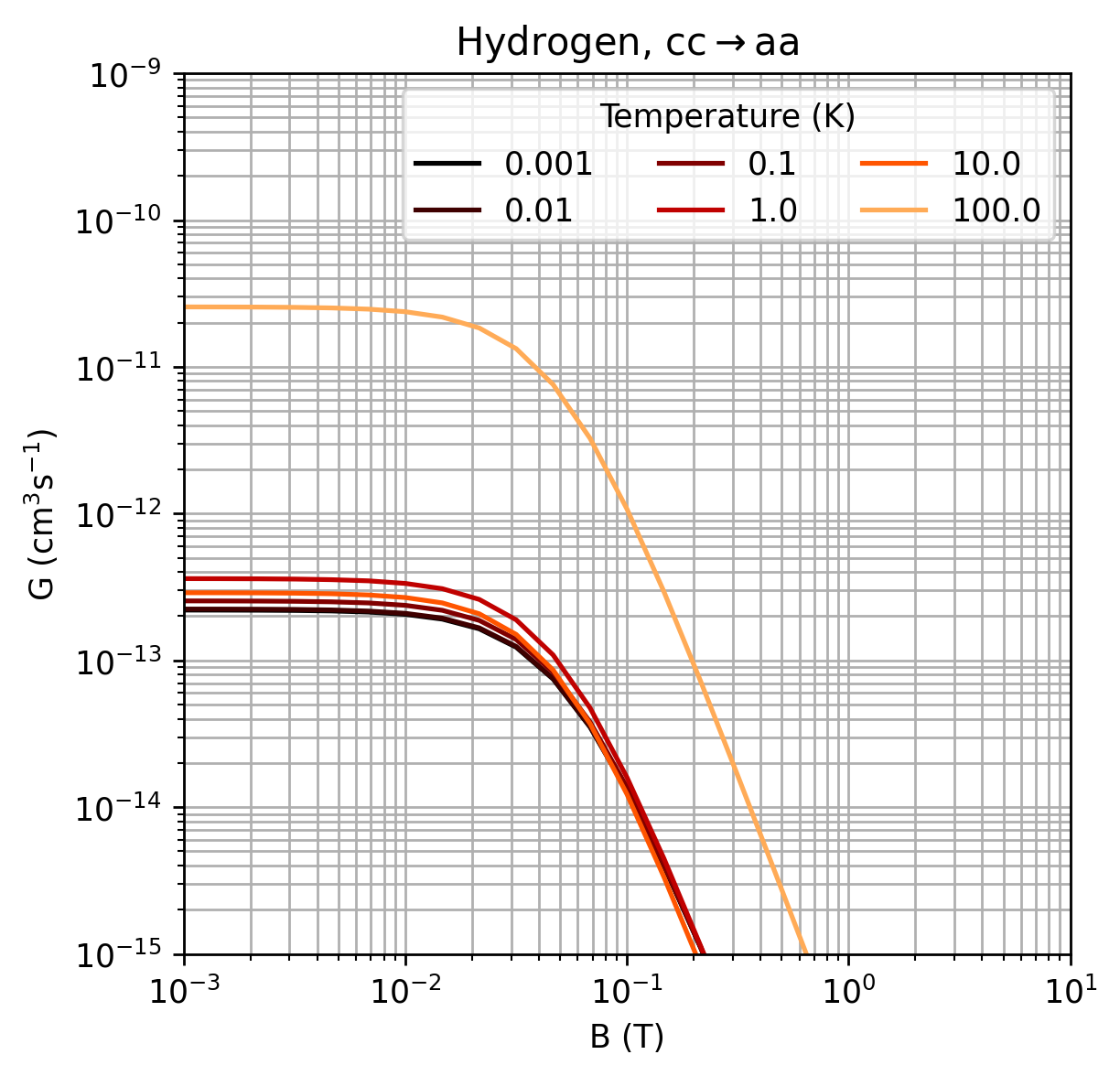}\includegraphics[width=0.3\linewidth]{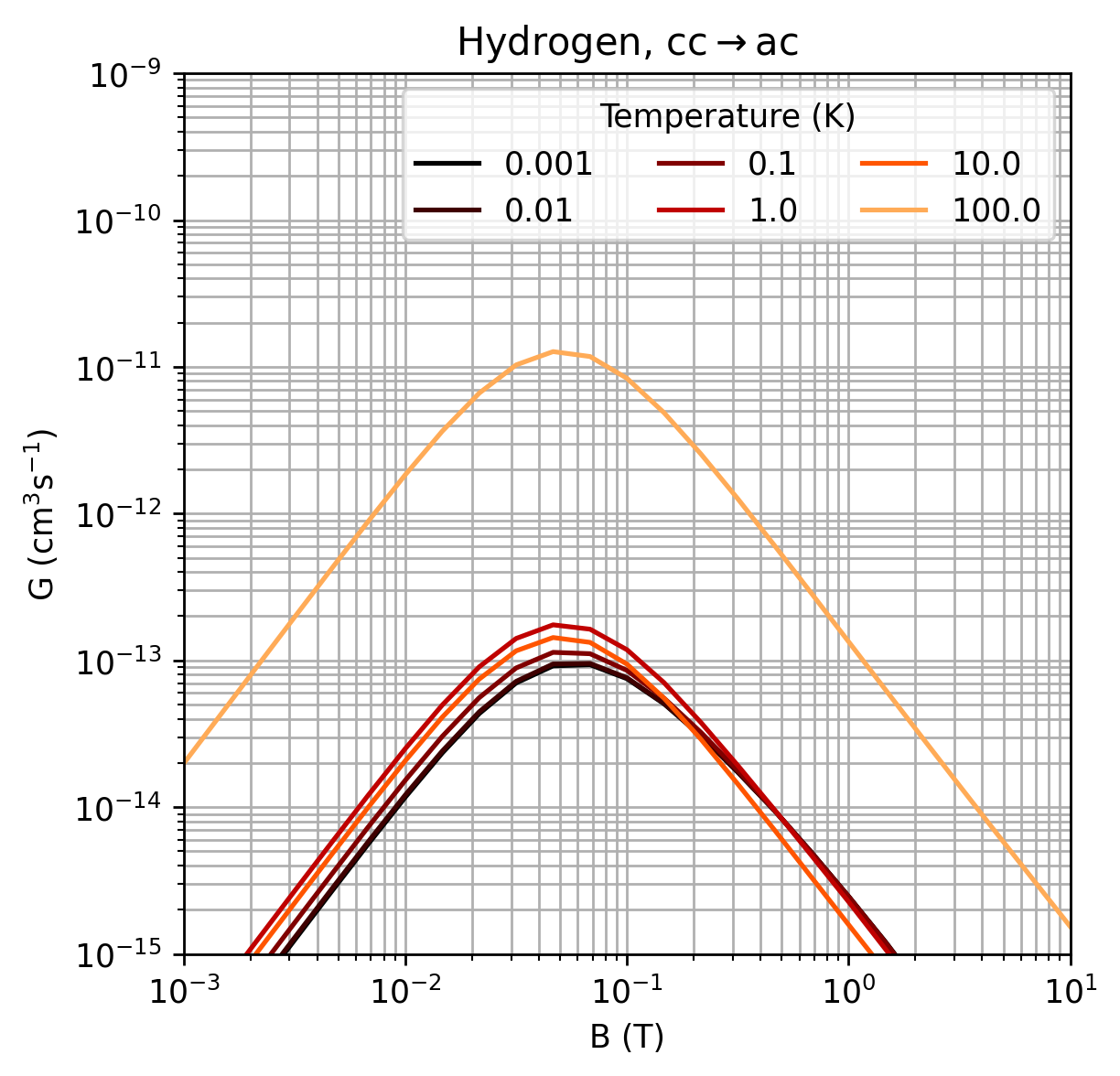}\includegraphics[width=0.3\linewidth]{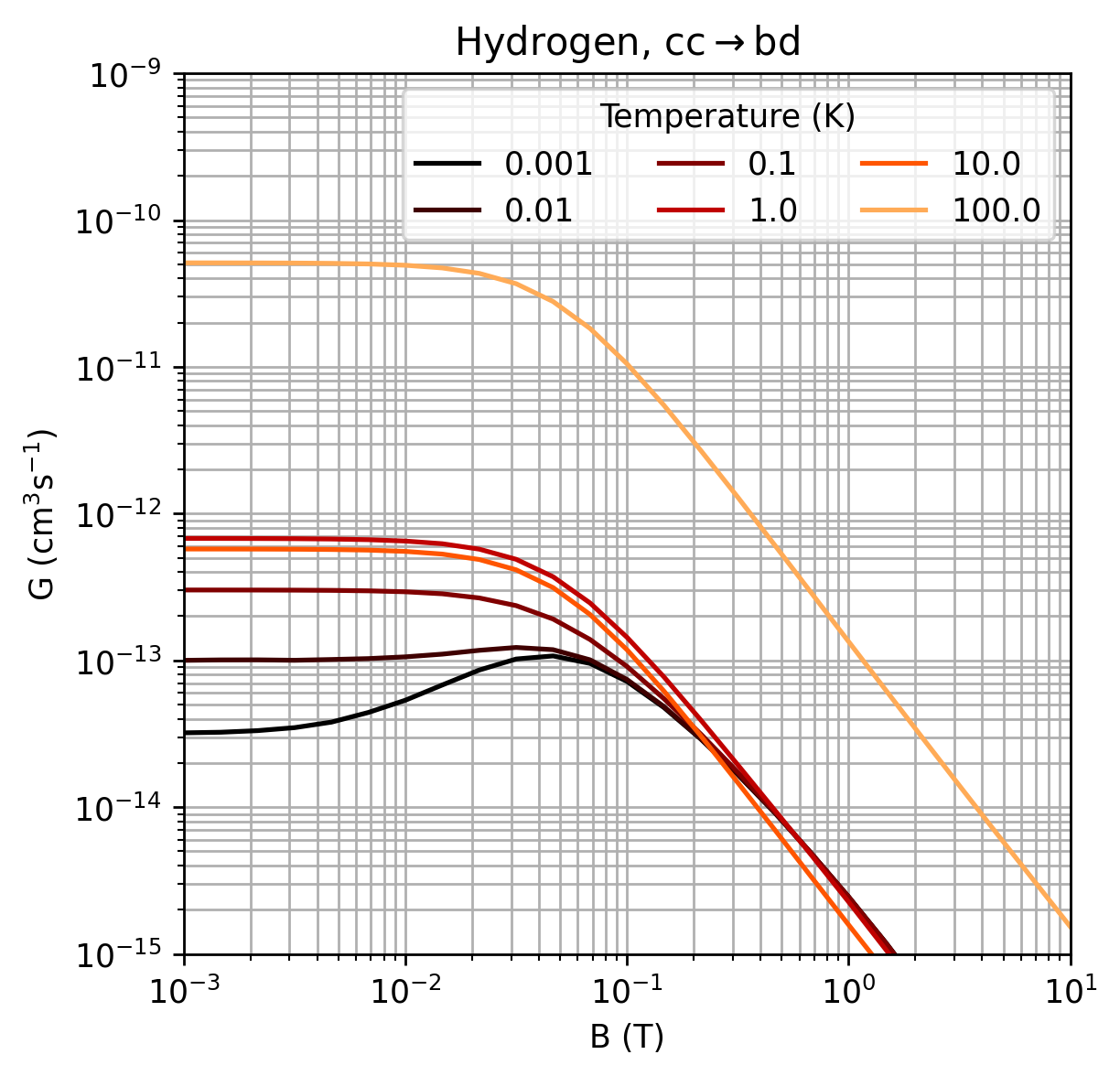}
    
    \includegraphics[width=0.3\linewidth]{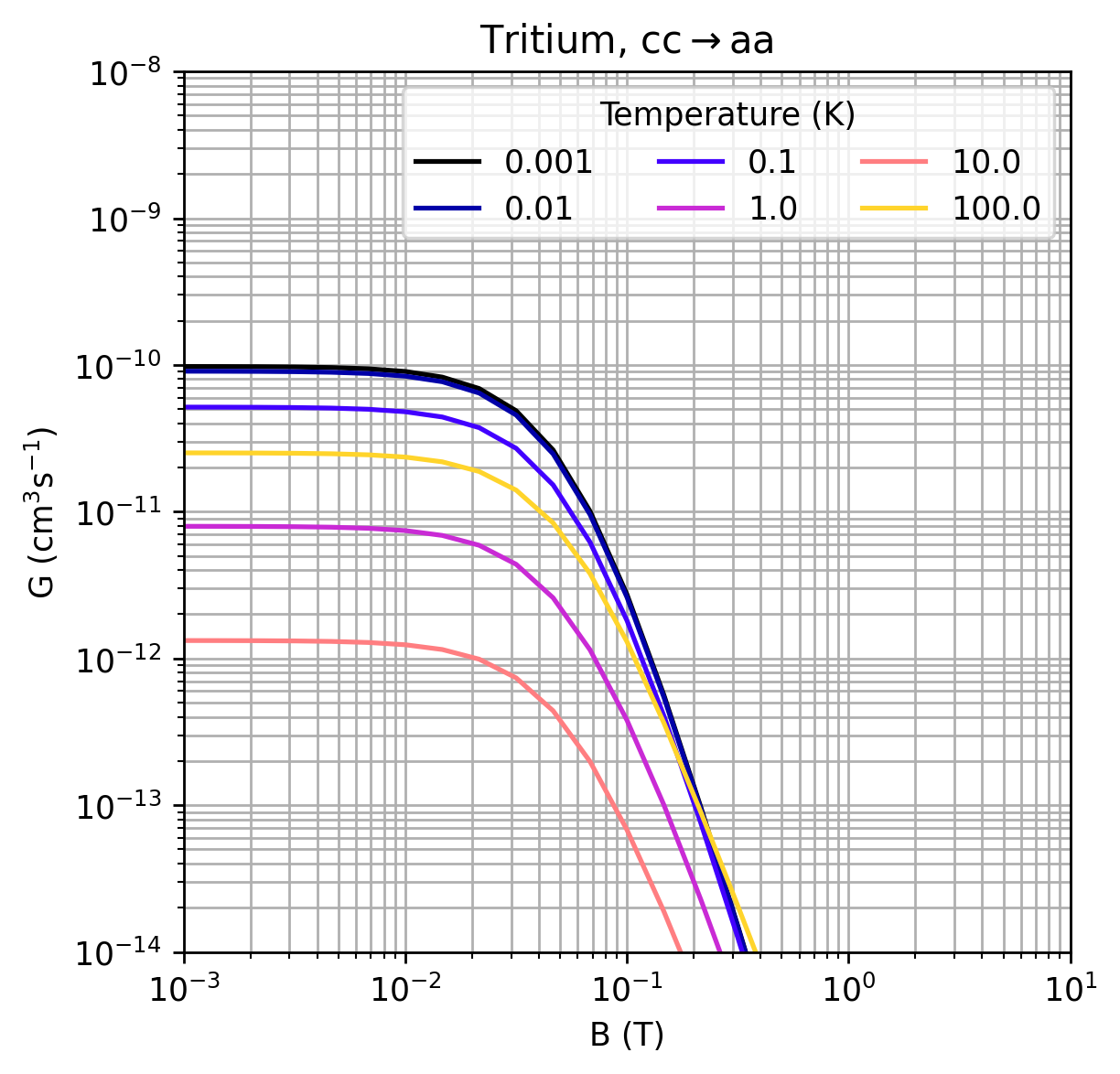}\includegraphics[width=0.3\linewidth]{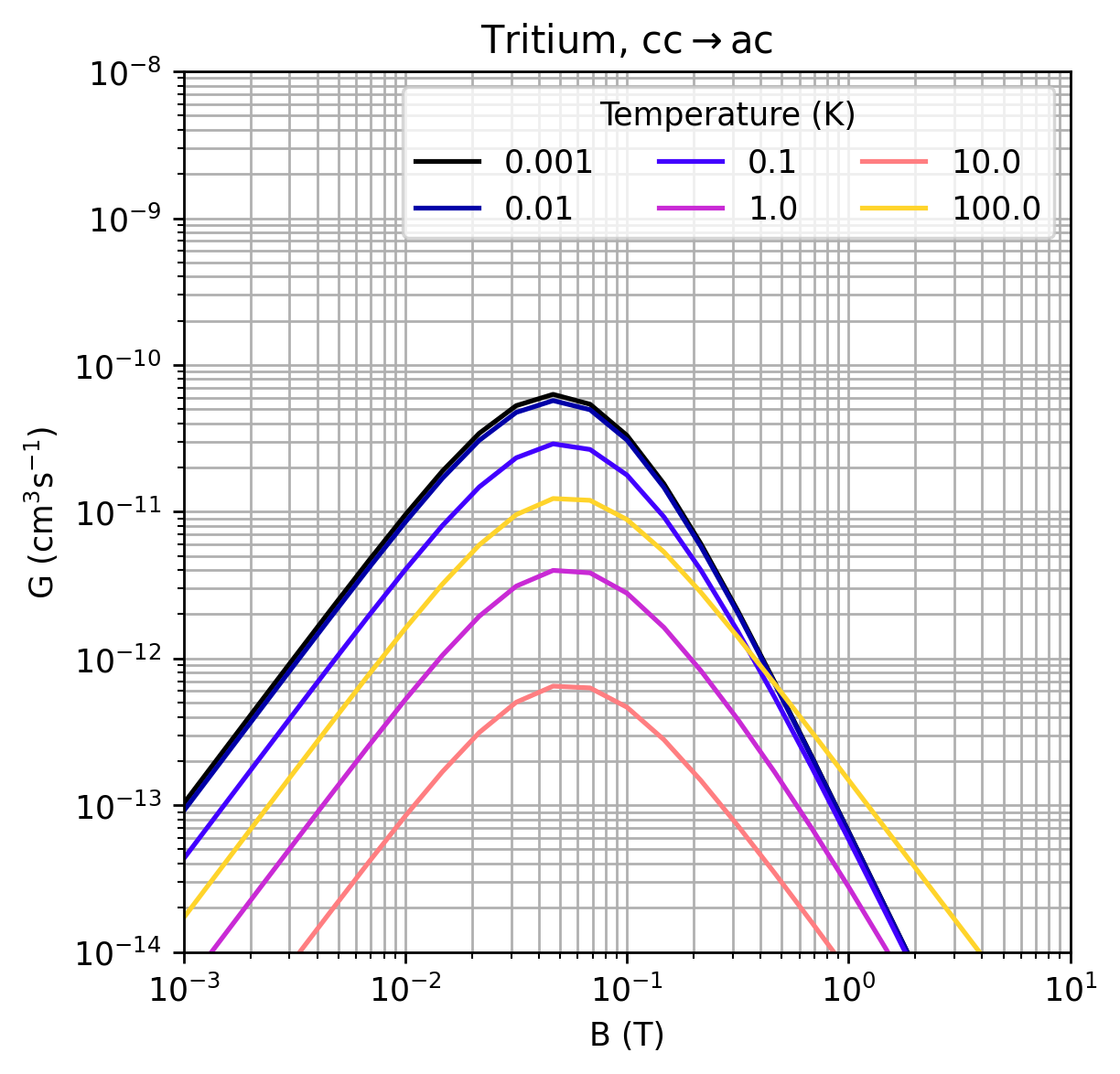}\includegraphics[width=0.3\linewidth]{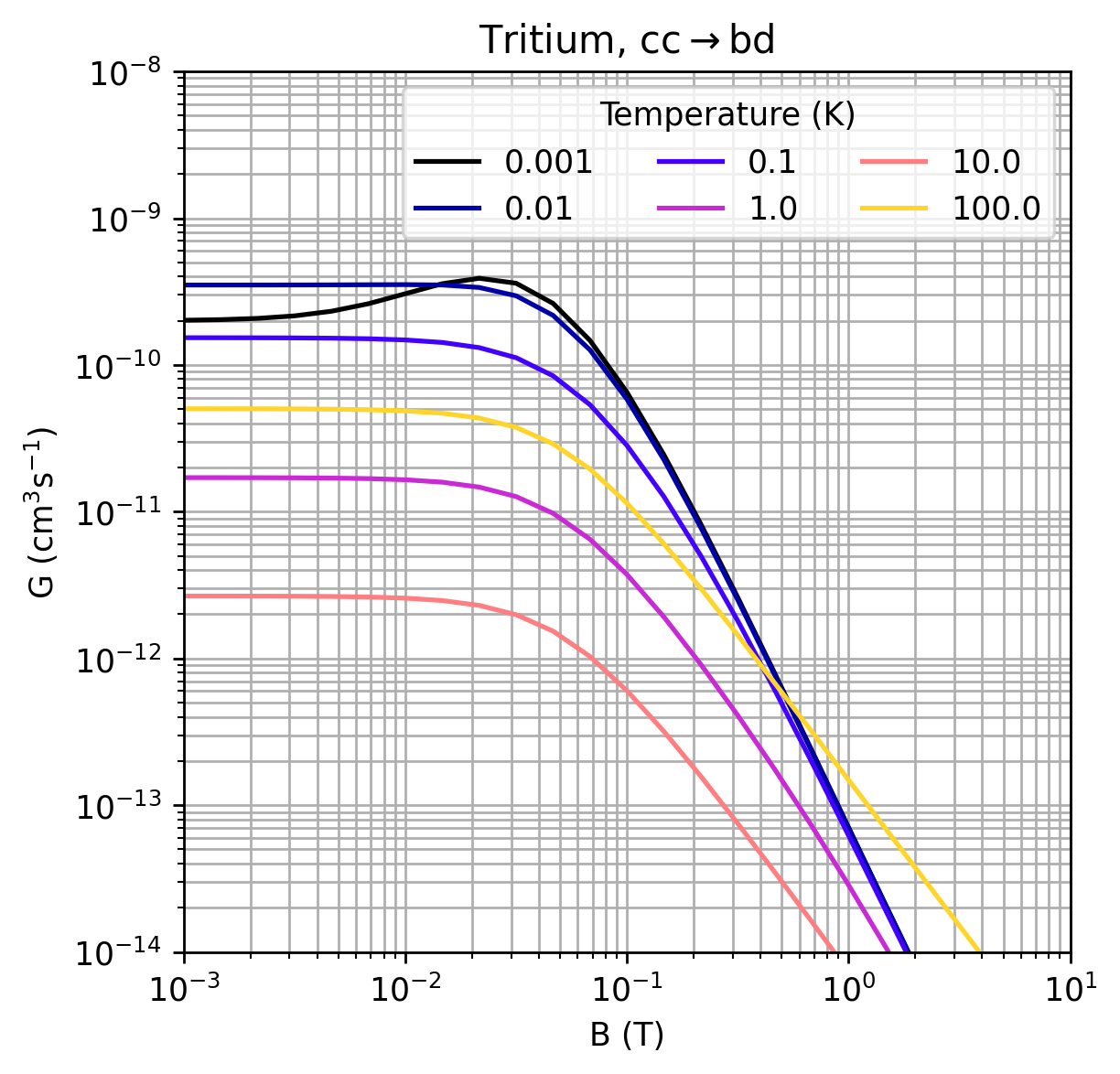}    \caption{Spin exchange loss rate constants, all channels, all B field, all temperatures.}
    \label{fig:AllSpinExRates}
\end{figure*}

\end{document}